\documentclass[table]{article}
\usepackage{booktabs}
\usepackage{ragged2e}
\usepackage{pgfplotstable}

\newcolumntype{L}[1]{>{\RaggedRight\arraybackslash}p{#1}}
\newcolumntype{R}[1]{>{\RaggedLeft\arraybackslash}p{#1}}
\newcolumntype{C}[1]{>{\centering\arraybackslash}p{#1}}

\def\pgfmathprintpmnumber#1#2#3#4{
    \pgfmathfloatparsenumber{\thisrow{#1}}
    \let\valueRe=\pgfmathresult
    \pgfmathfloatparsenumber{\thisrow{#2}}
    \let\valueIm=\pgfmathresult
    \edef\valueRe{\noexpand\pgfmathprintnumber[fixed zerofill, precision=#3]{\valueRe}}
    \edef\valueIm{\noexpand\pgfmathprintnumber[fixed zerofill, precision=#4]{\valueIm}}
    \toks0=\expandafter{\valueRe}
    \toks1=\expandafter{\valueIm}
	\edef\value{\the\toks0\,$\pm$\,\the\toks1}
}
\usepackage{arxiv}
\usepackage{graphicx}
\usepackage{subfigure}
\usepackage{icomma}
\usepackage{tikz}
\usepackage{amsmath}
\usepackage{amsfonts}
\usepackage{amsthm}
\usepackage[numbers]{natbib}
\usepackage{hyperref}
\usepackage{afterpage}
\usepackage{longtable}
\usepackage{xcolor}
\usepackage{caption}
\usetikzlibrary{shapes.geometric}
\graphicspath{{figs/}}

\newtheorem{definition}{Definition}

\title{Analysing Factorizations of Action-Value Networks for Cooperative Multi-Agent Reinforcement Learning}
\author{Jacopo Castellini\\
Dept. of Computer Science\\
University of Liverpool\\
\texttt{J.Castellini@liverpool.ac.uk}\\
\And
Frans A. Oliehoek\\
Interactive Intelligence Group\\
Delft University of Technology\\
\texttt{F.A.Oliehoek@tudelft.nl}\\
\And
Rahul Savani\\
Dept. of Computer Science\\
University of Liverpool\\
\texttt{rahul.savani@liverpool.ac.uk}\\
\And
Shimon Whiteson\\
Dept. of Computer Science\\
University of Oxford\\
\texttt{shimon.whiteson@cs.ox.ac.uk}}
\date{}

\begin{document}
\maketitle

\begin{abstract}
Recent years have seen the application of deep reinforcement learning techniques to cooperative multi-agent systems, with great empirical success. However, given the lack of theoretical insight, it remains unclear what the employed neural networks are learning, or how we should enhance their learning power to address the problems on which they fail. In this work, we empirically investigate the learning power of various network architectures on a series of one-shot games. Despite their simplicity, these games capture many of the crucial problems that arise in the multi-agent setting, such as an exponential number of joint actions or the lack of an explicit coordination mechanism. Our results extend those in \cite{ea} and quantify how well various approaches can represent the requisite value functions, and help us identify the reasons that can impede good performance, like sparsity of the values or too tight coordination requirements.
\end{abstract}


\section{Introduction}
Multi-agent reinforcement learning (MARL) uses reinforcement learning to train multiple agents for such systems, and can lead to flexible and robust solutions \cite{traffic,dilemmas,sensors,rubbish,marl}. In recent years, a variety of \emph{deep} MARL approaches have been developed and successfully applied \cite{madrl,communicate,maac}. While these approaches have shown good results, there is a general lack of theoretical insight, and often it remains unclear what the neural networks used by these approaches are learning, or how we should enhance their learning power to address the problems on which they fail.

Single-agent value-based reinforcement learning methods use (deep) neural networks to represent the discrete action-value function $Q(s,a;\theta)$ to select actions directly \cite{dqn} or as a `critic' in an actor-critic scheme \cite{a3c,ddpg}. A straightforward way to extend such methods to the multi-agent setting is by simply replacing the action by the joint action $\langle a_1,\dots ,a_n\rangle$ of all agents $Q(s,\langle a_1,\dots ,a_n\rangle;\theta)$. However, this approach heavily relies on the function approximation abilities of the neural network, since it must generalize across a discrete action space whose size is exponential in the number of agents. Moreover, selecting a joint action that maximizes the $Q$-function usually requires that, as in deep $Q$-networks \cite{dqn}, the (now joint) actions are output nodes of the network. As a result, the computational and sample costs scale poorly in the number of agents.

Another approach to extend single-agent reinforcement learning methods to multi-agent systems is to apply them to each agent independently \cite{independent}. This improves scalability at the expense of quality, e.g., independent deep $Q$-learners may not be able to accurately represent the value of coordination, as every agent learns on its own and ignores the others. Furthermore, the environment becomes non-stationary from the perspective of a single agent due to the other agents' simultaneous learning and thus their learning process may not converge \cite{dynamics,independent,classes}.

A middle ground is to learn a \emph{factored $Q$-value function} \cite{local,factored}, which represents the joint value but decomposes it as the sum of a number of local components, each involving only a subset of the agents. Compared to independent learning, a factored approach can better represent the value of coordination and eases the non-stationarity problem introduced by the other agents' changing policies. Compared to a naive joint approach, it has better scalability in the number of agents, thus reducing the number of samples required to learn a correct value function. Recently, factored approaches have shown success in deep MARL \cite{vdn,qmix,qtran}, although an in-depth analysis of the learned representations is not provided and the methods are mainly restricted to agent-wise factorizations.

In this paper, we focus on centralized learning of value-based MARL approaches for cooperative multi-agent systems. Although the usual focus of MARL is on decentralized execution \cite{vdn,qmix,coma,maven}, proper centralized learning is still useful in cases in which centralized execution is available or as a mean to inform decentralized agents under the centralized training-decentralized execution (CTDE) learning framework \cite{ctde}, for example when learning a centralized critic to inform decentralized policies \cite{coma} or in methods that involve bootstrapping \cite{qlearning} or message passing schemes \cite{maxsum}. We critically extend the work in \cite{ea} in examining the learning capacity of these various approaches by studying the accuracy of the learned $Q$-function approximations $\hat{Q}$. Our main contribution consists of a wide set of diverse experiments across different axes, and an in-depth analysis of the results to assess the potential benefits of these methods, as well as their possible drawbacks. Amongst the others, we aim at investigating the following points:

\begin{itemize}
\item How do factored methods compare to baseline algorithms,
\item Impact of factors size on the learned representations,
\item Scalability to larger systems,
\item Sample efficiency,
\item Effects of using an exploratory policy.
\end{itemize}

Furthermore, we extend the scope of this investigation beyond agent-wise factorizations usually investigated by deep MARL methods, showing the great advantages these ``higher-order'' factorization can bring both in terms of learning accuracy and sample efficiency, even when only few agents are comprised into each factor. To minimise confounding factors, we focus on one-shot (i.e., non-sequential) problems \cite{gt} and a stationary uniform sampling of the actions. Specifically, we investigate the learning power of various network architectures on a series of one-shot games that require a high level of coordination. Some of these games have an underlying factored structure (that we do not assume to be known in advance) and some do not. Despite their simplicity, these games capture many of the crucial problems that arise in the multi-agent setting, such as an exponential number of joint actions. As our results show, factored methods prove extremely effective on a variety of such games, achieving correct reconstruction even on those games that do not present a true underlying factored structure, and outperforms both independent learners and joint approaches in terms of learning speed. These benefits are even more apparent when the size of such systems grows larger, and a completely centralized solution proves impractical or even infeasible. Thus, an empirical evaluation to assess the accuracy of various representations in one-shot problems is key to understanding and improving deep MARL techniques, and our takeaways can help the community in taking informed decisions when developing solutions for multi-agent systems. We also discuss additional links to standard sequential MARL in Section \ref{sec:discussion}, as well as depicting some possible future directions to further clarify our understanding of action-value functions in this setting.

\section{Background}
In the following we provide some basilar notions required to understand the remainder of this work.

\subsection{One-Shot Games}
\label{sec:back}
\begin{definition}
\emph{One-shot games}: a one-shot game \cite{gt} consists of the tuple $$\mathcal{M}=\langle D,\{A_i\}_{i=1}^n,\{Q_i\}_{i=1}^n\rangle,$$ where $D=\{1,\ldots,n\}$ is the set of agents, $A_i$ is the set of actions for agent $i$, and $Q_i$ is the reward function for agent $i$ that depends only on the joint action $a\in A = \times_{i=1}^nA_i$ performed by the full team of agents, which expresses how much reward agent $i$ gets from the overall team decision\footnote{We write $Q_i(a)$ for the reward function in the one-shot problem to make the link with sequential MARL more apparent.}.
\end{definition}

\begin{definition}
\emph{Cooperative one-shot game}: a cooperative one-shot game is a game in which all agents share the same reward function $Q(a)$, so that the goal of the team is to maximize this shared reward by finding the optimal joint action $a\in A$ to perform.
\end{definition}

In this work, we focus on cooperative games. Our work aims at investigating the representations of the action-value function obtained with various neural network approaches and how close these are to the original one. Although we do not explicitly rely on any of the cooperative properties of these settings, and thus we could in principle extend our analysis to cooperative and mixed scenarios as well, we think that modelling agents with opposing or mixed interests into the same factor component would not make sense from a logical perspective, as these agents may not gain any benefit from sharing their own information locally with the other agents into the same factor.

\bigskip
\noindent\textbf{Problem Statement:} Given the original action-value function $Q(a)$ and a learned representation $\hat{Q}(a)$, we are interested in investigating the quality of this learned representation, both in terms of action ranking, i.e., $$\sigma(\mathfrak{R}(Q),\mathfrak{R}(\hat{Q})),$$ where $\sigma$ is a similarity measure and $\mathfrak{R}$ is a partial ordering of the joint actions according to their action-values, so that the learned function can reliably be used for decision making; and in terms of reconstruction error of the representation, computed using the mean squared error (MSE): $$MSE=\frac{1}{|A|}\sum_{a\in A}(Q(a)-\hat{Q}(a))^2.$$

A related, but different, setting is that of repeated games \cite{gt}, in which agents repeatedly play a
one-shot game over time, and can condition their strategies on the history of play in earlier rounds. In this work, we do \emph{not} investigate learning strategies for repeated play, but only for the one-shot game in isolation. Also related is the multi-agent bandit problem \cite{bandit}, in which a team of agents has to agree on which arm to choose in a classical multi-armed bandit problem. The main difference with our setting is that we are not interested in considering the regret during learning, but only in learning good approximations as close as possible to the original action-value function.

\subsection{Coordination Graphs}
In many problems, the decision of an agent is directly influenced by the decisions of only a small subset of other agents \cite{local}. The structure of the interactions between the agents can be represented with a (hyper-) graph called a \emph{coordination graph} \cite{graphs,maxsum}. A coordination graph has a node for each agent in the team and (hyper) edges $\mathcal{E}$ connecting agents in the same subset, called a \emph{factor}. Figure \ref{fig:graphs} shows some example coordination graphs. This locality of interaction means the joint action-value function $Q(a)$ can be represented as the sum of smaller reward functions, one for each factor $e\in\mathcal{E}$:
\begin{equation}
Q(a)=\sum_{e\in\mathcal{E}}Q_e(a_e),
\end{equation}
where $\mathcal{E}$ is the set of these factors and $a_e=\langle a_i\rangle_{i\in e}$ is the \emph{local joint action} of the agents that participate in factor $e$. Coordination graphs are a useful instrument to represent interactions between agents and many algorithms exploit such structure and require good approximations of the action-value function in order to efficiently select a maximizing joint action, e.g., \emph{variable elimination} \cite{local} or \emph{max-sum} \cite{maxsum,decentralized}.

However, there are many cases in which the problem itself is not factored, or the factorization is not known in advance and thus cannot be exploited. In these cases, however, it can still be useful to resort to an approximate factorization \cite{factored}:
\begin{equation}
Q(a)\approx\hat{Q}(a)=\sum_e\hat{Q}_e(a_e),
\end{equation}
obtained by decomposing the original function into a number of local approximate terms $\hat{Q}_e(a_e)$, thus approximating the original action-value function $Q(a)$.

\section{Investigated Action-Value Factorizations}
\label{sec:methods}
Most current value-based deep MARL approaches (of which a noticeable exception is \cite{dcg}) are either based on the assumption that the joint-action value function $Q(s,a)$ can be represented efficiently by a single neural network (when, in fact, the exponential number of joint actions can certainly make a good approximation hard to learn), or that it suffices to represent (approximated) individual action values $Q_i(s_i,a_i)$ \cite{survey}. Our aim is to investigate to what degree these assumptions are valid by exploring them in the one-shot case, as well as assessing if higher-order factorizations are an improved representations of such functions, while speeding learning (as only small factors need to be learned). When a problem presents an underlying factored structure, knowing such structure beforehand and being able to exploit it properly can be of the greatest benefit both in terms of learning speed and accuracy, but we argue that resorting to an approximate factorization can still be beneficial in many cases.

We use neural networks as function approximators to represent the various components of these factorizations. In our study, we vary two distinct aspects of the problem. Firstly, we study two learning algorithms, which we describe in Section \ref{sec:rules}. Secondly, we study different coordination graph structures, which capture how the team of agents are modelled, presented in Section \ref{sec:cg}. Finally, the full set of investigated games is presented in Section \ref{sec:games}.

\afterpage{
\begin{figure}[t!]
\centering
\subfigure[\label{sub:2r}]{
\resizebox{0.3\textwidth}{!}{
\begin{tikzpicture}[transform shape, font=\large]
\foreach \x in {1,...,6}{
\pgfmathparse{-(\x-1)*60-180}
\node[draw,circle,inner sep=0.2cm] (N-\x) at (\pgfmathresult:2cm) {\x};
}
\path[-,draw,thick] (N-1) edge[bend right=3] node[fill=white, anchor=center, pos=0.5] {$e_0$} (N-2);
\path[-,draw,thick] (N-3) edge[bend right=3] node[fill=white, anchor=center, pos=0.5] {$e_1$} (N-4);
\path[-,draw,thick] (N-5) edge[bend right=3] node[fill=white, anchor=center, pos=0.5] {$e_2$} (N-6);
\end{tikzpicture}
}
}
\hfill
\subfigure[\label{sub:2o}]{
\resizebox{0.3\textwidth}{!}{
\begin{tikzpicture}[transform shape, font=\large]
\foreach \x in {1,...,6}{
\pgfmathparse{-(\x-1)*60-180}
\node[draw,circle,inner sep=0.2cm] (N-\x) at (\pgfmathresult:2cm) {\x};
}
\path[-,draw,thick] (N-1) edge[bend right=3] node[fill=white, anchor=center, pos=0.35] {$e_0$} (N-4);
\path[-,draw,thick] (N-1) edge[bend right=3] node[fill=white, anchor=center, pos=0.35] {$e_1$} (N-5);
\path[-,draw,thick] (N-1) edge[bend right=3] node[fill=white, anchor=center, pos=0.35] {$e_2$} (N-6);
\path[-,draw,thick] (N-2) edge[bend right=3] node[fill=white, anchor=center, pos=0.35] {$e_3$} (N-3);
\path[-,draw,thick] (N-2) edge[bend right=3] node[fill=white, anchor=center, pos=0.35] {$e_4$} (N-5);
\path[-,draw,thick] (N-3) edge[bend right=3] node[fill=white, anchor=center, pos=0.35] {$e_5$} (N-4);
\end{tikzpicture}
}
}
\hfill
\subfigure[\label{sub:2c}]{
\resizebox{0.3\textwidth}{!}{
\begin{tikzpicture}[transform shape, font=\large]
\foreach \x in {1,...,6}{
\pgfmathparse{-(\x-1)*60-180}
\node[draw,circle,inner sep=0.2cm] (N-\x) at (\pgfmathresult:2cm) {\x};
}
\newcount\ga
\foreach \x [count=\xi from 1] in {2,...,6}{
\foreach \y in {\x,...,6}{
\path[-,draw,thick] (N-\xi) edge[bend right=3] node[fill=white, anchor=center, pos=0.35] {$e_{\the\ga}$} (N-\y);
\global\advance\ga by1;
}
}
\end{tikzpicture}
}
}
\caption{Example coordination graphs for: \protect\subref{sub:2r} random partition,
\protect\subref{sub:2o} overlapping factors, \protect\subref{sub:2c} complete factorization.}
\label{fig:graphs}
\end{figure}}

\subsection{Learning Algorithms}
\label{sec:rules}
Here we present the two different learning algorithms that are investigated in our experiments. We choose these two as these are highly related to many standard sequential MARL algorithms: the mixture of experts learning rule follow the same idea of standard independent learning approach used by early works \cite{independent}, while the factored $Q$-function rule uses a joint optimization process resembling that of value decomposition networks \cite{vdn}, but it is also similar to the QMIX algorithm \cite{qmix}, but uses a linear constant mixing rather than an additional mixing network.

\begin{itemize}
\item\emph{Mixture of experts} \cite{experts}: each factor network optimizes its own output $\hat{Q}_e$ individually to predict the global reward, thus becoming an ``expert'' on its own field of action. The loss for the network representing factor $e\in\mathcal{E}$ at training step $t$ is defined as:

\begin{equation}
\mathcal{L}_t^e(a_t^e)=\frac{1}{2}\left(Q(a_t)-\hat{Q}_e(a_t^e)\right)^2,
\label{eq:moe_train}
\end{equation}

where $Q(a_t)$ is the common reward signal received after selecting joint action $a_t$ and $\hat{Q}_e(a_t^e)$ is the output of the network for local joint action $a_t^e$. As we aim to assess how good the approximate action-value function $\hat{Q}$ is, \emph{after} training we compute the reconstruction obtained from the factors as the mean over the appropriate local $Q$-values (the ``opinion'' of each expert is weighted equally):

\begin{equation}
\hat{Q}(a)=\frac{1}{|\mathcal{E}|}\sum_{e\in\mathcal{E}}\hat{Q}_e(a_e)\quad\forall a\in A.
\label{eq:moe_rec}
\end{equation}

\item\emph{Factored $Q$-function} \cite{coordinated,vdn}: the algorithm jointly optimizes the factor networks to predict the global reward as a sum of their local $Q$-values $\hat{Q}_e$. The loss for the experience at time $t$ is identical for all factor networks:

\begin{equation}
\mathcal{L}_t(a_t)=\frac{1}{2}\left(Q(a_t)-\sum_{e\in\mathcal{E}}\hat{Q}_e(a_t^e)\right)^2.
\label{eq:fqf_train}
\end{equation}

Rather than learning proper action-value functions, the optimization problem in Equation \ref{eq:fqf_train} learns utility functions for each factor, that do not really represent the values of actions on their own, while Equation \ref{eq:moe_train} learns individual $Q$-values for each factor. After learning, the approximate joint action-value function $\hat{Q}$ is reconstructed by summing the appropriate local $Q$-values (the components collectively reconstruct the approximation):

\begin{equation}
\hat{Q}(a)=\sum_{e\in\mathcal{E}}\hat{Q}_e(a_e)\quad\forall a\in A.
\label{eq:fqf_rec}
\end{equation}
\end{itemize}

\subsection{Coordination Graphs}\
\label{sec:cg}
We study four different coordination graphs. Their structures differ both in the number of components and the degree of connection for each agent. Our empirical study considers all eight combinations of the two learning rules described above and the four coordination graphs described below.

\begin{itemize}
\item\emph{Single agent decomposition}: each agent $i$ is represented by an individual neural network and computes its own individual action-values $\hat{Q}_i(a_i)$, based on its local action $a_i$. Under the mixture of experts learning rule, this corresponds to the standard independent $Q$-learning approach in MARL \cite{independent}, in which we learn local agent-wise components, while under the factored $Q$-function approach this corresponds to value decomposition networks (VDN) \cite{vdn}.
\item\emph{Random partition}: agents are randomly partitioned to form factors of size $f$, with each agent $i$ involved in only one factor.\footnote{If $f$ is not a divisor of $n$, the random partition factorization would partition into factors that are all of size close to $f$.} Each of the $|\mathcal{E}|=\frac{n}{f}$ factors has a different neural network that represents local action-values $\hat{Q}_e(a_e)$ for that factor.
\item\emph{Overlapping factors}: a fixed number of factors $|\mathcal{E}|$ is picked at random from the set of all possible factors of size $f$. We require the sampled set to not include duplicate factors (we use only distinct components) and that every agent $i$ appears in at least one factor. Every factor $e\in\mathcal{E}$ is represented by a different neural network learning local action-values $\hat{Q}_e(a_e)$ for the local joint action $a_e$. In our experiments we choose $|\mathcal{E}|=n$, to keep the number of networks comparable to that of the single agent decomposition.
\item\emph{Complete factorization}: each agent $i$ is grouped with every possible combination of the other agents in the team $D\setminus i$ to form factors of size $f$, resulting in $|\mathcal{E}|=\binom{n}{f}$ factors, each represented by a network. Each of these networks learns local action values $\hat{Q}_e(a_e)$.
\end{itemize}

A fundamental problem in MARL is that there is currently no method capable of predicting the accuracy of a factored representation on a certain problem in advance (the problem is equivalent to predicting the result of a linear regression problem with a given set of basis functions). Therefore, assessing the performance and eventual advantages of different structures and approaches is a fundamental step for MARL research, as it can further improve our understanding of these settings and existing algorithms. In our empirical study, we mainly consider factors of size ${f\in\{2,3\}}$. The small size of these factors allow us to effectively explore the improvements in the complexity of learning; if the size of each factor is similar to the size of the full team of agents, we would not expect significant improvement over a full joint learner in terms of sample complexity and scalability (although we also conduct some experiments on this in Section \ref{sec:size}).

\subsection{Investigated Games}
\label{sec:games}
We investigate the proposed methods on a number of cooperative one-shot games that require a high degree of coordination. Some of these games do not present an underlying factored structure, while others are truly factored games. For the latter, none of the methods exploit prior knowledge of their true factored structure (but we also report results for the true underlying factorization to show the possible benefits when that is known beforehand).

\subsubsection{Non-Factored Games}
\emph{Dispersion Games}: In the Dispersion Game, also known as Anti-Coordination Game, the team of agents must divide as evenly as possible between the two local actions that each agent can perform \cite{dispersion}. Think of a town with two different pubs: the inhabitants like both the same, but the two are quite small and cannot contain all the people in the town at once, so the customers have to split up across the two pubs in order to enjoy the situation and not overcrowd them. This game requires explicit coordination, as none of the local actions is good per se, but the obtained reward depends on the decision of the whole team. We investigate two versions of this game: in the first one the agents obtain reward proportional to their \emph{dispersion coefficient} (i.e., how split the agents are in performing one of their two local actions). The reward function $Q(a)$ for this game with $n$ agents, each with a local action set $A_i=\{a_0,a_1\}$ is:
\begin{equation}
Q(a)=n-\max\{\# a_0, \#a_1\}.
\end{equation}

In the second version, which we call Sparse Dispersion Game, the agents receive a reward (which we set to the maximum dispersion coefficient with $n$ agents: $\frac{n}{2}$) only if they are perfectly split:
\begin{equation}
Q(a)=\begin{cases}
\frac{n}{2} & \mbox{if }\# a_0=\# a_1, \\
0 & \mbox{otherwise.}
\end{cases}
\end{equation}

\emph{Platonia Dilemma}: In the Platonia Dilemma \cite{platonia}, an eccentric trillionaire gathers $20$ people together and tells them that if one and only one of them sends him a telegram by noon the next day, that person will receive a billion dollars. In our cooperative version the reward is set to the number of agents $n$ and is received by the whole team, not just a single agent. Thus, the reward function for $n$ agents with local action sets $A_i=\{send,idle\}$ is:
\begin{equation}
Q(a)=\begin{cases}
n & \mbox{if }\# send=1, \\
0 & \mbox{otherwise.}
\end{cases}
\end{equation}

\emph{Climb Game}: In the Climb Game \cite{games}, each agent has three local actions $A_i=\{a_0,a_1,a_2\}$. Action $a_0$ yields a high reward if all the agents choose it, but no reward if only some do. The other two are suboptimal actions that give lower reward but do not require precise coordination. This game enforces a phenomenon called \emph{relative overgeneralization}, \cite{games} that pushes the agents to underestimate a certain action (in our example, $a_0$) because of the low rewards they usually receive, while they could get a higher reward by perfectly coordinating on it. The reward function $Q(a)$ is:
\begin{equation}
Q(a)=\begin{cases}
n & \mbox{if }\# a_0=n, \\
\frac{n}{2} & \# a_0=0, \\
0 & \mbox{otherwise.}
\end{cases}
\end{equation}

\emph{Penalty Game}: Similarly to the Climb Game, in the Penalty Game \cite{games} each agent has three local actions $A_i=\{a_0,a_1,a_2\}$. In this game, two local actions (for example, action $a_0$ and $a_2$) give a high reward if the agents perfectly coordinate on one of them, but also give a negative penalty if they mix them together. The third action $a_1$ is suboptimal and gives a lower reward when the team coordinates on it, but also no penalty if at least one of the agents uses it. This game could also lead to relative overgeneralization, as the suboptimal action is perceived as giving a higher reward than the optimal ones on average. We use the following reward function:
\begin{equation}
Q(a)=\begin{cases}
n & \mbox{if }\# a_0=n,\mbox{ or } \# a_2=n, \\
\frac{n}{2} & \mbox{if }\# a_1=n, \\
0 & \mbox{if }0<\# a_1<n, \\
-n & \mbox{otherwise.}
\end{cases}
\end{equation}

\subsubsection{Factored Games}
\emph{Generalized Firefighting}: The Generalized Firefighting problem \cite{cgbg} is an extension of the standard two-agent firefighting problem to $n$ agents. This is a cooperative graphical Bayesian game, so each agent $i$ has some private information, called its local type $\theta_i\in\Theta_i$, on which it can condition its decisions. The combination of the various agents types $\theta=\langle \theta_1,\ldots ,\theta_n\rangle$ determines the values of the reward function $Q(a,\theta)$. We have a team of $n$ firefighters that have to fight possible fires at $N_h$ different houses. Each house $j$ can be burning, $F_j$, or not, $N_j$. Each agent $i$ has a limited observation and action field: it can observe only $N_o$ houses (so its local type is $\theta_i\in\{F_j,N_j\}^{N_o}$) and can fight the fire only at $N_a$ houses (the sets of the observed and reachable houses are fixed beforehand and are part of the problem specification, with $N_o$ and $N_a$ being their cardinality respectively). Each house $h$ yields a reward component $q_h$: if one and only one agent fights the fire at a burning house, that house gives a positive reward $q_h=2$; if the house is not burning (or if it is burning but no-one is fighting the fire at it) it does not provide any reward $q_h=0$. The reward function is sub-additive: if two agents fight the fire at the same burning house, this gives a reward $q_h=3<2\cdot 2$. The overall value of the reward function $Q(a,\theta)$ experienced by agents for a given joint type $\theta$ and joint action $a$ is the sum of the rewards given by each house $q_h$:

\begin{equation}
Q(a,\theta)=\sum_{h\in N_h}q_h.
\end{equation}

Therefore, the optimal strategy for the $n$ agents is to split as evenly as possible across all the burning houses $F_j\in\theta$. If the number of burning houses is more than that of the agents, each agent should attend at a different house and fight the fire there, while if there are less burning houses than agents, the remaining agents should exploit sub-additivity and help their colleagues at already attended houses.

In our experiments we do not input the local types $\theta_i$ to the neural networks, but we instead use these to artificially inflate the size of the local action sets (and the joint one thereby) by considering the cardinal product $A_i\times\Theta_i$ as the new action set for agent $i$, where the agent choose the action $a_i\in A_i$ and the problem choose the local type $\theta_i\in\Theta_i$. In practice, this correspond to individually consider each local action for each possible local type, as if the agents are playing a different game (a different joint type) chosen by the environment every time, and they model the values of their actions on each game separately.

\emph{Aloha}: In Aloha \cite{aloha} there is a set of nearby islands, each provided with a radio station, trying to send messages to their inhabitants. We present a slightly altered one-shot version in which the ruler of each island wants to send a radio message to its inhabitants, but, given that some of the islands are near one to another, if they all send the message the radio frequencies interfere and the messages are not correctly received by the respective populations. Given that all the rulers are living in peace and they want to maximize the number of received messages by their populations, the reward signal is shared and thus the game is cooperative. It is a graphical game, as the result of each island transmission is affected only by the transmissions of nearby islands. Every ruler $i$ has two possible actions: send a message or not. If they do not send a message, they do not contribute to the total reward. If they send one and the message is correctly received by the population (no interference occurs) they get a reward $q_i=2$, but if they interfere with someone else, they get a penalty of $q_i=-1$. The common reward that all the rulers receive at the end is the sum of their local contributions:

\begin{equation}
Q(a)=\sum_{i\in n}q_i.
\end{equation}

\section{Experiments}
With our analysis, we aim at investigating the following research questions (RQs):

\begin{enumerate}
\item Comparisons to baselines: how well can the investigated methods represent the action-value function of different cooperative multi-agent systems (both truly factored or not)? How do these compare to both independent learners and joint learners?
\item Impact of factors size: how small can the factors of these methods be with respect to the team size? How is the factor size affecting the learned representations?
\item Scalability: how do the compared methods scale in the number of agents?
\item Sample efficiency: how is the sample efficiency of these methods compared to both independent learners and joint learners?
\item Exploratory policy: how do the same investigated methods behave with a non-uniform, time-varying policy used to select actions?
\end{enumerate}

The remainder of this Section is organized as follows: we address RQ1 in Section \ref{sec:results} by comparing the methods against both independent learners and a joint learner on a variety of different games, we then investigate RQ2 by selecting one of the games and comparing the effect of using small factors versus larger ones in Section \ref{sec:size}, a couple of games with an increasing number of agents is then investigated in Section \ref{sec:larger} to address RQ3, while RQ4 is tackled in Section \ref{sec:sample}. An initial step toward RQ5 is made in Section \ref{sec:exploration} and finally, a summary of the results and general takeaways are given in Section \ref{sec:summary}.

\subsection{Experimental Setup}
Table \ref{tab:combos} defines the abbreviations and acronyms of the combinations of learning approach, coordination graph structure, and factor size used throughout our analysis (other than where differently stated). In our empirical evaluation, we investigate these combinations on the one-shot coordination games presented above.

\begin{table}[htbp]
\centering
\begin{tabular}{ccc}
\toprule
 & Mix. of Experts & Factored $Q$ \\
\midrule
Single agent & M1(=IQL \cite{independent}) & F1(=VDN \cite{vdn}) \\
Random partition ($f=2,3$) & M2R, M3R & F2R, F3R \\
Complete factorization ($f=2,3$) & M2C, M3C & F2C, F3C \\
Overlapping factors ($f=2,3$) & M2O, M3O & F2O, F3O \\
True factorization (Factored games only) & MTF & FTF \\
\bottomrule
\end{tabular}
\vspace{\baselineskip}
\caption{Combinations of factorizations and learning rules.}
\label{tab:combos}
\end{table}

We hypothesize that factored representations, by avoiding the combinatorial explosion in the number of joint actions and allowing for some internal coordination inside each factor, are going to produce representations closer to the original action-value function for these multi-agent problems. We also expect them to be sample efficient due to this small size of the factors, speeding up the required training time and learning good representations faster than the other approaches.

We train the neural networks of the factored representations to reproduce action-value functions for the detailed cooperative one-shot games, using the loss functions and coordination graph structures described in Section \ref{sec:methods} as combined in Table \ref{tab:combos}. After training, the representation $\hat{Q}(a)$ is reconstructed from the factor components' outputs and compared with the original action value function $Q(a)$ (complete knowledge of this function is withheld from the networks during training, but only samples corresponding to the selected joint action $a$ are provided at every step) to assess the quality of the representation, both in terms of action ranking and reconstruction error, as defined in the Problem Statement in Section \ref{sec:back}.

We keep the same hyperparameters for all the investigated representations to favour a fair comparison of the learned representations: using the same learning rates ensures that no method can learn faster than the others, while using the same structure for all the neural network guarantees that none is given with more representational power. Every neural network has a single hidden layer with $16$ hidden units using the leaky ReLU activation function, while all output units are linear and output local action-values $\hat{Q}_e(a_e)$ for every local joint action $a_e$\footnote{We also did some preliminary experiments with deeper networks with $2$ and $3$ hidden layers, but did not find improvements for the considered problems.}. Given the absence of an environment state to feed to the networks as an input, at every time step they just receive a constant scalar value. We use the mean square error (MSE) defined in Section \ref{sec:back} as the loss function and the RMSprop training algorithm with a learning rate of $\eta=10^{-5}$. For every game, we train the networks with $100,000$ examples by sampling a joint action $a_t$ uniformly at random.\footnote{We do not use $\epsilon$-greedy because we are interested in representing the whole value function and not just the best performing action at every training step and collecting the reward $Q(a_t)$, as noted in Section \ref{sec:back}.} Then, we propagate the gradient update through each network $e$ from the output unit $\hat{Q}_e(a_t^e)$. The loss function minimizes the squared difference between the collected reward $Q(a_t)$ at each training step and the approximation computed by the networks. After training, the learned action-value function $\hat{Q}$ is compared to the original $Q$. We also consider a baseline joint learner (a single neural network with an exponential number $|A|=|A_i|^n$ of output units). Every experiment was repeated $10$ times with random initialization of weights, each time sampling different factors for the random partitions and the overlapping factors; we report the averages of these $10$ runs.

\subsection{Comparison to Baselines}
\label{sec:results}
Our aim here is to show that factored representation are suitable to represent a wide variety of games, including many that do not present any real underlying factorization, and that these can perform better than both independent learners and a joint learner baselines. We start by discussing the approximate value functions obtained by the investigated representations, with the following Table \ref{tab:games} summarizing the games that we use and their associated parameters.

\begin{table}[htbp]
\centering
\begin{tabular}{cccccc}
\toprule
Game & $n$ & $|A_i|$ & $|A|$ & Optimal & Factored \\
\midrule
Dispersion Game & $6$ & $2$ & $64$ & $20$ & No \\
Platonia Dilemma & $6$ & $2$ & $64$ & $6$ & No \\
Climb Game & $6$ & $3$ & $729$ & $1$ & No \\
Penalty Game & $6$ & $3$ & $729$ & $2$ & No \\
Generalized Firefighting & $6$ & $2$ (per type) & $64$ ($8192$ total) & $779$ & Yes \\
Aloha & $6$ & $2$ & $64$ & $2$ & Yes \\
\bottomrule
\end{tabular}
\vspace{\baselineskip}
\caption{Details of the investigated games in this Section.}
\label{tab:games}
\end{table}

In the following plots, the $x$-axis enumerates the joint actions $a\in A$ and the $y$-axis shows the corresponding values $\hat{Q}(a)$ for the reconstructed functions, with the heights of the bars encoding the magnitude of the action-values $\hat{Q}(a)$. As defined in the Problem Statement in Section \ref{sec:back}, we analyse the quality of the computed reconstructions considering two aspects: the total reconstruction error of $\hat{Q}(a)$ with respect to the true reward function $Q(a)$ $\forall a\in A$, and whether a reconstruction produces a correct ranking of the joint actions. For a good reconstruction, the bars have to have the same relative heights, indicating that the representation correctly ranks the joint actions with respect to their value, and to be of a similar value to those in the original one (the representation can reconstruct a correct value for that joint action). However, reconstruction error alone is not a good accuracy measure because lower reconstruction error does not imply better decision making, as a model could lower the total error by over- or underestimating the value of certain joint actions.

\begin{figure}[htbp]
\centering
\subfigure[\label{sub:dispersion}]{
\includegraphics[width=0.68\textwidth]{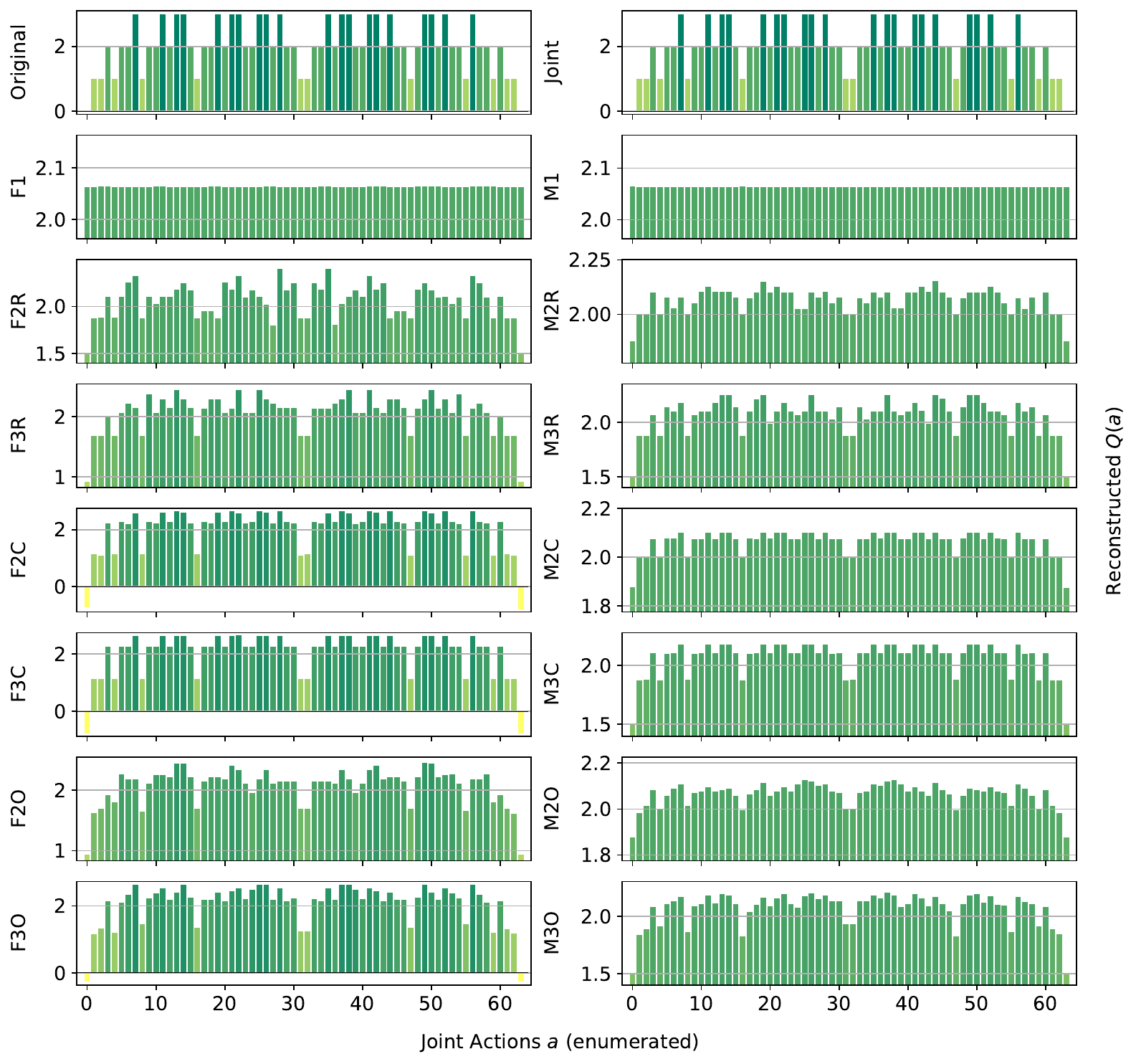}
}
\\
\subfigure[\label{sub:sparse}]{
\includegraphics[width=0.68\textwidth]{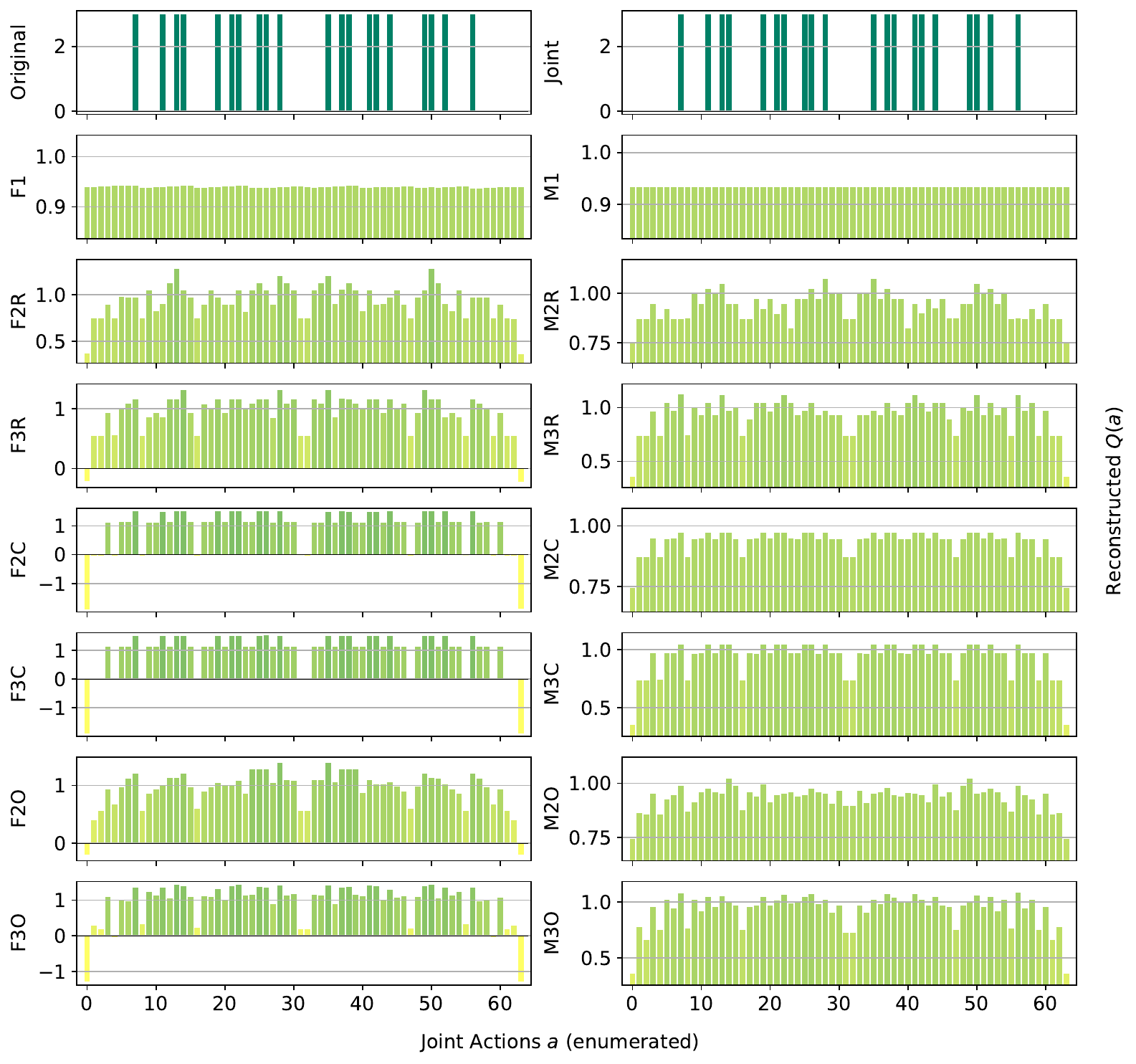}
}
\caption{Reconstructed $Q(a)$ for \protect\subref{sub:dispersion} the Dispersion Game, and \protect\subref{sub:sparse} its sparse variant.}
\label{fig:dispersion}
\end{figure}

\emph{Dispersion Games:} Figure \ref{fig:dispersion} shows the $Q$-function reconstructed by the proposed factorizations and learning approaches for the two variants of the Dispersion Game. Figure \ref{fig:dispersion}(a) shows that the proposed complete factorizations are able to almost perfectly reconstruct the relative ranking between the joint actions, meaning that these architectures can be reliably used for decision making. Moreover, the ones using the factored $Q$-function (F2C and F3C in the plot) are also able to produce a generally good approximation of the various values (expressed by the height of the bars), while those based on the mixture of experts produce a less precise reconstruction: the joint optimization of the former gives an advantage in this kind of extremely coordinated problems. 

Smaller factorizations, like the random pairings, are not sufficient to correctly represent this function, probably because a higher degree of connection is required to achieve good coordination. Figure \ref{fig:dispersion}(b) is similar but in this case the reconstruction is less accurate and the values of the bars are quite different from those of the original one. This is possibly due to the sparsity of the function, requiring the networks to correctly approximate quite different values with the same output components. In this case, the sparsity of the function to represent fools the representations into being similar to those of the non-sparse version.

In Table \ref{tab:dispersion} (as well as in the similar ones for the other games) we report the best and worst performing methods on the two variants of this game against a set of different measures that explore our results in terms of the Problem Statement presented in Section \ref{sec:back}: the mean squared error (MSE) tells us how far is the reconstructed action-value function with respect to the original one, while the number of optimal joint actions found (Opt. Found, i.e. how many of the true optimal actions are also ranked as optimal by the reconstruction), and the total number of correctly ranked actions (Ranked) points out how reliably these can reconstructions be used in decision making. Methods performing not too bad but also not the best are left out for compactness, for more data and measures for this game (and similarly for the following ones) please see the Appendix.

\begin{table}[htbp]
\centering
\pgfplotstableset{
alias/Model/.initial=0,
}

\begin{filecontents*}{csv/DispersionGame.csv}
% Dispersion Game n=6 (64 joint actions, 20 optimal)
Joint,0.0,0.0,0.0,0.0,20.0,0.0,0.0,0.0,0.0,0.0,64.0,0.0,1.0,0.0
F1,0.621,0.0,0.878,0.008,4.9,2.166,1.7,1.005,0.492,0.0,21.4,2.059,-0.001,0.029
F2R,0.516,0.0,0.816,0.005,8.0,0.0,0.0,0.0,0.381,0.0,23.7,1.005,0.265,0.033
F2C,0.094,0.0,0.141,0.003,20.0,0.0,0.0,0.0,0.158,0.0,64.0,0.0,1.0,0.0
F3C,0.094,0.0,0.14,0.003,20.0,0.0,0.0,0.0,0.158,0.001,64.0,0.0,1.0,0.0
F3O,0.185,0.028,0.296,0.044,13.0,1.342,0.2,0.4,0.199,0.015,47.4,2.835,0.701,0.046
M1,0.621,0.0,0.877,0.003,6.2,1.47,1.3,0.781,0.492,0.0,24.1,2.022,0.015,0.037
M2R,0.563,0.0,0.822,0.004,8.0,0.0,0.0,0.0,0.457,0.0,23.6,1.281,0.267,0.042
M2C,0.553,0.0,0.809,0.003,20.0,0.0,0.0,0.0,0.459,0.0,64.0,0.0,1.0,0.0
M3C,0.431,0.0,0.679,0.003,20.0,0.0,0.0,0.0,0.397,0.0,64.0,0.0,1.0,0.0
M2O,0.557,0.0,0.812,0.003,10.2,1.72,0.5,0.5,0.458,0.0,35.8,3.572,0.462,0.047
% Sparse Dispersion Game n=6 (64 joint actions, 20 optimal)
Joint,0.0,0.0,0.0,0.0,20.0,0.0,0.0,0.0,0.0,0.0,64.0,0.0,1.0,0.0
F1,1.934,0.0,4.245,0.02,6.5,0.671,1.5,1.5,1.766,0.0,37.0,1.342,0.018,0.049
F2R,1.828,0.0,3.95,0.012,8.0,0.0,0.0,0.0,1.636,0.001,40.0,0.0,0.127,0.0
F2C,1.406,0.0,2.253,0.025,20.0,0.0,0.0,0.0,1.32,0.001,64.0,0.0,1.0,0.0
F3C,1.407,0.0,2.26,0.019,20.0,0.0,0.0,0.0,1.32,0.002,64.0,0.0,1.0,0.0
M1,1.934,0.0,4.271,0.011,6.4,0.917,2.1,1.375,1.766,0.0,36.8,1.833,0.011,0.067
M2R,1.875,0.0,4.128,0.008,8.0,0.0,0.0,0.0,1.729,0.0,40.0,0.0,0.127,0.0
M2C,1.865,0.0,4.112,0.008,20.0,0.0,0.0,0.0,1.731,0.0,64.0,0.0,1.0,0.0
M3C,1.741,0.0,3.83,0.009,20.0,0.0,0.0,0.0,1.663,0.0,64.0,0.0,1.0,0.0
M2O,1.869,0.0,4.125,0.008,9.8,1.166,1.2,1.47,1.731,0.0,43.6,2.332,0.258,0.085
\end{filecontents*}
\pgfplotstabletypeset[col sep=comma,
	every head row/.style={before row=\toprule, after row=\midrule},
    every last row/.style={after row=\bottomrule},
	header=false,
	fixed,
	zerofill,
    columns/Model/.style={column type={C{0.9cm}},string type},
	columns/Mean square error/.style={string type, column type={C{1.5cm}}, column name={MSE}},
	create on use/Mean square error/.style={
		create col/assign/.code={
		\pgfmathprintpmnumber{1}{2}{2}{1}
		\pgfkeyslet{/pgfplots/table/create col/next content}\value
		}
	},
	columns/Optimal actions found/.style={string type, column type={C{1.6cm}}, column name={Opt. Found}},
	create on use/Optimal actions found/.style={
		create col/assign/.code={
		\pgfmathprintpmnumber{5}{6}{0}{0}
	    \pgfkeyslet{/pgfplots/table/create col/next content}\value
		}
	},
	columns/Correctly ranked/.style={string type, column type={C{1.8cm}}, column name={Ranked}},
	create on use/Correctly ranked/.style={
		create col/assign/.code={
		\pgfmathprintpmnumber{11}{12}{0}{0}
		\pgfkeyslet{/pgfplots/table/create col/next content}\value
		}
	},
	columns={Model,Mean square error,Optimal actions found,Correctly ranked},
	every row no 0/.style={before row={\multicolumn{4}{c}{\textbf{Dispersion Game}}\\\midrule}},
	every row no 11/.style={before row={\midrule\multicolumn{4}{c}{\textbf{Dispersion Game (sparse)}}\\\midrule}},
	every row no 0 column no 1/.style={postproc cell content/.append style={@cell content/.add={\cellcolor{green!25}}{}}},
	every row no 0 column no 2/.style={postproc cell content/.append style={@cell content/.add={\cellcolor{green!25}}{}}},
	every row no 0 column no 3/.style={postproc cell content/.append style={@cell content/.add={\cellcolor{green!25}}{}}},
	every row no 1 column no 1/.style={postproc cell content/.append style={@cell content/.add={\cellcolor{red!25}}{}}},
	every row no 1 column no 2/.style={postproc cell content/.append style={@cell content/.add={\cellcolor{red!25}}{}}},
	every row no 1 column no 3/.style={postproc cell content/.append style={@cell content/.add={\cellcolor{red!25}}{}}},
	every row no 2 column no 1/.style={postproc cell content/.append style={@cell content/.add={\cellcolor{red!25}}{}}},
	every row no 2 column no 2/.style={postproc cell content/.append style={@cell content/.add={\cellcolor{red!25}}{}}},
	every row no 2 column no 3/.style={postproc cell content/.append style={@cell content/.add={\cellcolor{red!25}}{}}},
	every row no 3 column no 1/.style={postproc cell content/.append style={@cell content/.add={\cellcolor{green!25}}{}}},
	every row no 3 column no 2/.style={postproc cell content/.append style={@cell content/.add={\cellcolor{green!25}}{}}},
	every row no 3 column no 3/.style={postproc cell content/.append style={@cell content/.add={\cellcolor{green!25}}{}}},
	every row no 4 column no 1/.style={postproc cell content/.append style={@cell content/.add={\cellcolor{green!25}}{}}},
	every row no 4 column no 2/.style={postproc cell content/.append style={@cell content/.add={\cellcolor{green!25}}{}}},
	every row no 4 column no 3/.style={postproc cell content/.append style={@cell content/.add={\cellcolor{green!25}}{}}},
	every row no 5 column no 1/.style={postproc cell content/.append style={@cell content/.add={\cellcolor{green!25}}{}}},
	every row no 6 column no 1/.style={postproc cell content/.append style={@cell content/.add={\cellcolor{red!25}}{}}},
	every row no 6 column no 2/.style={postproc cell content/.append style={@cell content/.add={\cellcolor{red!25}}{}}},
	every row no 6 column no 3/.style={postproc cell content/.append style={@cell content/.add={\cellcolor{red!25}}{}}},
	every row no 7 column no 1/.style={postproc cell content/.append style={@cell content/.add={\cellcolor{red!25}}{}}},
	every row no 7 column no 2/.style={postproc cell content/.append style={@cell content/.add={\cellcolor{red!25}}{}}},
	every row no 7 column no 3/.style={postproc cell content/.append style={@cell content/.add={\cellcolor{red!25}}{}}},
	every row no 8 column no 1/.style={postproc cell content/.append style={@cell content/.add={\cellcolor{red!25}}{}}},
	every row no 8 column no 2/.style={postproc cell content/.append style={@cell content/.add={\cellcolor{green!25}}{}}},
	every row no 8 column no 3/.style={postproc cell content/.append style={@cell content/.add={\cellcolor{green!25}}{}}},
	every row no 9 column no 2/.style={postproc cell content/.append style={@cell content/.add={\cellcolor{green!25}}{}}},
	every row no 9 column no 3/.style={postproc cell content/.append style={@cell content/.add={\cellcolor{green!25}}{}}},
	every row no 10 column no 1/.style={postproc cell content/.append style={@cell content/.add={\cellcolor{red!25}}{}}},
	every row no 11 column no 1/.style={postproc cell content/.append style={@cell content/.add={\cellcolor{green!25}}{}}},
	every row no 11 column no 2/.style={postproc cell content/.append style={@cell content/.add={\cellcolor{green!25}}{}}},
	every row no 11 column no 3/.style={postproc cell content/.append style={@cell content/.add={\cellcolor{green!25}}{}}},
	every row no 12 column no 1/.style={postproc cell content/.append style={@cell content/.add={\cellcolor{red!25}}{}}},
	every row no 12 column no 2/.style={postproc cell content/.append style={@cell content/.add={\cellcolor{red!25}}{}}},
	every row no 12 column no 3/.style={postproc cell content/.append style={@cell content/.add={\cellcolor{red!25}}{}}},
	every row no 13 column no 1/.style={postproc cell content/.append style={@cell content/.add={\cellcolor{red!25}}{}}},
	every row no 13 column no 2/.style={postproc cell content/.append style={@cell content/.add={\cellcolor{red!25}}{}}},
	every row no 14 column no 2/.style={postproc cell content/.append style={@cell content/.add={\cellcolor{green!25}}{}}},
	every row no 14 column no 3/.style={postproc cell content/.append style={@cell content/.add={\cellcolor{green!25}}{}}},
	every row no 15 column no 2/.style={postproc cell content/.append style={@cell content/.add={\cellcolor{green!25}}{}}},
	every row no 15 column no 3/.style={postproc cell content/.append style={@cell content/.add={\cellcolor{green!25}}{}}},
	every row no 16 column no 1/.style={postproc cell content/.append style={@cell content/.add={\cellcolor{red!25}}{}}},
	every row no 16 column no 2/.style={postproc cell content/.append style={@cell content/.add={\cellcolor{red!25}}{}}},
	every row no 16 column no 3/.style={postproc cell content/.append style={@cell content/.add={\cellcolor{red!25}}{}}},
	every row no 17 column no 1/.style={postproc cell content/.append style={@cell content/.add={\cellcolor{red!25}}{}}},
	every row no 17 column no 2/.style={postproc cell content/.append style={@cell content/.add={\cellcolor{red!25}}{}}},
	every row no 18 column no 1/.style={postproc cell content/.append style={@cell content/.add={\cellcolor{red!25}}{}}},
	every row no 18 column no 2/.style={postproc cell content/.append style={@cell content/.add={\cellcolor{green!25}}{}}},
	every row no 18 column no 3/.style={postproc cell content/.append style={@cell content/.add={\cellcolor{green!25}}{}}},
	every row no 19 column no 2/.style={postproc cell content/.append style={@cell content/.add={\cellcolor{green!25}}{}}},
	every row no 19 column no 3/.style={postproc cell content/.append style={@cell content/.add={\cellcolor{green!25}}{}}},
	every row no 20 column no 1/.style={postproc cell content/.append style={@cell content/.add={\cellcolor{red!25}}{}}},
]{csv/DispersionGame.csv}
\vspace{\baselineskip}
\caption{Best (green) and worst (red) performing methods on the two variants of the Dispersion Game.}
\label{tab:dispersion}
\end{table}

We observe how the joint learner can easily learn the entire action-value function for this small setting, resulting in a perfect ranking and a very small error. However, methods using the complete factorizations are also able to do so, with the mixture of experts achieving a larger reconstruction error but still a correct ranking of actions, including identifying all of the optimal ones, on both variants of this game. Independent learners instead do not seem able to correctly identify all of the optimal actions, also achieving a very large reconstruction error.

\begin{figure}[htbp]
\centering
\includegraphics[width=0.68\textwidth]{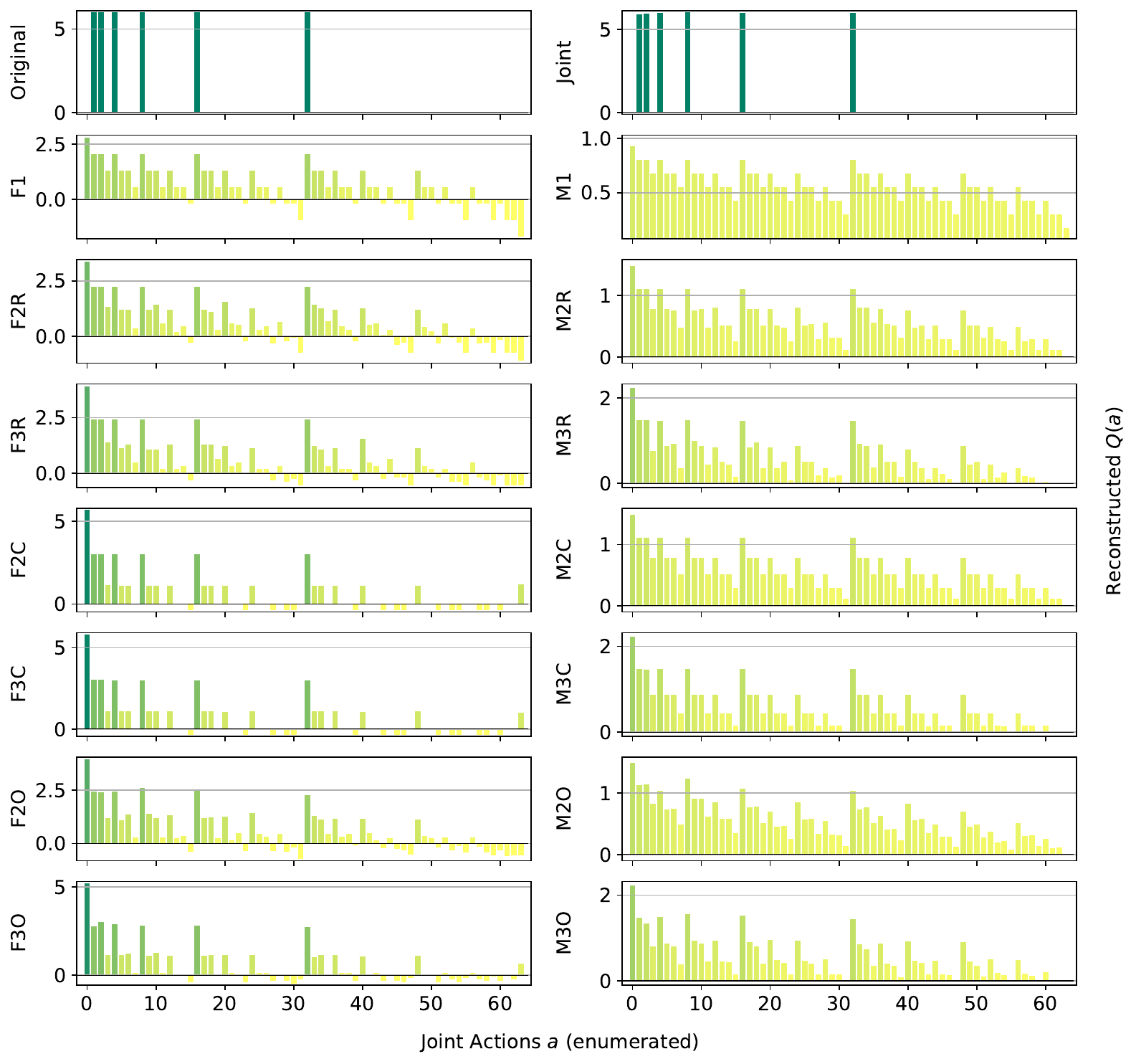}
\caption{Reconstructed $Q(a)$ for the Platonia Dilemma.}
\label{fig:platonia}
\end{figure}

\emph{Platonia Dilemma:} Figure \ref{fig:platonia} shows the reconstructed action-value functions for the Platonia Dilemma. For this problem, none of the proposed factorizations can correctly represent the action-value function. In fact, while they are perfectly able to correctly rank all the optimal actions (the ones in which only a single agent sends the telegram) at the same level, they all fail to correctly rank and reconstruct the same joint action (that is, the one in which none of the agents sends the telegram). In fact, the unique symmetric equilibrium for the team in this game is that each of them sends the telegram with probability $\frac{1}{n}$, so the agents usually gather more reward by not sending it themselves, but relying on someone else to do so. This results in an `imbalanced' action-value function in which the high reward is more often obtained, from an agent perspective, by choosing a certain action instead of the other, thus resulting in overestimating one of the actions (the one in which all the agents perform the same action, i.e., not sending the telegram).

This imbalance in the reward given by the two actions is probably the cause of the poor reconstruction. Thus, for this kind of tightly coupled coordination problem, none of the techniques to approximate action-values currently employed in deep MARL suffice to guarantee a good action is taken, even if the coordination problem is conceptually simple. Table \ref{tab:platonia} reports the best and worst performing methods on this game.

\begin{table}[htbp]
\centering
\pgfplotstableset{
alias/Model/.initial=0,
}

\begin{filecontents*}{csv/PlatoniaDilemma.csv}
Joint,0.003,0.006,0.029,0.061,6.0,0.0,0.0,0.0,0.006,0.01,64.0,0.0,1.0,0.0
M1,2.8,0.001,27.009,0.035,5.0,0.0,6.0,0.0,5.145,0.001,62.0,0.0,0.816,0.0
M2O,2.543,0.008,23.933,0.058,4.3,0.458,6.0,0.0,4.921,0.001,60.6,0.917,0.687,0.084
M3O,2.276,0.009,20.604,0.068,4.4,0.49,6.0,0.0,4.61,0.009,60.8,0.98,0.706,0.09
\end{filecontents*}
\pgfplotstabletypeset[col sep=comma,
	every head row/.style={before row=\toprule, after row=\midrule},
    every last row/.style={after row=\bottomrule},
	header=false,
	fixed,
	zerofill,
    columns/Model/.style={column type={C{0.9cm}},string type},
	columns/Mean square error/.style={string type, column type={C{1.5cm}}, column name={MSE}},
	create on use/Mean square error/.style={
		create col/assign/.code={
		\pgfmathprintpmnumber{1}{2}{2}{1}
		\pgfkeyslet{/pgfplots/table/create col/next content}\value
		}
	},
	columns/Optimal actions found/.style={string type, column type={C{1.6cm}}, column name={Opt. Found}},
	create on use/Optimal actions found/.style={
		create col/assign/.code={
		\pgfmathprintpmnumber{5}{6}{0}{0}
	    \pgfkeyslet{/pgfplots/table/create col/next content}\value
		}
	},
	columns/Correctly ranked/.style={string type, column type={C{1.8cm}}, column name={Ranked}},
	create on use/Correctly ranked/.style={
		create col/assign/.code={
		\pgfmathprintpmnumber{11}{12}{0}{0}
		\pgfkeyslet{/pgfplots/table/create col/next content}\value
		}
	},
	columns={Model,Mean square error,Optimal actions found,Correctly ranked},
	every row no 0 column no 1/.style={postproc cell content/.append style={@cell content/.add={\cellcolor{green!25}}{}}},
	every row no 0 column no 2/.style={postproc cell content/.append style={@cell content/.add={\cellcolor{green!25}}{}}},
	every row no 0 column no 3/.style={postproc cell content/.append style={@cell content/.add={\cellcolor{green!25}}{}}},
	every row no 1 column no 1/.style={postproc cell content/.append style={@cell content/.add={\cellcolor{red!25}}{}}},
	every row no 2 column no 2/.style={postproc cell content/.append style={@cell content/.add={\cellcolor{red!25}}{}}},
	every row no 2 column no 3/.style={postproc cell content/.append style={@cell content/.add={\cellcolor{red!25}}{}}},
	every row no 3 column no 2/.style={postproc cell content/.append style={@cell content/.add={\cellcolor{red!25}}{}}},
	every row no 3 column no 3/.style={postproc cell content/.append style={@cell content/.add={\cellcolor{red!25}}{}}},
]{csv/PlatoniaDilemma.csv}
\vspace{\baselineskip}
\caption{Best (green) and worst (red) performing methods on the Platonia Dilemma.}
\label{tab:platonia}
\end{table}

All of the methods using the factored $Q$-function learning approach are left out, as these achieve average performances. As already showed by Figure \ref{fig:platonia}, here only the joint learner is able to correctly identify all of the optimal actions. The mixture of experts with the overlapping factorization are the ones that perform the worst, possibly because the connectivity of this structure is too sparse to help in such an imbalanced reward game.

\begin{figure}[htbp]
\centering
\subfigure[\label{sub:climb_factored}]{
\includegraphics[width=0.68\textwidth]{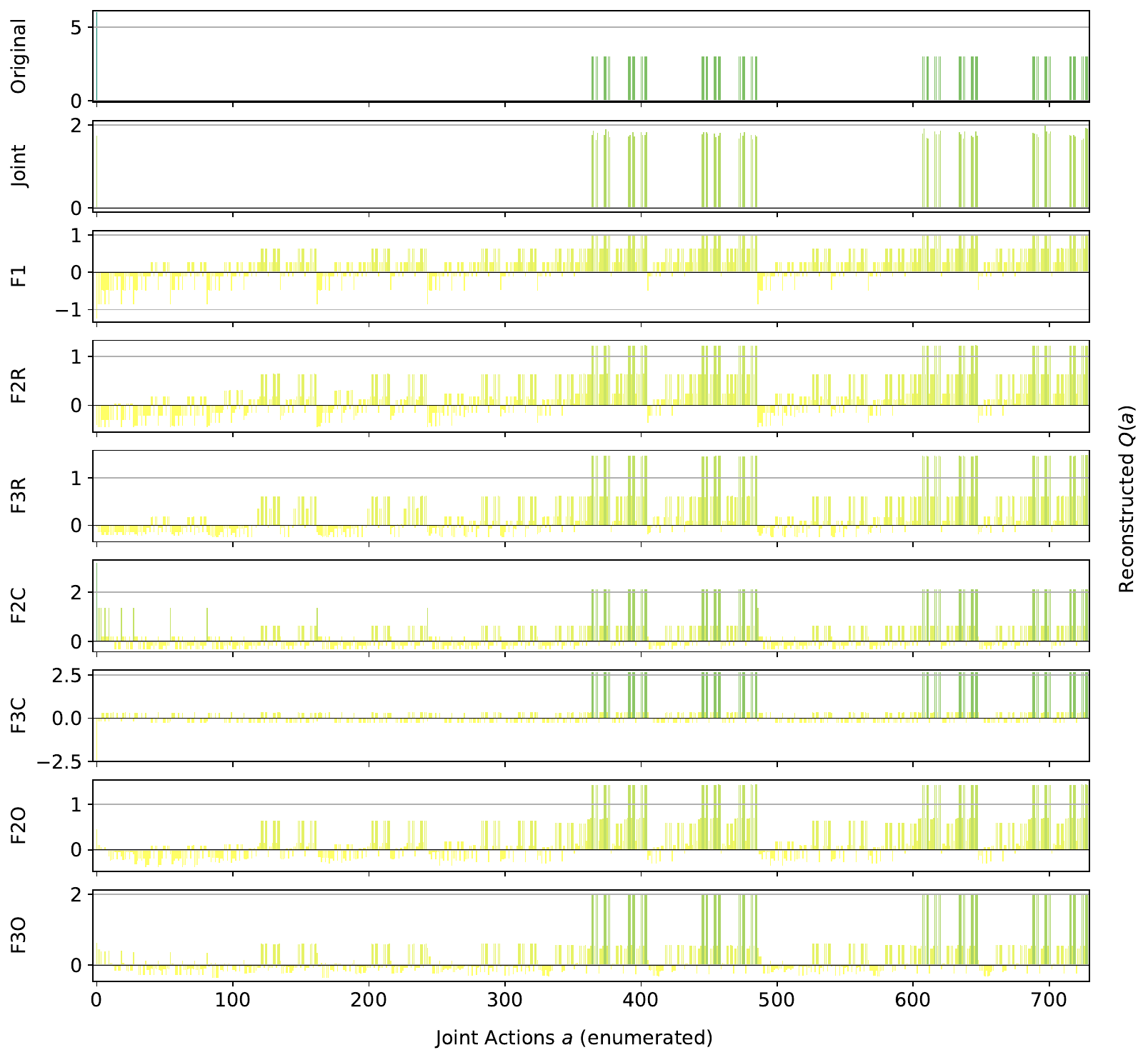}
}
\\
\subfigure[\label{sub:climb_moe}]{
\includegraphics[width=0.68\textwidth]{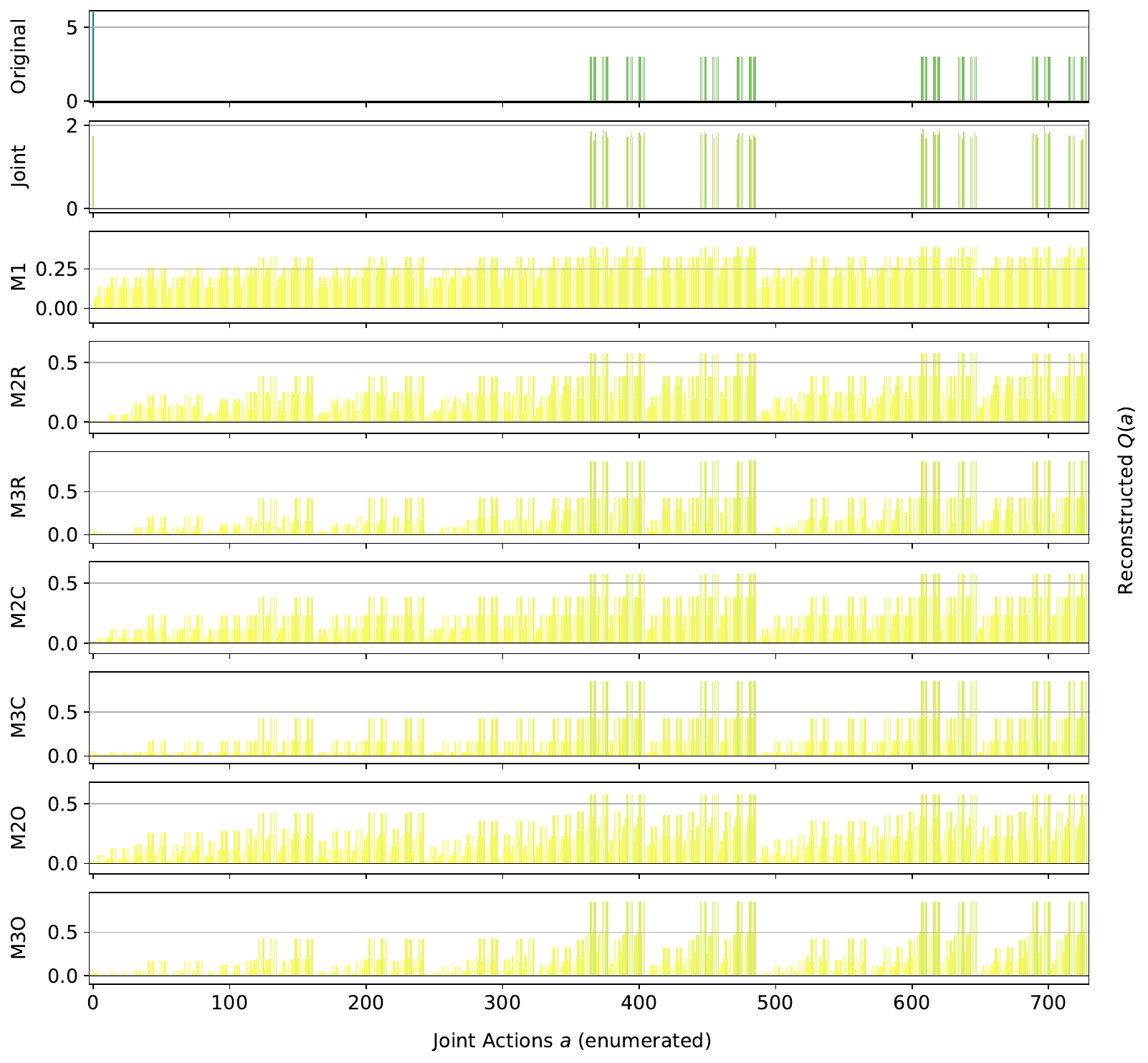}
}
\caption{Reconstructed $Q(a)$ for the Climb Game: \protect\subref{sub:climb_factored} factored $Q$-function learning approach, and \protect\subref{sub:climb_moe} mixture of experts learning approach.}
\label{fig:climb}
\end{figure}

\emph{Climb Game:} Figure \ref{fig:climb} shows the results obtained on the Climb Game. The joint network is not able to learn the correct action-value function in the given training time, due to the large number of joint actions. This highlights again how joint learners are not suited for this kind of even moderately large multi-agent system. By contrast, all the other architectures correctly rank the suboptimal actions.

The methods using the factored $Q$-function and a complete factorization are also able to correctly reconstruct the values for most of the joint actions, as can be seen from the bars. However, only F2C can correctly rank and reconstruct the optimal action (the coordinated one), while even F3C fails to do so and gives it a large negative value. A likely cause for this effect is that, where optimizing the loss function, assigning negative values to the components forming that joint action reduces the overall mean squared error, even if one of the reconstructed reward value is totally wrong. We can also observe how the mixture of experts plot looks somewhat comparable to the one for the factored $Q$-function learning approach, but more `compressed' and noisy. Table \ref{tab:climb} reports the best and worst performing methods on the Climb Game.

\begin{table}[htbp]
\centering
\pgfplotstableset{
alias/Model/.initial=0,
}

\begin{filecontents*}{csv/ClimbGame.csv}
Joint,0.173,0.073,18.449,4.908,0.1,0.3,2.7,0.9,1.517,0.275,727.2,0.6,0.999,0.0
F2C,0.254,0.0,7.864,0.106,1.0,0.0,0.0,0.0,1.395,0.002,729.0,0.0,1.0,0.0
F3C,0.169,0.001,70.767,0.669,0.0,0.0,3.0,0.0,0.961,0.002,726.0,0.0,0.98,0.0
M1,0.709,0.0,35.914,0.012,0.0,0.0,3.0,0.0,2.363,0.0,726.0,0.0,0.98,0.0
\end{filecontents*}
\pgfplotstabletypeset[col sep=comma,
	every head row/.style={before row=\toprule, after row=\midrule},
    every last row/.style={after row=\bottomrule},
	header=false,
	fixed,
	zerofill,
    columns/Model/.style={column type={C{0.9cm}},string type},
	columns/Mean square error/.style={string type, column type={C{1.5cm}}, column name={MSE}},
	create on use/Mean square error/.style={
		create col/assign/.code={
		\pgfmathprintpmnumber{1}{2}{2}{1}
		\pgfkeyslet{/pgfplots/table/create col/next content}\value
		}
	},
	columns/Optimal actions found/.style={string type, column type={C{1.6cm}}, column name={Opt. Found}},
	create on use/Optimal actions found/.style={
		create col/assign/.code={
		\pgfmathprintpmnumber{5}{6}{0}{0}
	    \pgfkeyslet{/pgfplots/table/create col/next content}\value
		}
	},
	columns/Correctly ranked/.style={string type, column type={C{1.8cm}}, column name={Ranked}},
	create on use/Correctly ranked/.style={
		create col/assign/.code={
		\pgfmathprintpmnumber{11}{12}{0}{0}
		\pgfkeyslet{/pgfplots/table/create col/next content}\value
		}
	},
	columns={Model,Mean square error,Optimal actions found,Correctly ranked},
	every row no 0 column no 1/.style={postproc cell content/.append style={@cell content/.add={\cellcolor{green!25}}{}}},
	every row no 1 column no 1/.style={postproc cell content/.append style={@cell content/.add={\cellcolor{green!25}}{}}},
	every row no 1 column no 2/.style={postproc cell content/.append style={@cell content/.add={\cellcolor{green!25}}{}}},
	every row no 1 column no 3/.style={postproc cell content/.append style={@cell content/.add={\cellcolor{green!25}}{}}},
	every row no 2 column no 1/.style={postproc cell content/.append style={@cell content/.add={\cellcolor{green!25}}{}}},
	every row no 3 column no 1/.style={postproc cell content/.append style={@cell content/.add={\cellcolor{red!25}}{}}},
]{csv/ClimbGame.csv}
\vspace{\baselineskip}
\caption{Best (green) and worst (red) performing methods on the Climb Game.}
\label{tab:climb}
\end{table}

With a larger joint action space, the joint learner begins to struggle and achieves a larger reconstruction error than some of the factored methods. Interestingly, F2C is the only method capable of identifying the optimal action of this game, when also F3C fails. We hypothesize that this happens because the larger factors push the overall representation to further improve the reconstructed values for the other local joint actions at the expense of these forming the optimal action itself. This points out how, although generally a larger factor size entails a better representation, it may not always be so. On the other hand however, it also shows how small factors can result in a good representation that is also easier and faster to be learned.

\begin{figure}[htbp]
\centering
\subfigure[\label{sub:penalty_factored}]{
\includegraphics[width=0.68\textwidth]{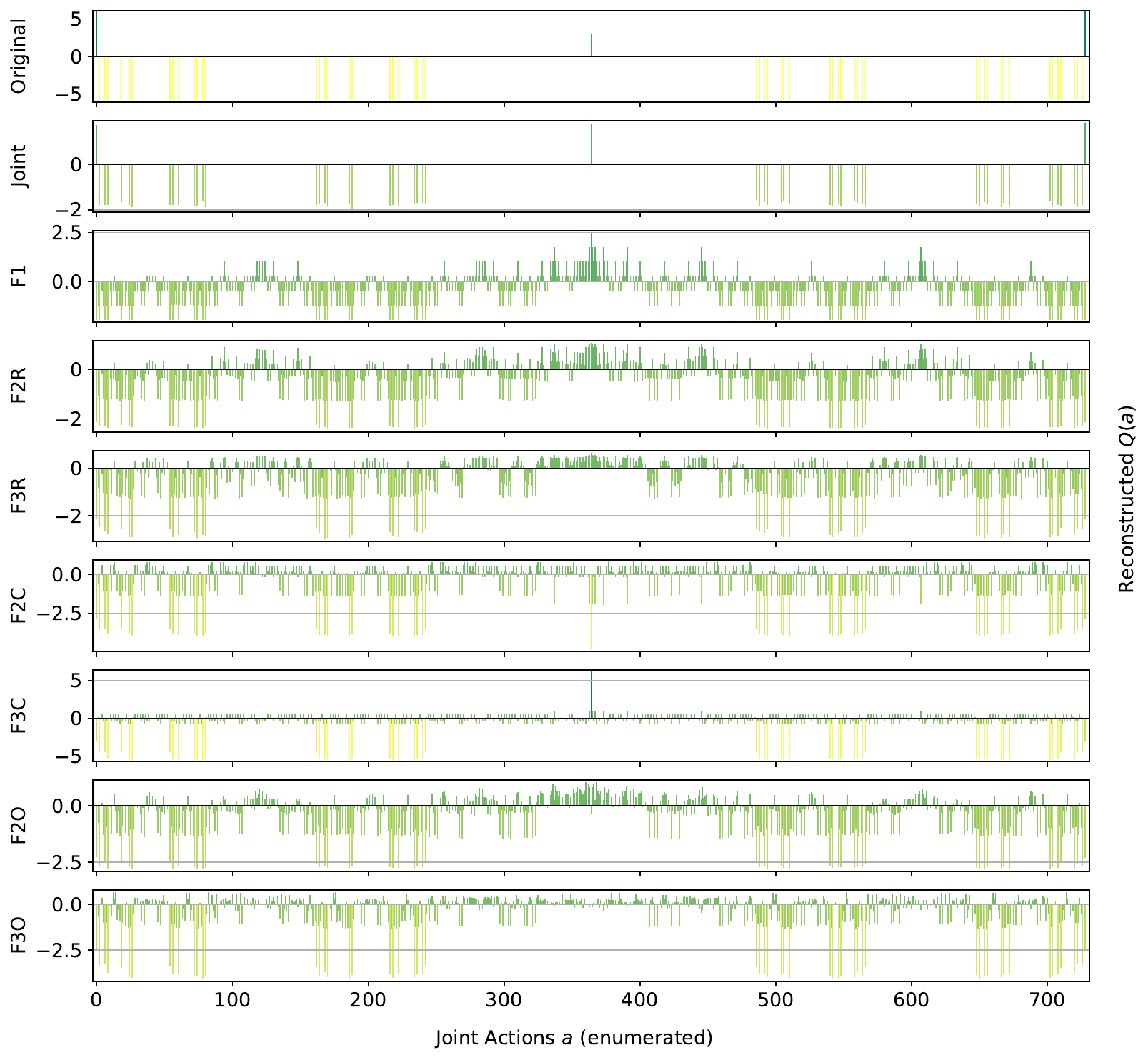}
}
\\
\subfigure[\label{sub:penalty_moe}]{
\includegraphics[width=0.68\textwidth]{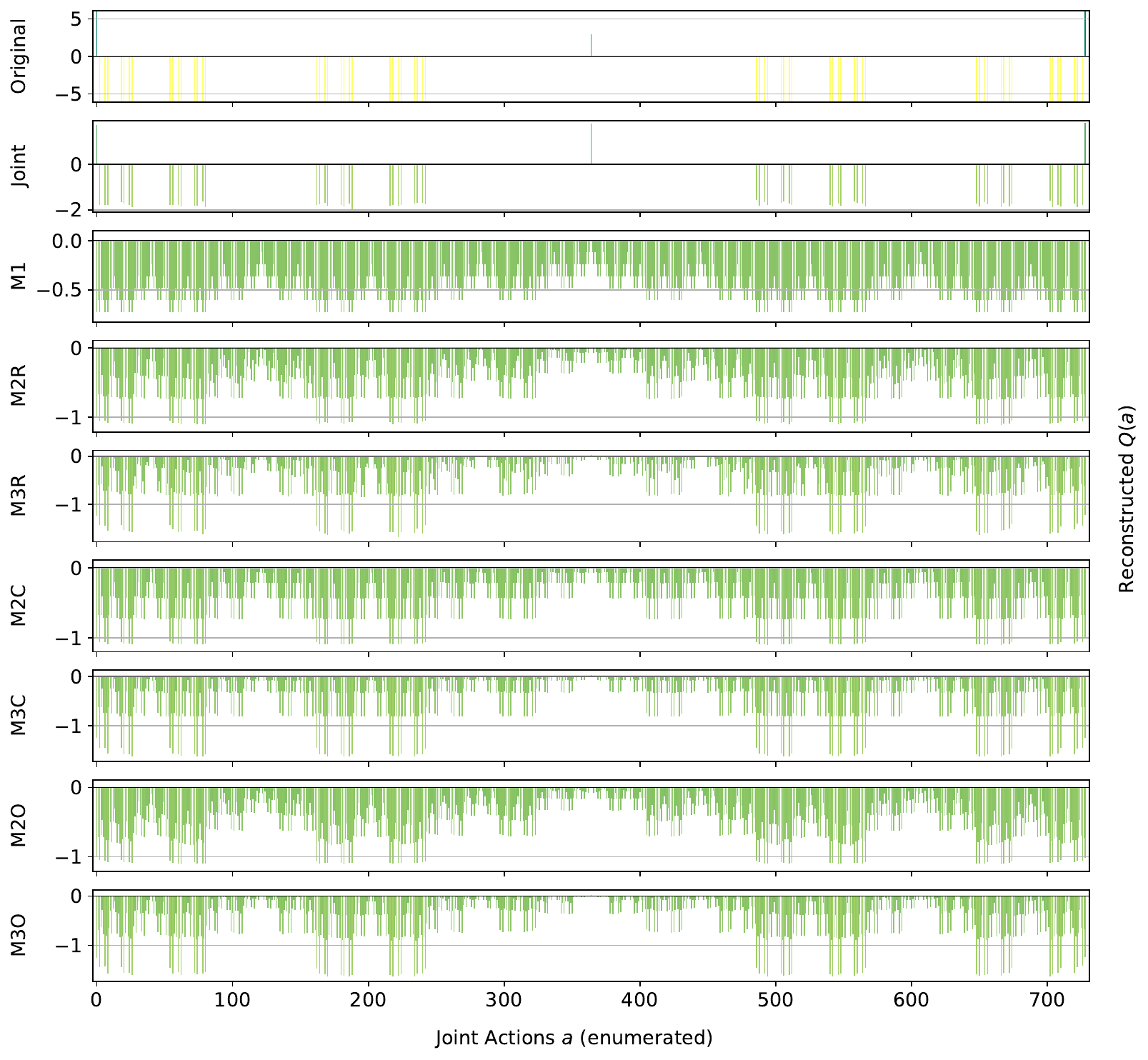}
}
\caption{Reconstructed $Q(a)$ for the Penalty Game: \protect\subref{sub:penalty_factored} factored $Q$-function learning approach, and \protect\subref{sub:penalty_moe} mixture of experts learning approach.}
\label{fig:penalty}
\end{figure}

\emph{Penalty Game:} Figure \ref{fig:penalty} presents the representations obtained by the investigated approximations. Given the high level of coordination required, all of the architectures using the mixture of experts learn a totally incorrect approximation, biased by the larger number of joint actions that yield a penalty rather than a positive reward.

For this game, none of the architectures can correctly reconstruct the whole structure of the action value function, and they all fail at the two optimal joint actions (at the two sides of the bar plots). This is probably due to the large gap in the reward values that the agents can receive when choosing one of their coordinated actions: they can get a high reward if all the agents perfectly coordinate, but it is more common for them to miscoordinate and receive a negative penalty, resulting in an approximation that ranks those two joint actions as bad in order to correctly reconstruct the other cases. Furthermore, the suboptimal action is hard to correctly approximate because, similarly to the optimal ones, it also usually results in a smaller reward than the one it gives when all the agents coordinate on it. Only F1 and F3C rank it as better than the other, but surprisingly only F1 is also able to reconstruct the correct value. Table \ref{tab:penalty} reports the best and worst performing methods on this game.

\begin{table}[htbp]
\centering
\pgfplotstableset{
alias/Model/.initial=0,
}

\begin{filecontents*}{csv/PenaltyGame.csv}
Joint,1.601,0.38,18.21,4.256,1.2,0.4,0.9,1.375,3.242,0.134,727.4,0.8,1.0,0.0
F1,2.177,0.0,63.71,0.148,0.0,0.0,3.0,0.0,3.306,0.001,722.0,0.0,0.905,0.0
F2C,1.288,0.002,79.664,0.001,0.0,0.0,6.0,0.0,3.297,0.001,722.0,0.0,0.922,0.0
F3C,0.535,0.001,82.768,0.0,0.0,0.0,3.0,0.0,2.055,0.0,724.0,0.0,0.969,0.0
F3O,1.273,0.039,73.484,1.365,0.0,0.0,6.0,0.0,3.292,0.0,723.0,0.0,0.954,0.0
M1,2.708,0.001,45.265,0.052,0.0,0.0,3.0,0.0,3.663,0.001,722.0,0.0,0.905,0.0
\end{filecontents*}
\pgfplotstabletypeset[col sep=comma,
	every head row/.style={before row=\toprule, after row=\midrule},
    every last row/.style={after row=\bottomrule},
	header=false,
	fixed,
	zerofill,
    columns/Model/.style={column type={C{0.9cm}},string type},
	columns/Mean square error/.style={string type, column type={C{1.5cm}}, column name={MSE}},
	create on use/Mean square error/.style={
		create col/assign/.code={
		\pgfmathprintpmnumber{1}{2}{2}{1}
		\pgfkeyslet{/pgfplots/table/create col/next content}\value
		}
	},
	columns/Optimal actions found/.style={string type, column type={C{1.6cm}}, column name={Opt. Found}},
	create on use/Optimal actions found/.style={
		create col/assign/.code={
		\pgfmathprintpmnumber{5}{6}{0}{0}
	    \pgfkeyslet{/pgfplots/table/create col/next content}\value
		}
	},
	columns/Correctly ranked/.style={string type, column type={C{1.8cm}}, column name={Ranked}},
	create on use/Correctly ranked/.style={
		create col/assign/.code={
		\pgfmathprintpmnumber{11}{12}{0}{0}
		\pgfkeyslet{/pgfplots/table/create col/next content}\value
		}
	},
	columns={Model,Mean square error,Optimal actions found,Correctly ranked},
	every row no 0 column no 2/.style={postproc cell content/.append style={@cell content/.add={\cellcolor{green!25}}{}}},
	every row no 0 column no 3/.style={postproc cell content/.append style={@cell content/.add={\cellcolor{green!25}}{}}},
	every row no 1 column no 3/.style={postproc cell content/.append style={@cell content/.add={\cellcolor{red!25}}{}}},
	every row no 2 column no 1/.style={postproc cell content/.append style={@cell content/.add={\cellcolor{green!25}}{}}},
	every row no 2 column no 3/.style={postproc cell content/.append style={@cell content/.add={\cellcolor{red!25}}{}}},
	every row no 3 column no 1/.style={postproc cell content/.append style={@cell content/.add={\cellcolor{green!25}}{}}},
	every row no 4 column no 1/.style={postproc cell content/.append style={@cell content/.add={\cellcolor{green!25}}{}}},
	every row no 5 column no 1/.style={postproc cell content/.append style={@cell content/.add={\cellcolor{red!25}}{}}},
	every row no 5 column no 3/.style={postproc cell content/.append style={@cell content/.add={\cellcolor{red!25}}{}}},
]{csv/PenaltyGame.csv}
\vspace{\baselineskip}
\caption{Best (green) and worst (red) performing methods on the Penalty Game.}
\label{tab:penalty}
\end{table}

For this setting as well, the joint learner is struggling to represent the entire action-value function, although it is the only method capable of correctly identify one of the optimal joint actions. All the other methods fail in doing so, even though some of those that use the factored $Q$-function learning approach achieve a very small MSE.

\begin{figure}[htbp]
\centering
\begin{tikzpicture}
\tikzset{dot/.style 2 args={fill, rectangle, inner sep=5pt, label={#1:\scriptsize #2}}}
\foreach \x in {0,...,6}{
\pgfmathtruncatemacro\h{\x+1}
\node[dot={90}{$H_{\h}$}] at (\x,0) {};
}
\tikzset{dot/.style 2 args={fill, circle, inner sep=3pt, label={#1:\scriptsize #2}}}
\foreach \x in {0,...,5}{
\pgfmathtruncatemacro\n{\x+1}
\draw[line width=.5pt] (\x+0.5,-1) -- (\x,0);
\draw[line width=.5pt] (\x+0.5,-1) -- (\x+1,0);
\node[dot={-90}{\n}] at (\x+0.5,-1) {};
}
\end{tikzpicture}
\caption{Firefighters formation with $n=6$ agents and $N_h=7$ houses.}
\label{fig:formation}
\end{figure}

\begin{figure}[htbp]
\centering
\includegraphics[width=0.68\textwidth]{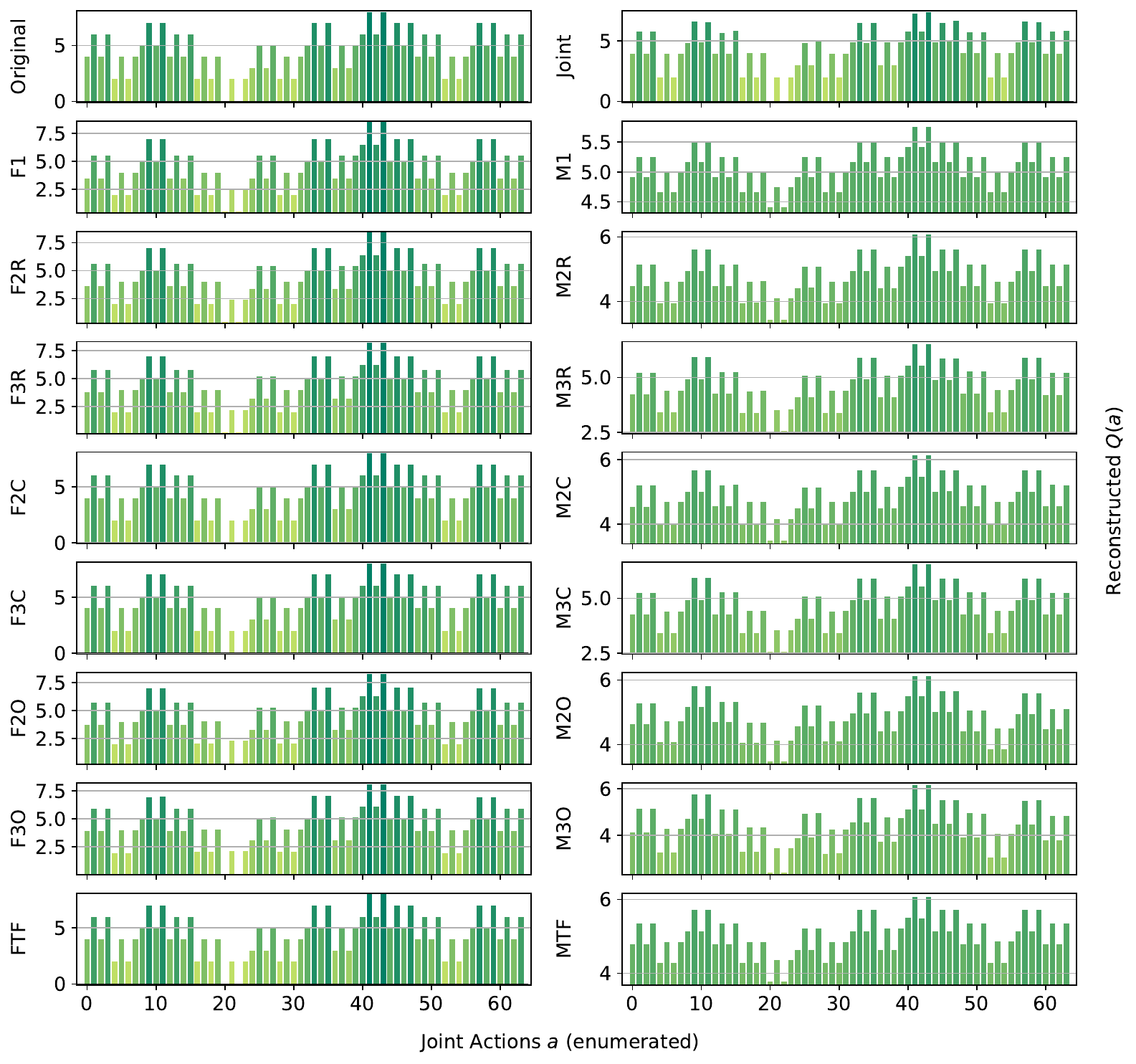}
\caption{Reconstructed $Q(a)$ for a single joint type of the Generalized Firefighting problem.}
\label{fig:generalized_ff1}
\end{figure}

\emph{Generalized Firefighting}: In our experiments, a team of $n=6$ agents have to fight fire at $N_h=7$ houses. Each agent can observe $N_o=2$ houses and can fight fire at the same set of locations ($N_a=2$), disposed as shown in Figure \ref{fig:formation}. Figure \ref{fig:generalized_ff1} shows the representations learned for the joint type $\theta=\{N_1,F_2,N_3,F_4,N_5,N_6,F_7\}$.

This game requires less coordination than those studied earlier (agents have to coordinate only with other agents that can fight fire at the same locations), and every investigated architecture correctly ranks all the joint actions, even the single agent factorizations F1 and M1. However, while those using the factored $Q$-function can also correctly reconstruct the reward value of each action, those using the mixture of experts are less precise in their reconstructions. Overall, this experiment demonstrates that there exist non-trivial coordination problems that can effectively be tackled using small factors, including even individual learning approaches. Also, it is to note how both learning approaches, when coupled with the true underlying factorization, are achieving very good reconstruction and can rank all of the joint actions correctly.

Figure \ref{fig:generalized_ff2} shows the results for a different joint type, $\theta=\{F_1,F_2,F_3,F_4,F_5,N_6,F_7\}$:

\begin{figure}[htbp]
\centering
\includegraphics[width=0.68\textwidth]{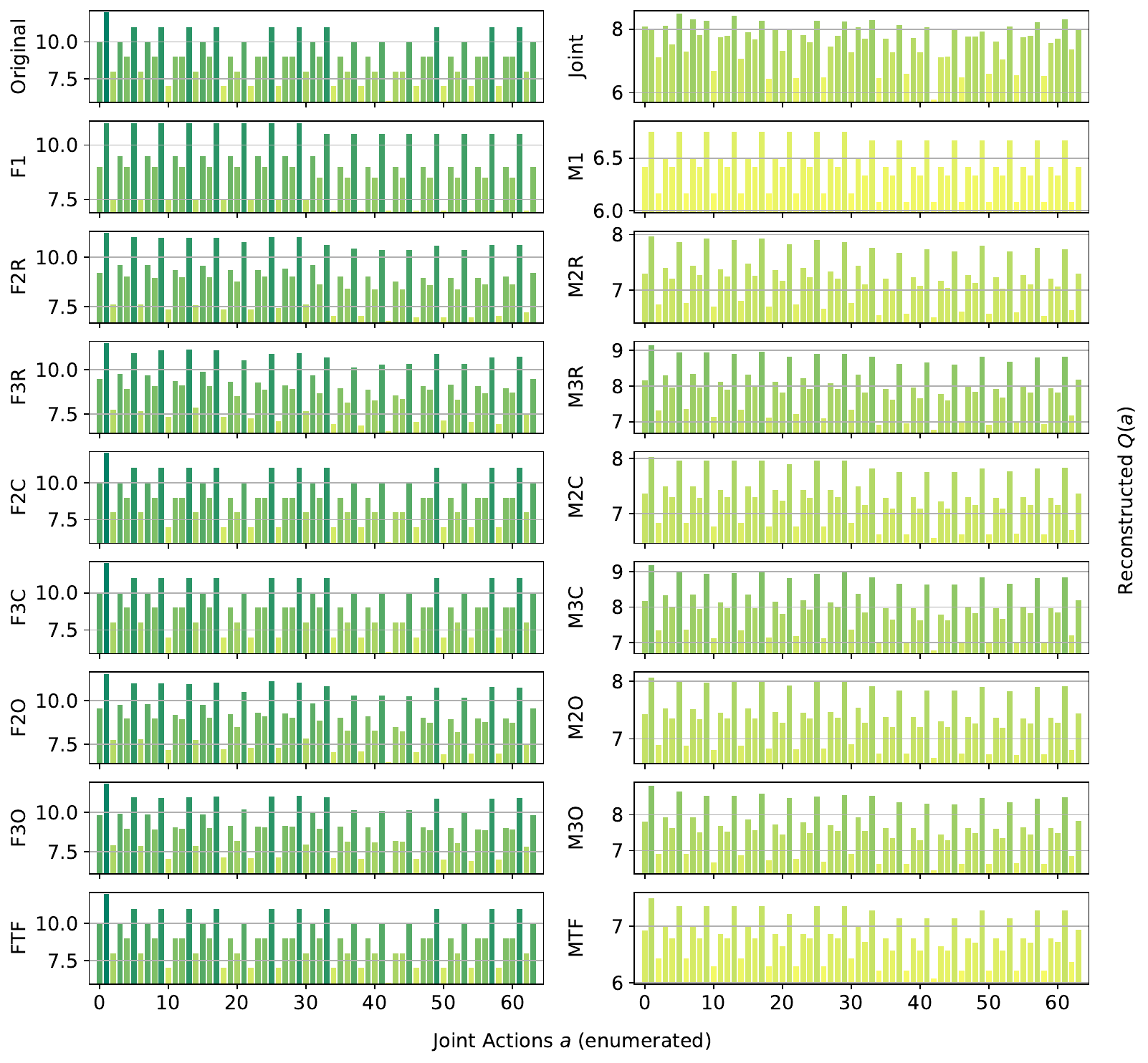}
\caption{Reconstructed $Q(a)$ for a different joint type of the Generalized Firefighting problem.}
\label{fig:generalized_ff2}
\end{figure}

This type presents multiple adjacent houses burning at the same time, so the agents have to correctly estimate the value of fighting fire at a certain location both on their own or collaborating with other agents. The joint learner is not able to correctly learn the values for this type in the given training time, thus resulting in ranking as optimal actions that are not. Simpler factorizations like F1 or M1 on the other hand fail as well, ranking suboptimal actions as optimal. However, the other factored representations are quite accurate and correctly represent the value of coordination: even simpler factorization using overlapping factors with both learning approaches or random pairing coupled with the factored $Q$-function learning approach, can correctly identify the optimal joint action. Again, the representations obtained with the factored $Q$-function learning approach are more accurate in terms of values of the actions. Table \ref{tab:generalized_ff} is showing the best and worst performing methods on this game. On this type as well, both FTF and MTF are achieving very good reconstructions, with FTF also approximating the values of the joint actions correctly.

\begin{table}[htbp]
\centering
\pgfplotstableset{
alias/Model/.initial=0,
}

\begin{filecontents*}{csv/GeneralizedFireFighting.csv}
Joint,1.293,2.452,4.958,7.21,656.2,123.081,61.6,68.194,46.416,48.364,6892.8,1475.467,0.852,0.196
F3R,0.091,0.036,0.107,0.05,743.0,24.548,11.1,10.153,3.42,1.703,7287.6,557.77,0.944,0.034
F2C,0.0,0.0,0.0,0.0,779.0,0.0,0.0,0.0,0.0,0.0,8192.0,0.0,1.0,0.0
F3C,0.0,0.0,0.0,0.0,779.0,0.0,0.0,0.0,0.0,0.0,8192.0,0.0,1.0,0.0
F2O,0.088,0.023,0.101,0.033,746.5,17.862,8.0,7.099,3.75,1.125,7333.1,381.521,0.947,0.023
F3O,0.028,0.017,0.029,0.018,777.8,3.6,0.2,0.6,1.137,0.949,8148.6,130.2,0.997,0.008
FTF,0.0,0.0,0.0,0.0,779.0,0.0,0.0,0.0,0.0,0.0,8192.0,0.0,1.0,0.0
M1,3.548,0.004,8.347,0.027,700.4,6.344,27.8,6.4,163.836,0.073,6219.9,30.386,0.88,0.001
M2C,1.821,0.002,4.896,0.011,777.0,0.0,0.0,0.0,124.415,0.068,7826.0,0.0,0.974,0.0
M3C,0.848,0.001,2.577,0.008,777.5,0.5,0.0,0.0,88.094,0.104,8151.0,4.025,0.997,0.0
M2O,1.971,0.106,5.078,0.111,740.8,12.742,11.7,5.001,127.214,1.909,5627.8,249.237,0.835,0.026
M3O,1.73,1.189,4.958,3.322,737.8,54.54,8.7,14.015,97.667,10.894,5866.8,682.394,0.853,0.05
MTF,2.602,0.004,6.135,0.015,779.0,0.0,0.0,0.0,133.786,0.141,8177.2,2.227,0.999,0.0
\end{filecontents*}
\pgfplotstabletypeset[col sep=comma,
	every head row/.style={before row=\toprule, after row=\midrule},
    every last row/.style={after row=\bottomrule},
	header=false,
	fixed,
	zerofill,
    columns/Model/.style={column type={C{0.9cm}},string type},
	columns/Mean square error/.style={string type, column type={C{1.5cm}}, column name={MSE}},
	create on use/Mean square error/.style={
		create col/assign/.code={
		\pgfmathprintpmnumber{1}{2}{2}{1}
		\pgfkeyslet{/pgfplots/table/create col/next content}\value
		}
	},
	columns/Optimal actions found/.style={string type, column type={C{1.6cm}}, column name={Opt. Found}},
	create on use/Optimal actions found/.style={
		create col/assign/.code={
		\pgfmathprintpmnumber{5}{6}{0}{0}
	    \pgfkeyslet{/pgfplots/table/create col/next content}\value
		}
	},
	columns/Correctly ranked/.style={string type, column type={C{1.8cm}}, column name={Ranked}},
	create on use/Correctly ranked/.style={
		create col/assign/.code={
		\pgfmathprintpmnumber{11}{12}{0}{0}
		\pgfkeyslet{/pgfplots/table/create col/next content}\value
		}
	},
	columns={Model,Mean square error,Optimal actions found,Correctly ranked},
	every row no 0 column no 2/.style={postproc cell content/.append style={@cell content/.add={\cellcolor{red!25}}{}}},
	every row no 1 column no 1/.style={postproc cell content/.append style={@cell content/.add={\cellcolor{green!25}}{}}},
	every row no 2 column no 1/.style={postproc cell content/.append style={@cell content/.add={\cellcolor{green!25}}{}}},
	every row no 2 column no 2/.style={postproc cell content/.append style={@cell content/.add={\cellcolor{green!25}}{}}},
	every row no 2 column no 3/.style={postproc cell content/.append style={@cell content/.add={\cellcolor{green!25}}{}}},
	every row no 3 column no 1/.style={postproc cell content/.append style={@cell content/.add={\cellcolor{green!25}}{}}},
	every row no 3 column no 2/.style={postproc cell content/.append style={@cell content/.add={\cellcolor{green!25}}{}}},
	every row no 3 column no 3/.style={postproc cell content/.append style={@cell content/.add={\cellcolor{green!25}}{}}},
	every row no 4 column no 1/.style={postproc cell content/.append style={@cell content/.add={\cellcolor{green!25}}{}}},
	every row no 5 column no 1/.style={postproc cell content/.append style={@cell content/.add={\cellcolor{green!25}}{}}},
	every row no 5 column no 2/.style={postproc cell content/.append style={@cell content/.add={\cellcolor{green!25}}{}}},
	every row no 5 column no 3/.style={postproc cell content/.append style={@cell content/.add={\cellcolor{green!25}}{}}},
	every row no 6 column no 1/.style={postproc cell content/.append style={@cell content/.add={\cellcolor{green!25}}{}}},
	every row no 6 column no 2/.style={postproc cell content/.append style={@cell content/.add={\cellcolor{green!25}}{}}},
	every row no 6 column no 3/.style={postproc cell content/.append style={@cell content/.add={\cellcolor{green!25}}{}}},
	every row no 7 column no 1/.style={postproc cell content/.append style={@cell content/.add={\cellcolor{red!25}}{}}},
	every row no 8 column no 2/.style={postproc cell content/.append style={@cell content/.add={\cellcolor{green!25}}{}}},
	every row no 8 column no 3/.style={postproc cell content/.append style={@cell content/.add={\cellcolor{green!25}}{}}},
	every row no 9 column no 2/.style={postproc cell content/.append style={@cell content/.add={\cellcolor{green!25}}{}}},
	every row no 9 column no 3/.style={postproc cell content/.append style={@cell content/.add={\cellcolor{green!25}}{}}},
	every row no 10 column no 3/.style={postproc cell content/.append style={@cell content/.add={\cellcolor{red!25}}{}}},
	every row no 11 column no 3/.style={postproc cell content/.append style={@cell content/.add={\cellcolor{red!25}}{}}},
	every row no 12 column no 1/.style={postproc cell content/.append style={@cell content/.add={\cellcolor{red!25}}{}}},
	every row no 12 column no 2/.style={postproc cell content/.append style={@cell content/.add={\cellcolor{green!25}}{}}},
	every row no 12 column no 3/.style={postproc cell content/.append style={@cell content/.add={\cellcolor{green!25}}{}}},
]{csv/GeneralizedFireFighting.csv}
\vspace{\baselineskip}
\caption{Best (green) and worst (red) performing methods on the Generalized Firefighting problem.}
\label{tab:generalized_ff}
\end{table}

Although the joint action space is very large here (more than $8000$ joint actions), most of the factored methods achieves very good performance both in terms of MSE (factored $Q$-function learning approach methods) and action ranking. Also smaller factorizations like the overlapping factors ones are able to identify almost all of the optimal actions and produce a very good ranking. Both methods using the true factorizations are doing very well, with also the mixture of experts one identifying all of the optimal actions. On the other hand, the joint learner is failing in this task, being outperformed even by M1 (that has a higher MSE but a better ranking).

\begin{figure}[htbp]
\centering
\begin{tikzpicture}
\tikzset{dot/.style 2 args={fill, regular polygon, regular polygon sides=7, inner sep=3pt, label={#1:\scriptsize #2}}}
\def\islands{{4,5,6,1,2,3}} 
\def\cols{2}
\def\rows{1}
\def\scale{1.5}
\foreach \x [count=\n] in {0,...,\cols}{
\foreach \y in {0,...,\rows}{
\pgfmathtruncatemacro\tmp{\n+(\cols+1)*\y}
\pgfmathtruncatemacro\num{\islands[\tmp-1]}
\draw[line width=.5pt] (\x*\scale,0) -- (\x*\scale,\rows*\scale);
\draw[line width=.5pt] (0,\y*\scale) -- (\cols*\scale,\y*\scale);
\node[dot={45}{\num}] at (\x*\scale,\y*\scale) {};
}
}
\end{tikzpicture}
\caption{Islands configuration with $n=6$ agents.}
\label{fig:islands}
\end{figure}

\begin{figure}[htbp]
\centering
\includegraphics[width=0.68\textwidth]{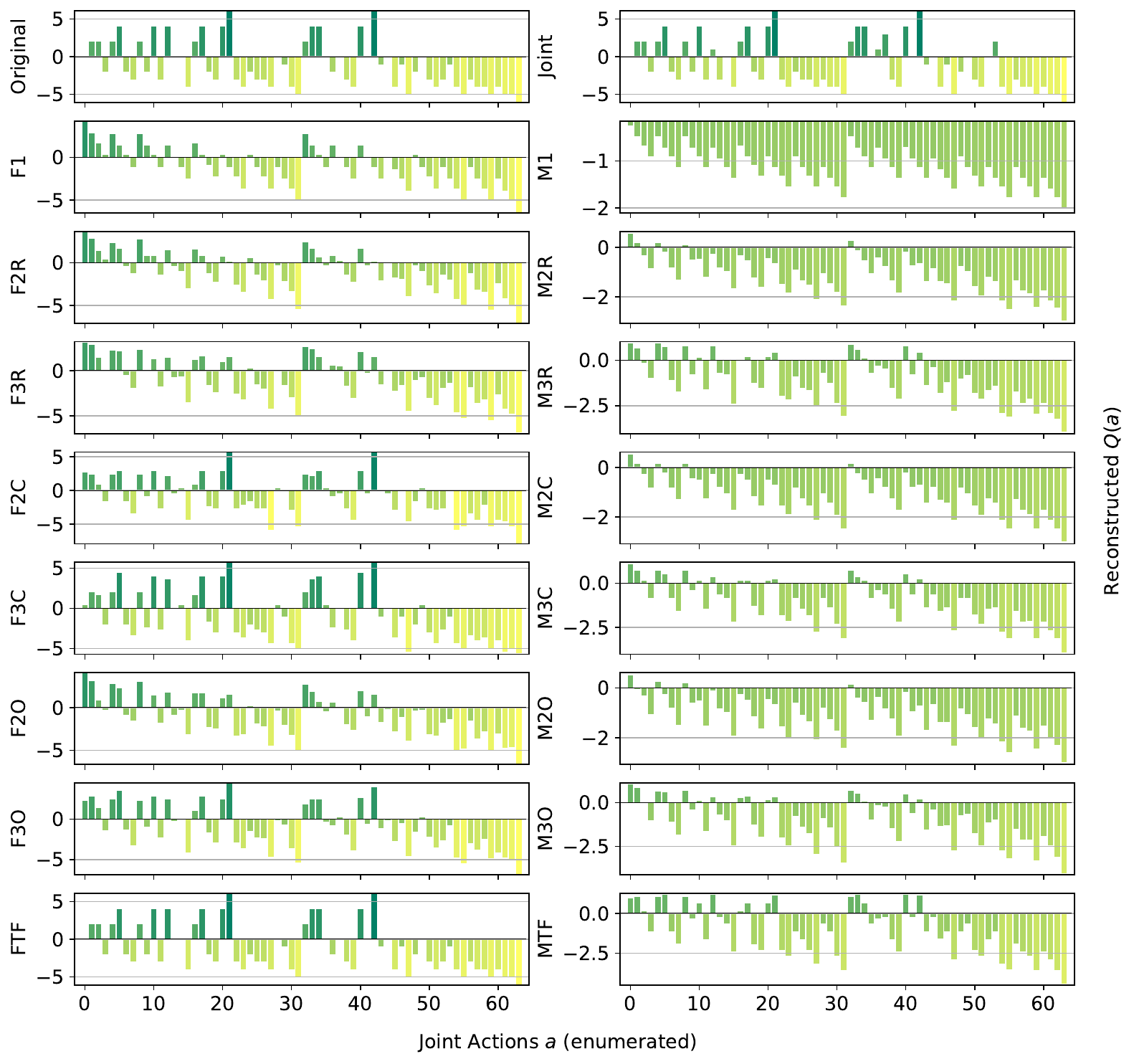}
\caption{Reconstructed $Q(a)$ for Aloha.}
\label{fig:aloha}
\end{figure}

\emph{Aloha:} Our experiment uses a set of $n=6$ islands disposed in a $2\times 3$ grid as in Figure \ref{fig:islands}, with each island affected only by the transmissions of the islands on their sides and in front of them (islands on the corner of the grid miss one of their side neighbours). Representations learned for this game are reported in Figure \ref{fig:aloha}.

The plot shows clearly how this game is challenging for the proposed factorizations to learn, with only three of them (plus the joint learner) able to correctly represent the action-value function. The structure of the game is similar to that of Generalized Firefighting, with an agent depending directly only on a small subset of the others, but the different properties of its $Q$-function make it more challenging to correctly represent. This is possibly due to the large difference between the two rewards an agent can get when transmitting the radio message, depending on a potential interference. Observing only the total reward, this action looks neutral per se, similarly to what happens for the two actions in the Dispersion Game, its outcome depending on the action of the neighbouring agents, thus possibly fooling many of the proposed factorizations, especially those using the mixture of experts approach. Table \ref{tab:aloha} is showing the best and worst performing methods on this game.

\begin{table}[htbp]
\centering
\pgfplotstableset{
alias/Model/.initial=0,
}

\begin{filecontents*}{csv/Aloha.csv}
Joint,1.125,0.0,0.0,0.0,2.0,0.0,0.0,0.0,0.076,0.0,50.7,0.9,0.876,0.011
F2R,4.054,0.439,35.002,6.975,0.0,0.0,5.0,1.342,3.69,0.375,22.2,4.445,0.696,0.025
F3R,3.164,0.531,20.644,4.614,0.0,0.0,4.2,1.4,3.234,0.856,25.7,3.874,0.741,0.04
F2C,0.914,0.0,0.138,0.007,2.0,0.0,0.0,0.0,-0.038,0.003,42.0,0.0,0.893,0.0
F3C,0.07,0.0,0.138,0.005,2.0,0.0,0.0,0.0,0.22,0.003,64.0,0.0,1.0,0.0
F2O,3.266,0.267,20.63,3.049,0.0,0.0,4.4,1.2,3.236,0.531,22.7,4.451,0.743,0.028
F3O,1.456,0.319,3.554,1.349,1.2,0.6,0.8,1.327,1.185,0.394,28.7,4.562,0.832,0.027
FTF,0.0,0.0,0.0,0.0,2.0,0.0,0.0,0.0,-0.0,0.0,64.0,0.0,1.0,0.0
M1,8.262,0.005,50.836,0.061,0.0,0.0,6.0,0.0,5.466,0.002,26.5,0.5,0.67,0.01
M2R,6.522,0.156,44.165,1.535,0.0,0.0,5.0,1.342,4.569,0.109,24.5,3.956,0.704,0.021
M2O,6.631,0.389,44.51,1.136,0.0,0.0,5.2,0.98,4.624,0.183,22.0,4.775,0.663,0.043
M3O,4.711,0.287,33.363,1.186,0.0,0.0,5.2,0.98,3.647,0.163,25.4,4.2,0.738,0.046
\end{filecontents*}
\pgfplotstabletypeset[col sep=comma,
	every head row/.style={before row=\toprule, after row=\midrule},
    every last row/.style={after row=\bottomrule},
	header=false,
	fixed,
	zerofill,
    columns/Model/.style={column type={C{0.9cm}},string type},
	columns/Mean square error/.style={string type, column type={C{1.5cm}}, column name={MSE}},
	create on use/Mean square error/.style={
		create col/assign/.code={
		\pgfmathprintpmnumber{1}{2}{2}{1}
		\pgfkeyslet{/pgfplots/table/create col/next content}\value
		}
	},
	columns/Optimal actions found/.style={string type, column type={C{1.6cm}}, column name={Opt. Found}},
	create on use/Optimal actions found/.style={
		create col/assign/.code={
		\pgfmathprintpmnumber{5}{6}{0}{0}
	    \pgfkeyslet{/pgfplots/table/create col/next content}\value
		}
	},
	columns/Correctly ranked/.style={string type, column type={C{1.8cm}}, column name={Ranked}},
	create on use/Correctly ranked/.style={
		create col/assign/.code={
		\pgfmathprintpmnumber{11}{12}{0}{0}
		\pgfkeyslet{/pgfplots/table/create col/next content}\value
		}
	},
	columns={Model,Mean square error,Optimal actions found,Correctly ranked},
	every row no 0 column no 2/.style={postproc cell content/.append style={@cell content/.add={\cellcolor{green!25}}{}}},
	every row no 1 column no 3/.style={postproc cell content/.append style={@cell content/.add={\cellcolor{red!25}}{}}},
	every row no 2 column no 3/.style={postproc cell content/.append style={@cell content/.add={\cellcolor{red!25}}{}}},
	every row no 3 column no 2/.style={postproc cell content/.append style={@cell content/.add={\cellcolor{green!25}}{}}},
	every row no 4 column no 1/.style={postproc cell content/.append style={@cell content/.add={\cellcolor{green!25}}{}}},
	every row no 4 column no 2/.style={postproc cell content/.append style={@cell content/.add={\cellcolor{green!25}}{}}},
	every row no 4 column no 3/.style={postproc cell content/.append style={@cell content/.add={\cellcolor{green!25}}{}}},
	every row no 5 column no 3/.style={postproc cell content/.append style={@cell content/.add={\cellcolor{red!25}}{}}},
	every row no 6 column no 2/.style={postproc cell content/.append style={@cell content/.add={\cellcolor{green!25}}{}}},
	every row no 7 column no 1/.style={postproc cell content/.append style={@cell content/.add={\cellcolor{green!25}}{}}},
	every row no 7 column no 2/.style={postproc cell content/.append style={@cell content/.add={\cellcolor{green!25}}{}}},
	every row no 7 column no 3/.style={postproc cell content/.append style={@cell content/.add={\cellcolor{green!25}}{}}},
	every row no 8 column no 1/.style={postproc cell content/.append style={@cell content/.add={\cellcolor{red!25}}{}}},
	every row no 9 column no 3/.style={postproc cell content/.append style={@cell content/.add={\cellcolor{red!25}}{}}},
	every row no 10 column no 3/.style={postproc cell content/.append style={@cell content/.add={\cellcolor{red!25}}{}}},
	every row no 11 column no 3/.style={postproc cell content/.append style={@cell content/.add={\cellcolor{red!25}}{}}},
]{csv/Aloha.csv}
\vspace{\baselineskip}
\caption{Best (green) and worst (red) performing methods on Aloha.}
\label{tab:aloha}
\end{table}

On this more difficult game, all of the mixture of experts methods are not able to identify the optimal actions and achieve a very large MSE. However, the complete factorizations using the factored $Q$-function learning approach are able to do so, with F3C also ranking correctly all of the other joint actions. Again, FTF is performing the best, with a perfect ranking and a very low reconstruction error, outperforming even the joint learner. This once more shows how beneficial would it be to exploit an appropriate factored structure when that is known beforehand.

\subsection{Impact of Factors Size}
\label{sec:size}
Although we mainly focus on factors of small size, we are also interested in investigating how the size of the factors is affecting the final representation, and if using factors of larger size can help to overcome some of the issues encountered with small factors. To investigate this, we test the methods defined in Table \ref{tab:larger} on the Platonia Dilemma and Penalty Game with $n=6$ agents, two of the games that proved more problematic to correctly represent.

\begin{table}[htbp]
\centering
\begin{tabular}{ccc}
\toprule
 & Mix. of Experts & Factored $Q$ \\
\midrule
Random partition ($f=4,5$) & M4R, M5R & F4R, F5R \\
Complete factorization ($f=4,5$) & M4C, M5C & F4C, F5C \\
Overlapping factors ($f=4,5$) & M4O, M5O & F4O, F5O \\
\bottomrule
\end{tabular}
\vspace{\baselineskip}
\caption{Combinations of factorizations and learning rules with larger factors.}
\label{tab:larger}
\end{table}

\begin{figure}[htbp]
\centering
\includegraphics[width=0.68\textwidth]{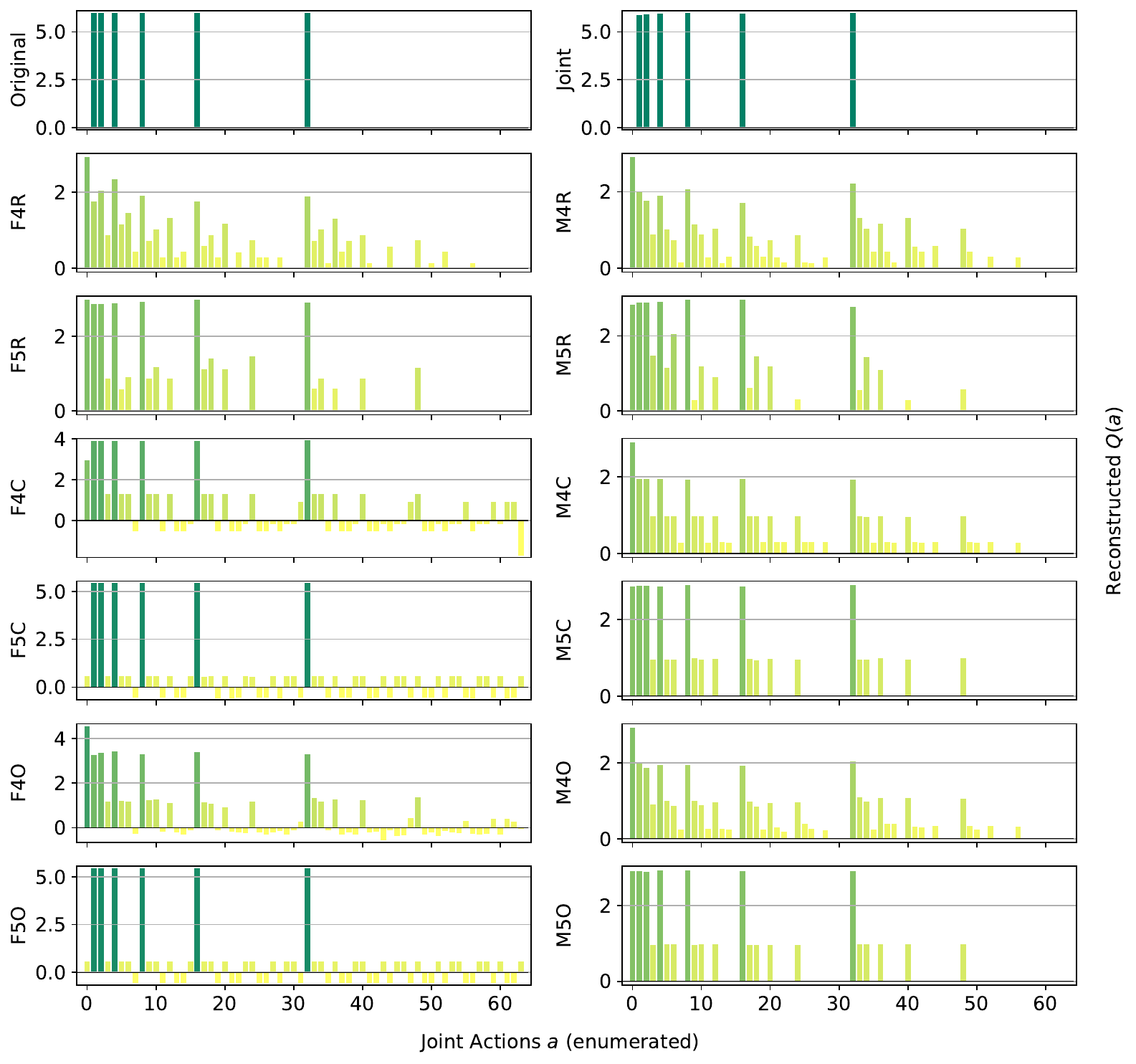}
\caption{Reconstructed $Q(a)$ for the Platonia Dilemma.}
\label{fig:platonia_lf}
\end{figure}

\emph{Platonia Dilemma:} Figure \ref{fig:platonia_lf} shows the reconstructed action-value functions for the Platonia Dilemma. We can see how this game, that none of the factored methods in Figure \ref{fig:platonia} was able to solve, remains very challenging even with factors comprising more agents. Indeed, only methods with a factor size $f=5$, thus very close to the entire team size $n=6$, and using the factored $Q$-function learning approach (F5C and F5O), are able to correctly reconstruct the action-value function. The same factorizations using the mixture of experts learning approach are instead consistently ranking one of the suboptimal actions (the one in which none of the agents is sending the telegram) as an optimal one, the same as with factors of smaller size.

\begin{figure}[htbp]
\centering
\subfigure[\label{sub:penalty_factored_lf}]{
\includegraphics[width=0.68\textwidth]{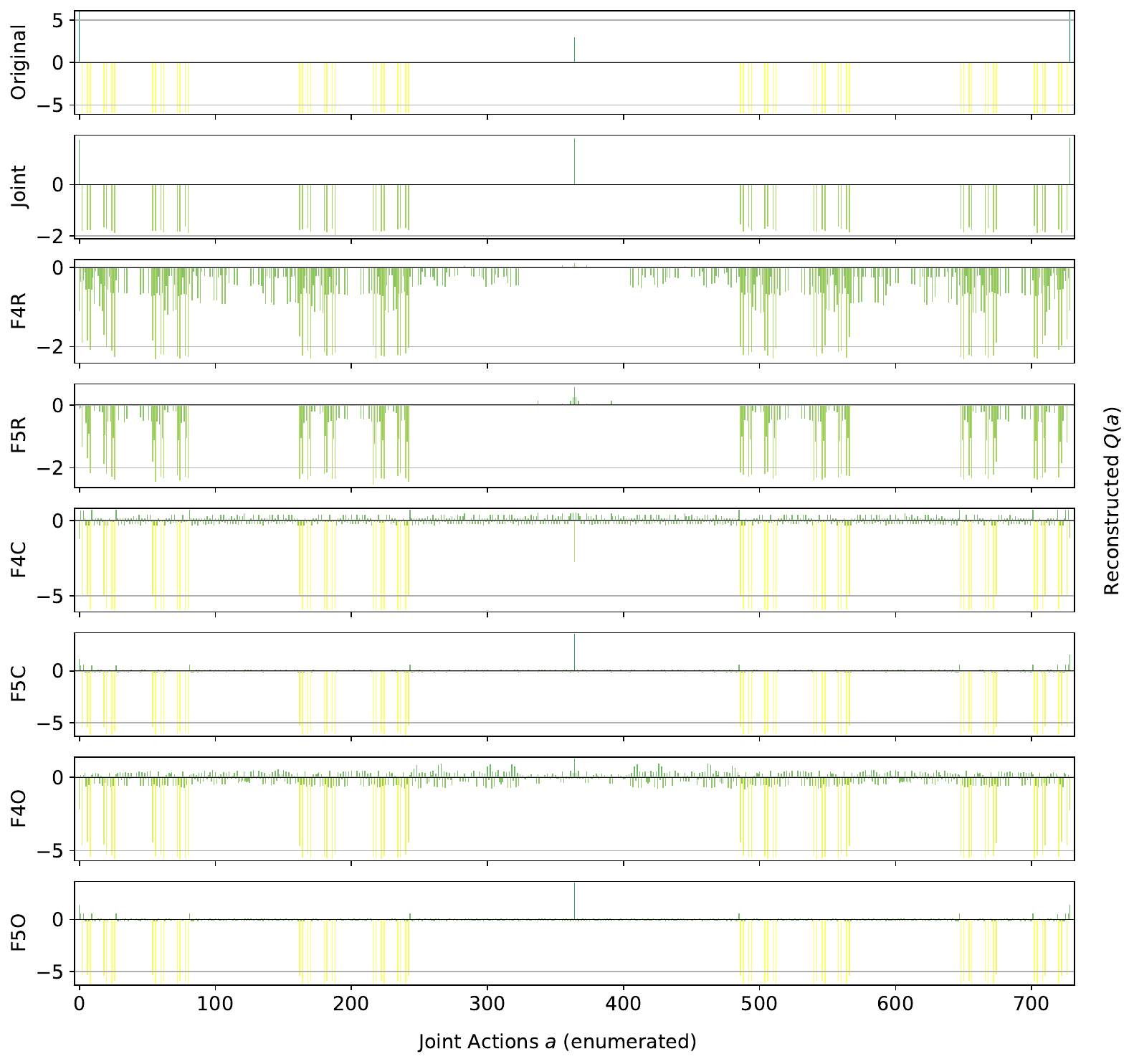}
}
\\
\subfigure[\label{sub:penalty_moe_lf}]{
\includegraphics[width=0.68\textwidth]{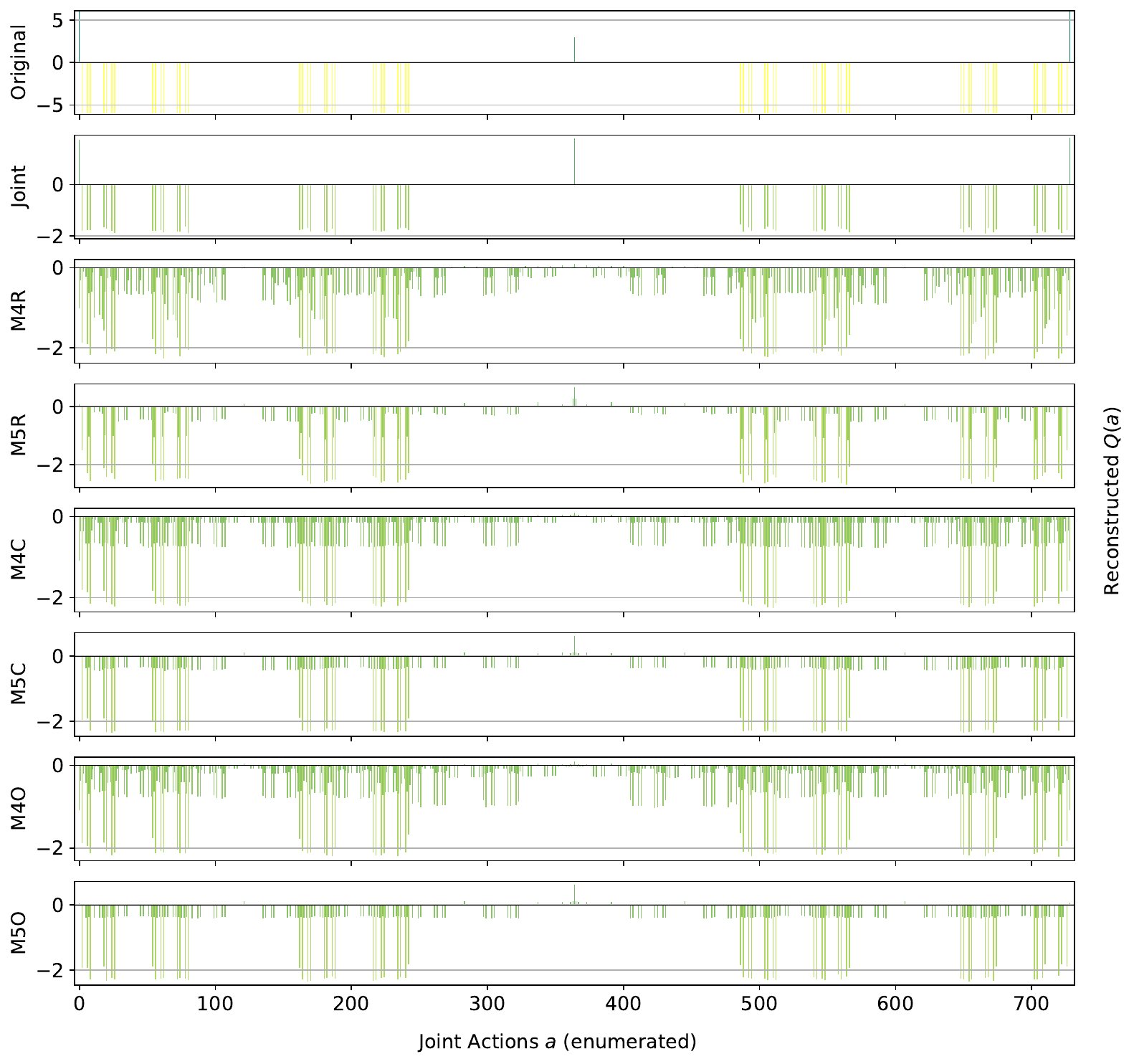}
}
\caption{Reconstructed $Q(a)$ for the Penalty Game \protect\subref{sub:penalty_factored_lf} factored $Q$-function learning approach, and \protect\subref{sub:penalty_moe_lf} the mixture of experts learning approach.}
\label{fig:penalty_lf}
\end{figure}

\emph{Penalty Game:} Figure \ref{fig:penalty_lf} presents the representations obtained by the new investigated approximations. Even with larger factors, none of the methods is able to reconstruct any of the optimal actions, but they only are able to discern the value of the sub-optimal one like F1 and F3C in Figure \ref{fig:penalty} (that is seen as optimal). The same kind of problems that arose with smaller factors are also present here, with the mixture of experts methods tending to underestimate values for all the joint actions and generally none of the methods being able to represent the true value of coordination for this problem. However, the methods using a complete factorization coupled the factored $Q$-function learning approach are reconstructing small yet positive values for these optimal actions, meaning that the resulting reconstruction is at least identifying these as good actions that the agents may desire to perform.

\subsection{Scalability}
\label{sec:larger}
A fundamental aspect for a multi-agent algorithm is how well it can scale with the size of the system, i.e. when more agents are introduced and therefore the size of the joint action set exponentially increases. In this section we investigate how using a factored representation helps when such systems get larger, as well as analyse how this affects the performance of both independent learners and joint learners. Table \ref{tab:larger_games} illustrates the game we use to investigate this:

\begin{table}[htbp]
\centering
\begin{tabular}{cccccc}
\toprule
Game & $n$ & $|A_i|$ & $|A|$ & Optimal & Factored \\
\midrule
Dispersion Game & $9$ & $2$ & $512$ & $252$ & No \\
Dispersion Game & $12$ & $2$ & $4096$ & $924$ & No \\
Generalized Firefighting & $9$ & $2$ (per type) & $512$ ($524.288$ total) & $17.682$ & Yes \\
Aloha & $9$ & $2$ & $512$ & $1$ & Yes \\
Aloha & $12$ & $2$ & $4096$ & $2$ & Yes \\
\bottomrule
\end{tabular}
\vspace{\baselineskip}
\caption{Details of the investigated games in this Section.}
\label{tab:larger_games}
\end{table}

\begin{figure}[htbp]
\centering
\includegraphics[width=0.68\textwidth]{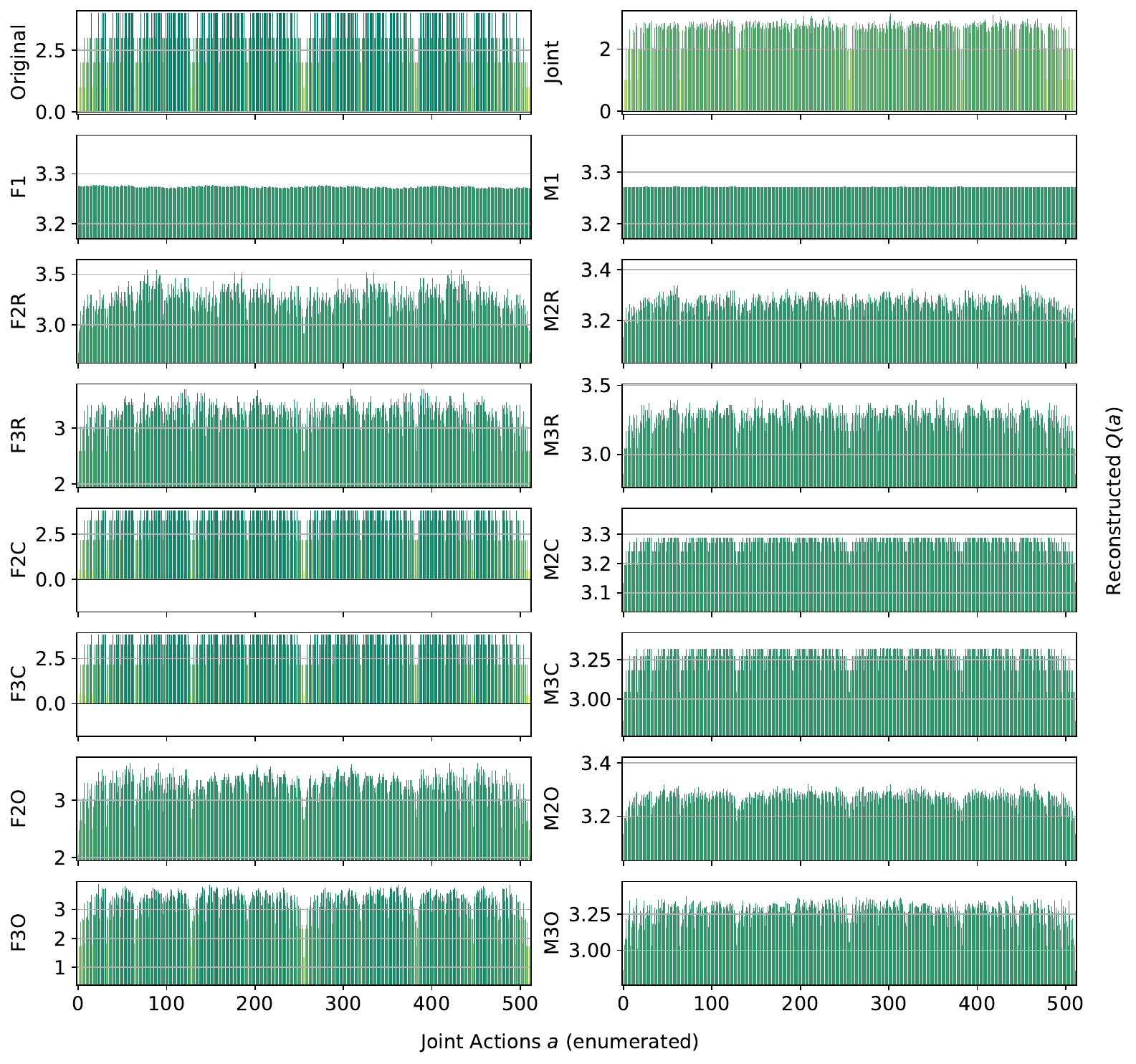}
\caption{Reconstructed $Q(a)$ for the Dispersion Game with $n=9$ agents.}
\label{fig:dispersion_large}
\end{figure}

\emph{Dispersion Games:} Figure \ref{fig:dispersion_large} shows the action-value function reconstructed by the proposed factorizations and learning approaches for the Dispersion Game with $n=9$ agents (a similar figure for the case when $n=12$ would have rendered unreadable and is thereby not included). We can observe how the complete factorizations are able to almost perfectly reconstruct the relative ranking between the joint actions even in this larger setting, showing how reliable and general can this kind of approach be. As usual, the ones using the factored $Q$-function are also able to produce a generally good approximation of the various components, while those based on the mixture of experts produce a less precise reconstruction: the joint optimization of the former seems to have an even bigger benefit when more agents are present. 

It is interesting to note how both independent learners and the joint learner are failing here, but for different reasons: both types of independent learners seem not able to correctly learn the value of coordination with the others (something already appearing on the smaller instance shown in Figure \ref{fig:dispersion}), while the latter is struggling because of the increased number of agents that makes the function it has to represent too big to be reliably learned in the given training time. The other factored approaches instead are capturing the value of such coordination up to some extent (especially these using the overlapping factors), but the small number of factors is probably not sufficient to completely represent such a function. However, the resulting MSE is still lower than the joint learner and some of the optimal actions are still ranked correctly, making these approaches still viable for decision making. Table \ref{tab:dispersion_large} reports the best and worst performing methods on the two instances of this game, both in terms of action ranking and reconstruction error.

\begin{table}[htbp]
\centering
\pgfplotstableset{
alias/Model/.initial=0,
}

\begin{filecontents*}{csv/DispersionGameLarge.csv}
% Dispersion Game n=9 (512 joint actions, 252 optimal)
Joint,0.933,0.488,1.743,0.873,162.2,30.318,0.0,0.0,0.247,0.094,307.3,77.121,0.488,0.188
F1,0.736,0.0,0.528,0.006,123.9,0.7,1.1,0.943,0.471,0.0,191.3,2.648,0.017,0.032
F2C,0.063,0.0,0.032,0.002,252.0,0.0,0.0,0.0,0.089,0.0,512.0,0.0,1.0,0.0
F3C,0.063,0.0,0.033,0.002,252.0,0.0,0.0,0.0,0.088,0.001,512.0,0.0,1.0,0.0
M1,0.736,0.0,0.531,0.005,124.3,0.781,0.9,0.831,0.471,0.0,191.8,3.628,0.005,0.027
M2C,0.699,0.0,0.508,0.004,252.0,0.0,0.0,0.0,0.453,0.0,512.0,0.0,1.0,0.0
M3C,0.628,0.0,0.463,0.003,252.0,0.0,0.0,0.0,0.418,0.0,512.0,0.0,1.0,0.0
% Dispersion Game n=12 (4096 joint actions, 924 optimal)
Joint,19.477,0.688,31.715,0.897,209.6,8.476,2.3,1.005,0.796,0.001,1107.8,33.973,0.003,0.011
F1,1.168,0.0,1.818,0.016,186.0,39.426,2.4,1.281,0.797,0.0,1136.9,36.621,0.006,0.009
F3R,0.986,0.0,1.665,0.009,350.5,6.515,0.9,0.7,0.629,0.001,1363.5,19.941,0.28,0.009
F2C,0.169,0.0,0.373,0.011,924.0,0.0,0.0,0.0,0.268,0.001,4096.0,0.0,1.0,0.0
F3C,0.198,0.044,0.433,0.096,773.5,193.822,0.0,0.0,0.267,0.003,3711.0,510.039,0.923,0.101
M1,1.168,0.0,1.827,0.01,186.9,30.373,2.2,1.536,0.797,0.0,1134.3,36.081,0.004,0.008
M3R,1.088,0.0,1.746,0.006,350.3,9.306,1.0,0.632,0.752,0.0,1362.9,23.543,0.276,0.008
M2C,1.138,0.0,1.796,0.005,813.2,69.452,0.0,0.0,0.782,0.0,3831.4,245.751,0.946,0.049
M3C,1.079,0.0,1.739,0.005,920.4,5.004,0.0,0.0,0.753,0.0,4088.8,10.008,0.999,0.002
\end{filecontents*}
\pgfplotstabletypeset[col sep=comma,
	every head row/.style={before row=\toprule, after row=\midrule},
    every last row/.style={after row=\bottomrule},
	header=false,
	fixed,
	zerofill,
    columns/Model/.style={column type={C{0.9cm}},string type},
	columns/Mean square error/.style={string type, column type={C{1.5cm}}, column name={MSE}},
	create on use/Mean square error/.style={
		create col/assign/.code={
		\pgfmathprintpmnumber{1}{2}{2}{1}
		\pgfkeyslet{/pgfplots/table/create col/next content}\value
		}
	},
	columns/Optimal actions found/.style={string type, column type={C{1.6cm}}, column name={Opt. Found}},
	create on use/Optimal actions found/.style={
		create col/assign/.code={
		\pgfmathprintpmnumber{5}{6}{0}{0}
	    \pgfkeyslet{/pgfplots/table/create col/next content}\value
		}
	},
	columns/Correctly ranked/.style={string type, column type={C{1.8cm}}, column name={Ranked}},
	create on use/Correctly ranked/.style={
		create col/assign/.code={
		\pgfmathprintpmnumber{11}{12}{0}{0}
		\pgfkeyslet{/pgfplots/table/create col/next content}\value
		}
	},
	columns={Model,Mean square error,Optimal actions found,Correctly ranked},
	every row no 0/.style={before row={\multicolumn{4}{c}{\textbf{Dispersion Game $n=9$}}\\\midrule}},
	every row no 7/.style={before row={\midrule\multicolumn{4}{c}{\textbf{Dispersion Game $n=12$}}\\\midrule}},
	every row no 0 column no 1/.style={postproc cell content/.append style={@cell content/.add={\cellcolor{red!25}}{}}},
	every row no 1 column no 2/.style={postproc cell content/.append style={@cell content/.add={\cellcolor{red!25}}{}}},
	every row no 1 column no 3/.style={postproc cell content/.append style={@cell content/.add={\cellcolor{red!25}}{}}},
	every row no 2 column no 1/.style={postproc cell content/.append style={@cell content/.add={\cellcolor{green!25}}{}}},
	every row no 2 column no 2/.style={postproc cell content/.append style={@cell content/.add={\cellcolor{green!25}}{}}},
	every row no 2 column no 3/.style={postproc cell content/.append style={@cell content/.add={\cellcolor{green!25}}{}}},
	every row no 3 column no 1/.style={postproc cell content/.append style={@cell content/.add={\cellcolor{green!25}}{}}},
	every row no 3 column no 2/.style={postproc cell content/.append style={@cell content/.add={\cellcolor{green!25}}{}}},
	every row no 3 column no 3/.style={postproc cell content/.append style={@cell content/.add={\cellcolor{green!25}}{}}},
	every row no 4 column no 2/.style={postproc cell content/.append style={@cell content/.add={\cellcolor{red!25}}{}}},
	every row no 4 column no 3/.style={postproc cell content/.append style={@cell content/.add={\cellcolor{red!25}}{}}},
	every row no 5 column no 2/.style={postproc cell content/.append style={@cell content/.add={\cellcolor{green!25}}{}}},
	every row no 5 column no 3/.style={postproc cell content/.append style={@cell content/.add={\cellcolor{green!25}}{}}},
	every row no 6 column no 2/.style={postproc cell content/.append style={@cell content/.add={\cellcolor{green!25}}{}}},
	every row no 6 column no 3/.style={postproc cell content/.append style={@cell content/.add={\cellcolor{green!25}}{}}},
	every row no 7 column no 1/.style={postproc cell content/.append style={@cell content/.add={\cellcolor{red!25}}{}}},
	every row no 7 column no 2/.style={postproc cell content/.append style={@cell content/.add={\cellcolor{red!25}}{}}},
	every row no 7 column no 3/.style={postproc cell content/.append style={@cell content/.add={\cellcolor{red!25}}{}}},
	every row no 8 column no 2/.style={postproc cell content/.append style={@cell content/.add={\cellcolor{red!25}}{}}},
	every row no 8 column no 3/.style={postproc cell content/.append style={@cell content/.add={\cellcolor{red!25}}{}}},
	every row no 9 column no 3/.style={postproc cell content/.append style={@cell content/.add={\cellcolor{red!25}}{}}},
	every row no 10 column no 1/.style={postproc cell content/.append style={@cell content/.add={\cellcolor{green!25}}{}}},
	every row no 10 column no 2/.style={postproc cell content/.append style={@cell content/.add={\cellcolor{green!25}}{}}},
	every row no 10 column no 3/.style={postproc cell content/.append style={@cell content/.add={\cellcolor{green!25}}{}}},
	every row no 11 column no 1/.style={postproc cell content/.append style={@cell content/.add={\cellcolor{green!25}}{}}},
	every row no 11 column no 2/.style={postproc cell content/.append style={@cell content/.add={\cellcolor{green!25}}{}}},
	every row no 11 column no 3/.style={postproc cell content/.append style={@cell content/.add={\cellcolor{green!25}}{}}},
	every row no 12 column no 2/.style={postproc cell content/.append style={@cell content/.add={\cellcolor{red!25}}{}}},
	every row no 12 column no 3/.style={postproc cell content/.append style={@cell content/.add={\cellcolor{red!25}}{}}},
	every row no 13 column no 3/.style={postproc cell content/.append style={@cell content/.add={\cellcolor{red!25}}{}}},
	every row no 14 column no 2/.style={postproc cell content/.append style={@cell content/.add={\cellcolor{green!25}}{}}},
	every row no 14 column no 3/.style={postproc cell content/.append style={@cell content/.add={\cellcolor{green!25}}{}}},
	every row no 15 column no 2/.style={postproc cell content/.append style={@cell content/.add={\cellcolor{green!25}}{}}},
	every row no 15 column no 3/.style={postproc cell content/.append style={@cell content/.add={\cellcolor{green!25}}{}}},
]{csv/DispersionGameLarge.csv}
\vspace{\baselineskip}
\caption{Best (green) and worst (red) performing methods on the two instances of the Dispersion Game.}
\label{tab:dispersion_large}
\end{table}

As already stated, when the size of the system increase both independent learners and the joint learner struggle in representing the corresponding action-value function correctly. The latter especially, that was achieving a perfect reconstruction for the same game with only $n=6$ agents, is now resulting in a higher reconstruction error and fail in identifying all of the optimal joint actions. Methods using a complete factorization with both learning approaches instead are still able to identify most of them (all, when $n=9$), while at the same time reducing the MSE considerably. Smaller factorizations are not reported because these aBLA BLA BLAre not achieving such good performances (as in the smaller case with $n=6$), showing that on this kind of very tightly coordinated problems these may not suffice for a completely correct representation.

\begin{figure}[htbp]
\centering
\begin{tikzpicture}
\tikzset{dot/.style 2 args={fill, rectangle, inner sep=5pt, label={#1:\scriptsize #2}}}
\foreach \x in {0,...,9}{
\pgfmathtruncatemacro\h{\x+1}
\node[dot={90}{$H_{\h}$}] at (\x,0) {};
}
\tikzset{dot/.style 2 args={fill, circle, inner sep=3pt, label={#1:\scriptsize #2}}}
\foreach \x in {0,...,8}{
\pgfmathtruncatemacro\n{\x+1}
\draw[line width=.5pt] (\x+0.5,-1) -- (\x,0);
\draw[line width=.5pt] (\x+0.5,-1) -- (\x+1,0);
\node[dot={-90}{\n}] at (\x+0.5,-1) {};
}
\end{tikzpicture}
\caption{Firefighters formation with $n=9$ agents and $N_h=10$ houses.}
\label{fig:formation_large}
\end{figure}

\begin{figure}[htbp]
\centering
\includegraphics[width=0.68\textwidth]{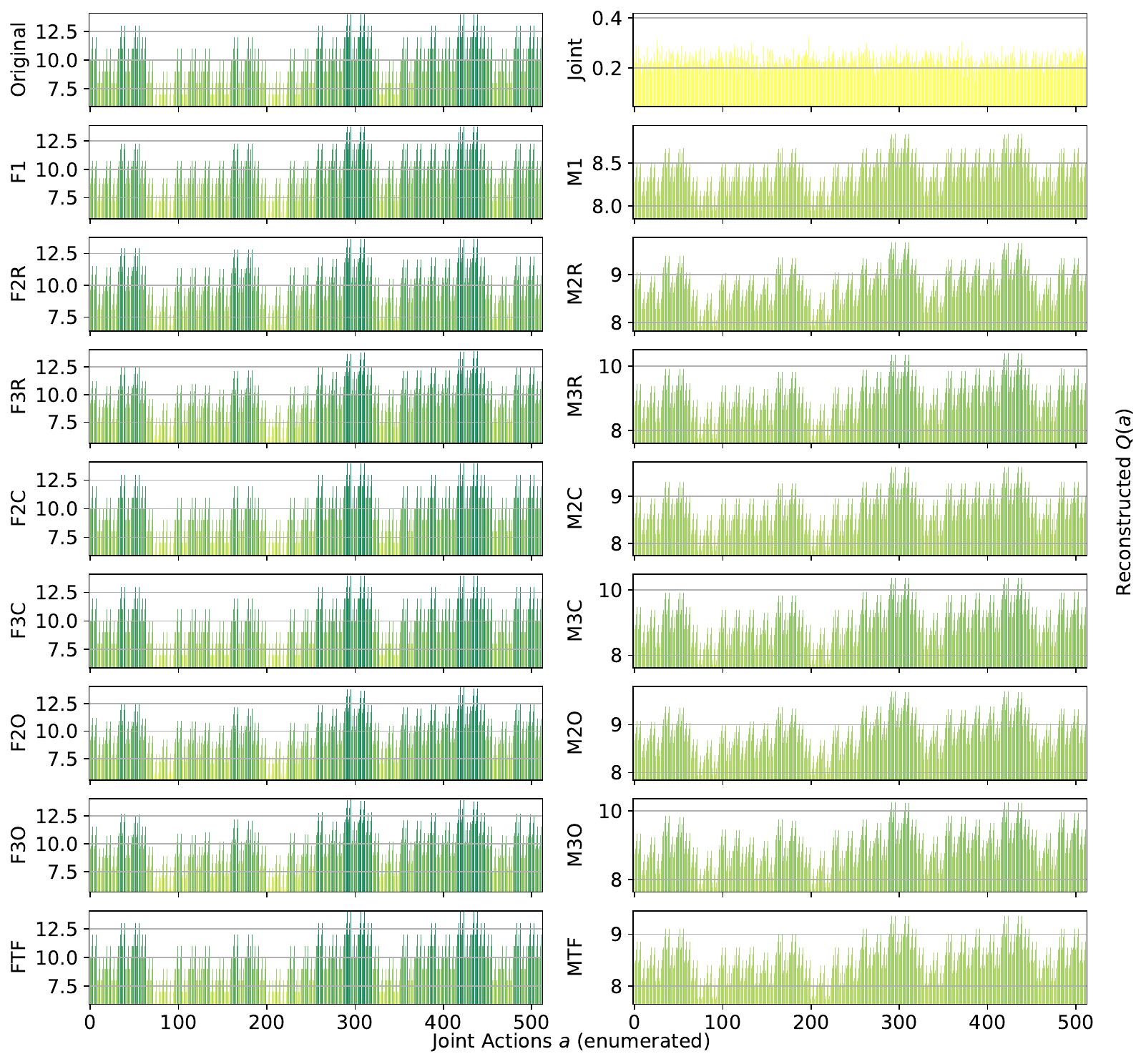}
\caption{Reconstructed $Q(a)$ for a single joint type of the Generalized Firefighting problem with $n=9$ agents.}
\label{fig:generalized_ff1_large}
\end{figure}

\emph{Generalized Firefighting:} In this larger experiments, a team of $n=9$ agents is fighting fire at $N_h=10$ houses. As in the previous setting, each agent can observe $N_o=2$ houses and can fight fire at the same set of locations ($N_a=2$), as shown in Figure \ref{fig:formation_large}. Reconstruction results for the joint type $\theta=\{N_1,F_2,F_3,N_4,F_5,F_6,N_7,N_8,\\F_9,F_{10}\}$ are reported in Figure \ref{fig:generalized_ff1_large}.

From these results, we can observe how, although the problem is very large (with more than half a million total joint actions in our formulation) most of the factored methods are perfectly representing the corresponding $Q$-function. While methods using the complete factorization or exploiting the true underlying structure with both learning approaches are capable of achieving a perfect reconstruction, even simpler methods like random pairing with the factored $Q$-function learning approach are capable or almost perfectly reconstrut the values for this joint type. Conversely, the joint learner seems not capable of doing so, resulting in a totally wrong representation that is not close to the original function. Things are similar for a second joint type, $\theta=\{F_1,F_2,N_3,N_4,N_5,N_6,N_7,F_8,F_9,N_{10}\}$, whose resulting learned representations are shown in Figure \ref{fig:generalized_ff_large2}:

\begin{figure}[htbp]
\centering
\includegraphics[width=0.68\textwidth]{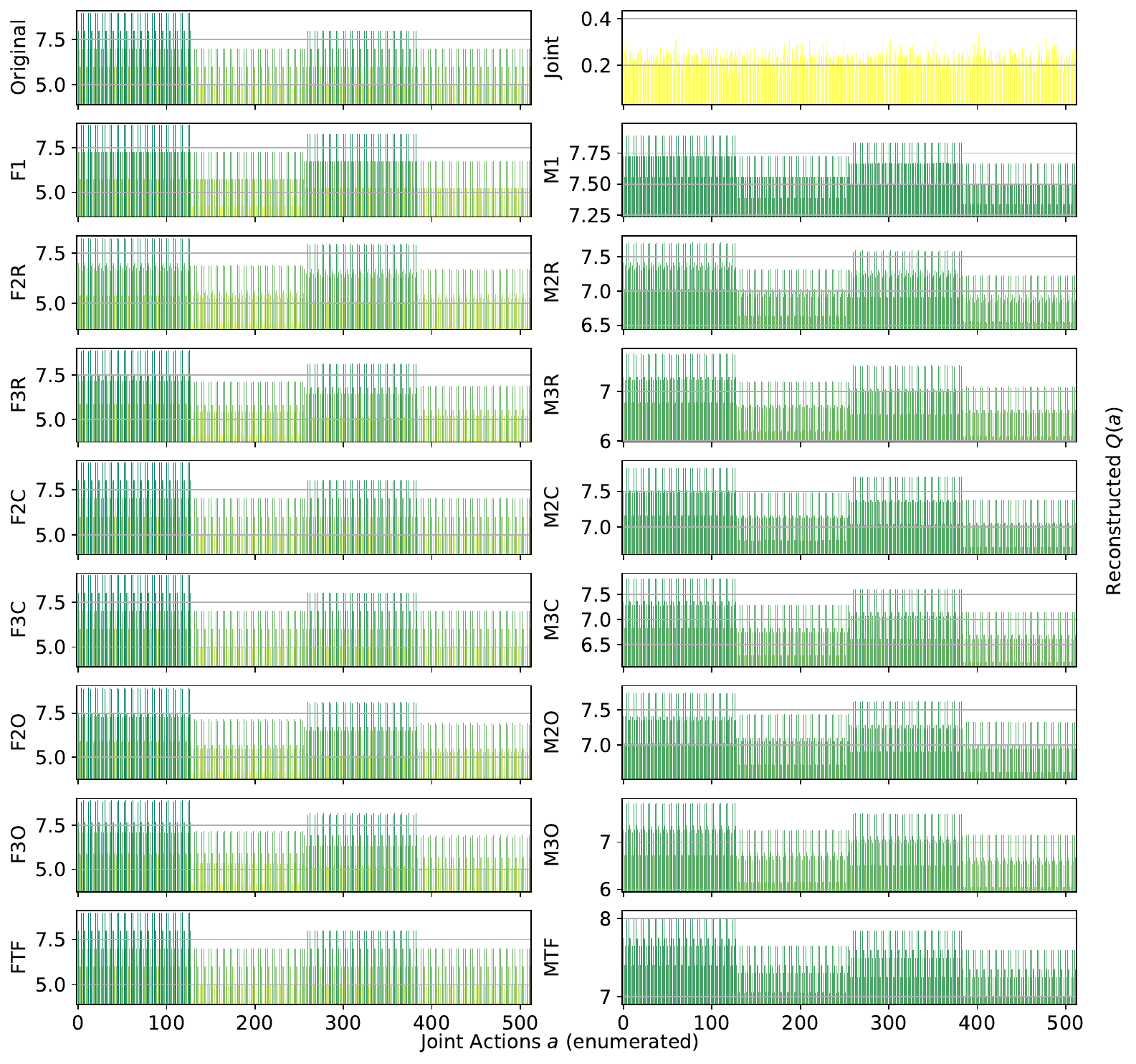}
\caption{Reconstructed $Q(a)$ for a different joint type of the Generalized Firefighting problem with $n=9$ agents.}
\label{fig:generalized_ff_large2}
\end{figure}

Again, the joint learner is not capable of achieving a good representation, but also some of the simpler factorizations are not resulting in a perfect reconstruction, although still capable of correctly identifying the optimal joint actions. Complete factorizations are instead perfectly representing the original $Q$-function for this joint type as well, even with the mixture of experts learning approach. General metrics and results for the best and worst performing methods on this problem are reported in Table \ref{tab:generalized_ff_large}.

\begin{table}[htbp]
\centering
\pgfplotstableset{
alias/Model/.initial=0,
}

\begin{filecontents*}{csv/GeneralizedFireFightingLarge.csv}
Joint,69.977,0.863,142.274,1.291,2556.3,50.257,3808.4,60.81,2339.16,1.281,94559.1,522.156,0.019,0.001
F2C,0.0,0.0,0.0,0.0,17682.0,0.0,0.0,0.0,0.0,0.0,524288.0,0.0,1.0,0.0
F3C,0.0,0.0,0.0,0.0,17682.0,0.0,0.0,0.0,0.0,0.0,524288.0,0.0,1.0,0.0
F3O,0.091,0.026,0.099,0.032,16939.0,414.632,51.4,38.663,22.715,8.577,464893.1,26118.502,0.957,0.018
FTF,0.0,0.0,0.0,0.0,17682.0,0.0,0.0,0.0,0.0,0.0,524288.0,0.0,1.0,0.0
M2C,4.296,0.004,12.833,0.024,17680.0,0.0,0.0,0.0,1721.178,0.878,465018.6,2.375,0.954,0.0
M3C,2.706,0.003,8.99,0.014,17680.0,0.0,0.0,0.0,1422.875,0.729,465089.8,2.088,0.954,0.0
M3O,2.958,0.104,9.238,0.134,16783.2,302.897,54.0,23.921,1431.611,23.036,298182.0,26363.664,0.817,0.032
MTF,5.304,0.004,14.338,0.023,17682.0,0.0,0.0,0.0,1765.642,0.978,512022.4,810.2,0.99,0.001
\end{filecontents*}
\pgfplotstabletypeset[col sep=comma,
	every head row/.style={before row=\toprule, after row=\midrule},
    every last row/.style={after row=\bottomrule},
	header=false,
	fixed,
	zerofill,
    columns/Model/.style={column type={C{0.9cm}},string type},
	columns/Mean square error/.style={string type, column type={C{1.5cm}}, column name={MSE}},
	create on use/Mean square error/.style={
		create col/assign/.code={
		\pgfmathprintpmnumber{1}{2}{2}{1}
		\pgfkeyslet{/pgfplots/table/create col/next content}\value
		}
	},
	columns/Optimal actions found/.style={string type, column type={C{1.6cm}}, column name={Opt. Found}},
	create on use/Optimal actions found/.style={
		create col/assign/.code={
		\pgfmathprintpmnumber{5}{6}{0}{0}
	    \pgfkeyslet{/pgfplots/table/create col/next content}\value
		}
	},
	columns/Correctly ranked/.style={string type, column type={C{1.8cm}}, column name={Ranked}},
	create on use/Correctly ranked/.style={
		create col/assign/.code={
		\pgfmathprintpmnumber{11}{12}{0}{0}
		\pgfkeyslet{/pgfplots/table/create col/next content}\value
		}
	},
	columns={Model,Mean square error,Optimal actions found,Correctly ranked},
	every row no 0 column no 1/.style={postproc cell content/.append style={@cell content/.add={\cellcolor{red!25}}{}}},
	every row no 0 column no 2/.style={postproc cell content/.append style={@cell content/.add={\cellcolor{red!25}}{}}},
	every row no 0 column no 3/.style={postproc cell content/.append style={@cell content/.add={\cellcolor{red!25}}{}}},
	every row no 1 column no 1/.style={postproc cell content/.append style={@cell content/.add={\cellcolor{green!25}}{}}},
	every row no 1 column no 2/.style={postproc cell content/.append style={@cell content/.add={\cellcolor{green!25}}{}}},
	every row no 1 column no 3/.style={postproc cell content/.append style={@cell content/.add={\cellcolor{green!25}}{}}},
	every row no 2 column no 1/.style={postproc cell content/.append style={@cell content/.add={\cellcolor{green!25}}{}}},
	every row no 2 column no 2/.style={postproc cell content/.append style={@cell content/.add={\cellcolor{green!25}}{}}},
	every row no 2 column no 3/.style={postproc cell content/.append style={@cell content/.add={\cellcolor{green!25}}{}}},
	every row no 3 column no 1/.style={postproc cell content/.append style={@cell content/.add={\cellcolor{green!25}}{}}},
	every row no 3 column no 2/.style={postproc cell content/.append style={@cell content/.add={\cellcolor{green!25}}{}}},
	every row no 3 column no 3/.style={postproc cell content/.append style={@cell content/.add={\cellcolor{green!25}}{}}},
	every row no 4 column no 1/.style={postproc cell content/.append style={@cell content/.add={\cellcolor{green!25}}{}}},
	every row no 4 column no 2/.style={postproc cell content/.append style={@cell content/.add={\cellcolor{green!25}}{}}},
	every row no 4 column no 3/.style={postproc cell content/.append style={@cell content/.add={\cellcolor{green!25}}{}}},
	every row no 5 column no 2/.style={postproc cell content/.append style={@cell content/.add={\cellcolor{green!25}}{}}},
	every row no 5 column no 3/.style={postproc cell content/.append style={@cell content/.add={\cellcolor{green!25}}{}}},
	every row no 6 column no 2/.style={postproc cell content/.append style={@cell content/.add={\cellcolor{green!25}}{}}},
	every row no 6 column no 3/.style={postproc cell content/.append style={@cell content/.add={\cellcolor{green!25}}{}}},
	every row no 7 column no 2/.style={postproc cell content/.append style={@cell content/.add={\cellcolor{green!25}}{}}},
	every row no 7 column no 3/.style={postproc cell content/.append style={@cell content/.add={\cellcolor{green!25}}{}}},
	every row no 8 column no 2/.style={postproc cell content/.append style={@cell content/.add={\cellcolor{green!25}}{}}},
	every row no 8 column no 3/.style={postproc cell content/.append style={@cell content/.add={\cellcolor{green!25}}{}}},
]{csv/GeneralizedFireFightingLarge.csv}
\vspace{\baselineskip}
\caption{Best (green) and worst (red) performing methods on the larger instance of the Generalized Firefighting problem.}
\label{tab:generalized_ff_large}
\end{table}

As expected, the methods provided with the true underlying factorizations are performing best, with that using the factored $Q$-function learning approach capable of achieving a perfect reconstruction and ranking of the actions even on this very large problem. Also complete factorizations are always identifying all of the optimal joint actions and producing correct ranking (perfect for those using the factored $Q$-function learning approach). It is interesting to note that even overlapping factorizations, when coupled with larger factors, are performing very well, and can produce good rankings of the actions. As expected, the mixture of experts methods are resulting in a larger MSE, although being comparable on the other metrics with their counterparts, but are still capable of learning more accutare representations than the joint learner, that is instead achieving the highest MSE and worst ranking among all the compared methods.

\begin{figure}[htbp]
\centering
\subfigure[$n=9$]{
\begin{tikzpicture}
\tikzset{dot/.style 2 args={fill, regular polygon, regular polygon sides=7, inner sep=3pt, label={#1:\scriptsize #2}}}
\def\islands{{7,8,9,4,5,6,1,2,3}} 
\def\cols{2}
\def\rows{2}
\def\scale{1.5}
\foreach \x [count=\n] in {0,...,\cols}{
\foreach \y in {0,...,\rows}{
\pgfmathtruncatemacro\tmp{\n+(\cols+1)*\y}
\pgfmathtruncatemacro\num{\islands[\tmp-1]}
\draw[line width=.5pt] (\x*\scale,0) -- (\x*\scale,\rows*\scale);
\draw[line width=.5pt] (0,\y*\scale) -- (\cols*\scale,\y*\scale);
\node[dot={45}{\num}] at (\x*\scale,\y*\scale) {};
}
}
\end{tikzpicture}
}
\hspace{50pt}
\subfigure[$n=12$]{
\begin{tikzpicture}
\tikzset{dot/.style 2 args={fill, regular polygon, regular polygon sides=7, inner sep=3pt, label={#1:\scriptsize #2}}}
\def\islands{{10,11,12,7,8,9,4,5,6,1,2,3}} 
\def\cols{2}
\def\rows{3}
\def\scale{1}
\foreach \x [count=\n] in {0,...,\cols}{
\foreach \y in {0,...,\rows}{
\pgfmathtruncatemacro\tmp{\n+(\cols+1)*\y}
\pgfmathtruncatemacro\num{\islands[\tmp-1]}
\draw[line width=.5pt] (\x*\scale,0) -- (\x*\scale,\rows*\scale);
\draw[line width=.5pt] (0,\y*\scale) -- (\cols*\scale,\y*\scale);
\node[dot={45}{\num}] at (\x*\scale,\y*\scale) {};
}
}
\end{tikzpicture}
}
\caption{Islands configuration for the two instances of Aloha.}
\label{fig:islands_large}
\end{figure}

\begin{figure}[htbp]
\centering
\includegraphics[width=0.68\textwidth]{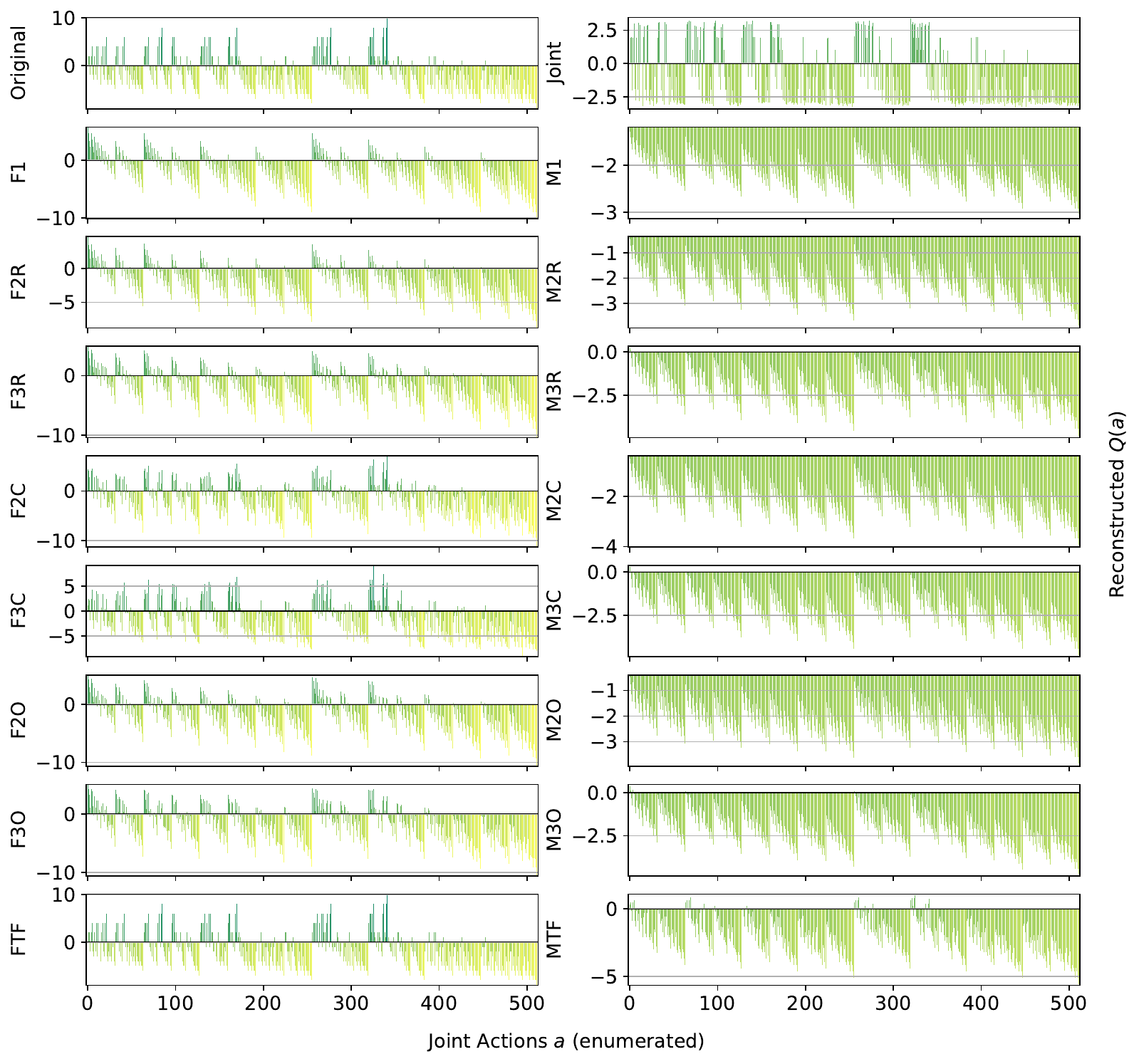}
\caption{Reconstructed $Q(a)$ for Aloha with $n=9$ agents.}
\label{fig:aloha_large}
\end{figure}

\emph{Aloha:} Our experiments here use $n=9$ and $n=12$ islands disposed in a $3\times 3$ and $4\times 3$ grid respectively, as shown in Figure \ref{fig:islands_large}. Representations learned for this game with $n=9$ islands are reported in Figure \ref{fig:aloha_large} (for identical reasons to those of the Dispersion Game, the figure with $n=12$ agents is not included).

Again, this game proves to be challenging for almost all of the proposed factorizations. Indeed, other than the true underlying factorization coupled with the factored $Q$-function learning approach (that achieve a perfect reconstruction, showing how beneficial would it be to know and exploit such an underlying factorization in advance), only the complete factorizations seems able to learn something useful. All the other methods struggle to correctly identify the optimal action, probably because not enough coordination is achieved in order to discriminate between the two local actions for each agent (that seems similar from an agent perspective). Also for this game, the joint learner is not capable of correctly approximating the action-value function because of the increasing number of agents. Table \ref{tab:aloha_large} is showing the best and worst performing methods on this game.

\begin{table}[htbp]
\centering
\pgfplotstableset{
alias/Model/.initial=0,
}

\begin{filecontents*}{csv/AlohaLarge.csv}
% Aloha n=9 (512 joint actions, 1 optimal)
F2R,6.968,0.446,123.345,26.454,0.0,0.0,9.0,1.844,6.733,0.345,100.4,11.038,0.581,0.023
F2C,2.25,0.007,9.027,0.2,1.0,0.0,0.0,0.0,2.06,0.025,186.6,1.114,0.817,0.001
FTF,0.0,0.0,0.0,0.0,1.0,0.0,0.0,0.0,0.0,0.0,512.0,0.0,1.0,0.0
M1,11.929,0.003,145.967,0.115,0.0,0.0,10.0,0.0,9.684,0.002,125.9,1.044,0.635,0.0
M2R,10.449,0.249,140.152,5.66,0.0,0.0,9.0,1.612,8.849,0.124,105.2,10.591,0.596,0.026
M2C,10.311,0.004,138.522,0.133,0.0,0.0,10.0,0.0,8.793,0.002,140.4,1.428,0.681,0.001
M2O,10.478,0.186,138.601,5.831,0.0,0.0,9.6,0.8,8.862,0.14,114.4,10.423,0.615,0.036
% Aloha n=12 (4096 joint actions, 2 optimal)
Joint,23.98,0.639,135.364,2.538,0.0,0.0,7.4,3.105,12.115,0.273,700.4,13.463,0.418,0.005
F2C,2.181,0.004,10.364,0.142,2.0,0.0,0.0,0.0,2.062,0.024,1379.8,4.578,0.832,0.0
F3C,0.809,0.005,12.45,0.171,1.0,0.0,2.0,0.0,1.1,0.01,2487.5,9.521,0.914,0.0
FTF,0.0,0.0,0.0,0.0,2.0,0.0,0.0,0.0,0.0,0.0,4096.0,0.0,1.0,0.0
M1,15.374,0.029,230.648,0.367,0.0,0.0,12.0,0.0,12.449,0.015,845.0,25.838,0.632,0.01
M2O,13.938,0.182,225.436,1.462,0.0,0.0,11.6,0.8,11.614,0.105,670.6,57.535,0.555,0.031
M3O,12.562,0.308,214.012,3.247,0.0,0.0,11.4,0.917,10.846,0.193,752.3,87.455,0.609,0.04
\end{filecontents*}
\pgfplotstabletypeset[col sep=comma,
	every head row/.style={before row=\toprule, after row=\midrule},
    every last row/.style={after row=\bottomrule},
	header=false,
	fixed,
	zerofill,
    columns/Model/.style={column type={C{0.9cm}},string type},
	columns/Mean square error/.style={string type, column type={C{1.5cm}}, column name={MSE}},
	create on use/Mean square error/.style={
		create col/assign/.code={
		\pgfmathprintpmnumber{1}{2}{2}{1}
		\pgfkeyslet{/pgfplots/table/create col/next content}\value
		}
	},
	columns/Optimal actions found/.style={string type, column type={C{1.6cm}}, column name={Opt. Found}},
	create on use/Optimal actions found/.style={
		create col/assign/.code={
		\pgfmathprintpmnumber{5}{6}{0}{0}
	    \pgfkeyslet{/pgfplots/table/create col/next content}\value
		}
	},
	columns/Correctly ranked/.style={string type, column type={C{1.8cm}}, column name={Ranked}},
	create on use/Correctly ranked/.style={
		create col/assign/.code={
		\pgfmathprintpmnumber{11}{12}{0}{0}
		\pgfkeyslet{/pgfplots/table/create col/next content}\value
		}
	},
	columns={Model,Mean square error,Optimal actions found,Correctly ranked},
	every row no 0/.style={before row={\multicolumn{4}{c}{\textbf{Aloha $n=9$}}\\\midrule}},
	every row no 7/.style={before row={\midrule\multicolumn{4}{c}{\textbf{Aloha $n=12$}}\\\midrule}},
	every row no 0 column no 3/.style={postproc cell content/.append style={@cell content/.add={\cellcolor{red!25}}{}}},
	every row no 1 column no 2/.style={postproc cell content/.append style={@cell content/.add={\cellcolor{green!25}}{}}},
	every row no 2 column no 1/.style={postproc cell content/.append style={@cell content/.add={\cellcolor{green!25}}{}}},
	every row no 2 column no 2/.style={postproc cell content/.append style={@cell content/.add={\cellcolor{green!25}}{}}},
	every row no 2 column no 3/.style={postproc cell content/.append style={@cell content/.add={\cellcolor{green!25}}{}}},
	every row no 3 column no 1/.style={postproc cell content/.append style={@cell content/.add={\cellcolor{red!25}}{}}},
	every row no 4 column no 1/.style={postproc cell content/.append style={@cell content/.add={\cellcolor{red!25}}{}}},
	every row no 4 column no 3/.style={postproc cell content/.append style={@cell content/.add={\cellcolor{red!25}}{}}},
	every row no 5 column no 1/.style={postproc cell content/.append style={@cell content/.add={\cellcolor{red!25}}{}}},
	every row no 6 column no 1/.style={postproc cell content/.append style={@cell content/.add={\cellcolor{red!25}}{}}},
	every row no 6 column no 3/.style={postproc cell content/.append style={@cell content/.add={\cellcolor{red!25}}{}}},
	every row no 7 column no 1/.style={postproc cell content/.append style={@cell content/.add={\cellcolor{red!25}}{}}},
	every row no 7 column no 3/.style={postproc cell content/.append style={@cell content/.add={\cellcolor{red!25}}{}}},
	every row no 8 column no 2/.style={postproc cell content/.append style={@cell content/.add={\cellcolor{green!25}}{}}},
	every row no 9 column no 1/.style={postproc cell content/.append style={@cell content/.add={\cellcolor{green!25}}{}}},
	every row no 9 column no 2/.style={postproc cell content/.append style={@cell content/.add={\cellcolor{green!25}}{}}},
	every row no 9 column no 3/.style={postproc cell content/.append style={@cell content/.add={\cellcolor{green!25}}{}}},
	every row no 10 column no 1/.style={postproc cell content/.append style={@cell content/.add={\cellcolor{green!25}}{}}},
	every row no 10 column no 2/.style={postproc cell content/.append style={@cell content/.add={\cellcolor{green!25}}{}}},
	every row no 10 column no 3/.style={postproc cell content/.append style={@cell content/.add={\cellcolor{green!25}}{}}},
	every row no 11 column no 1/.style={postproc cell content/.append style={@cell content/.add={\cellcolor{red!25}}{}}},
	every row no 12 column no 3/.style={postproc cell content/.append style={@cell content/.add={\cellcolor{red!25}}{}}},
	every row no 13 column no 3/.style={postproc cell content/.append style={@cell content/.add={\cellcolor{red!25}}{}}},
]{csv/AlohaLarge.csv}
\vspace{\baselineskip}
\caption{Best (green) and worst (red) performing methods on the two instances of Aloha.}
\label{tab:aloha_large}
\end{table}

The table shows how, except for FTF (always capable of correctly represent the entire $Q$-function), all the methods start deteriorating their performance when the system size increases on this particular problem. Particularly, the joint learner achieves a very high reconstruction error and is not able to identify any of the optimal joint actions. On the other hand, although the corresponding ranking is not perfect, complete factorizations using the factored $Q$-function learning approach can identify such optimal actions. The mixture of experts instead are performing worse here, probably because the benefits of a coordinated optimization is crucial to correctly represent this problem.

\subsection{Sample Complexity}
\label{sec:sample}
\begin{figure}[htbp]
\centering
\subfigure[Dispersion Game, $n=6$\label{sub:training_dispersion}]{
\includegraphics[width=0.48\textwidth]{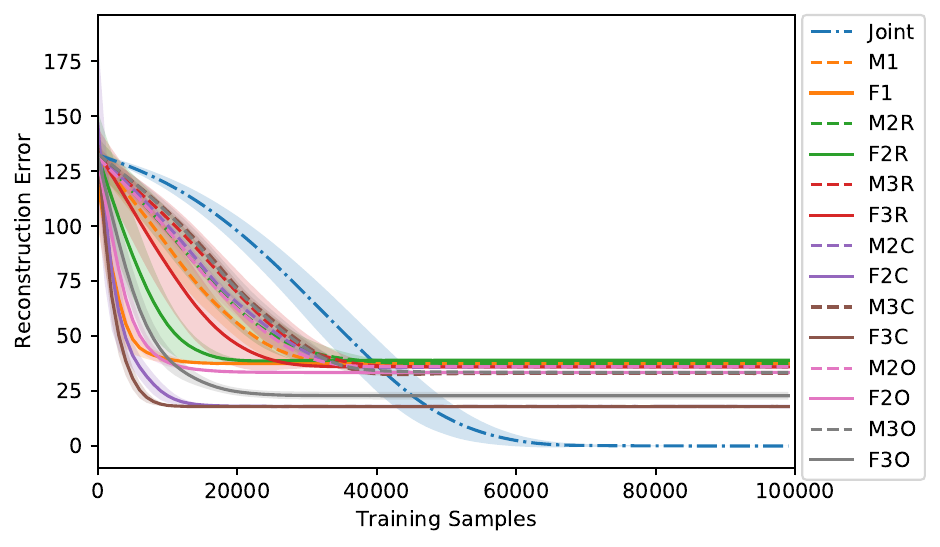}
}
\hfill
\subfigure[Generalized Firefighting, $n=6$\label{sub:training_generalized_ff}]{
\includegraphics[width=0.48\textwidth]{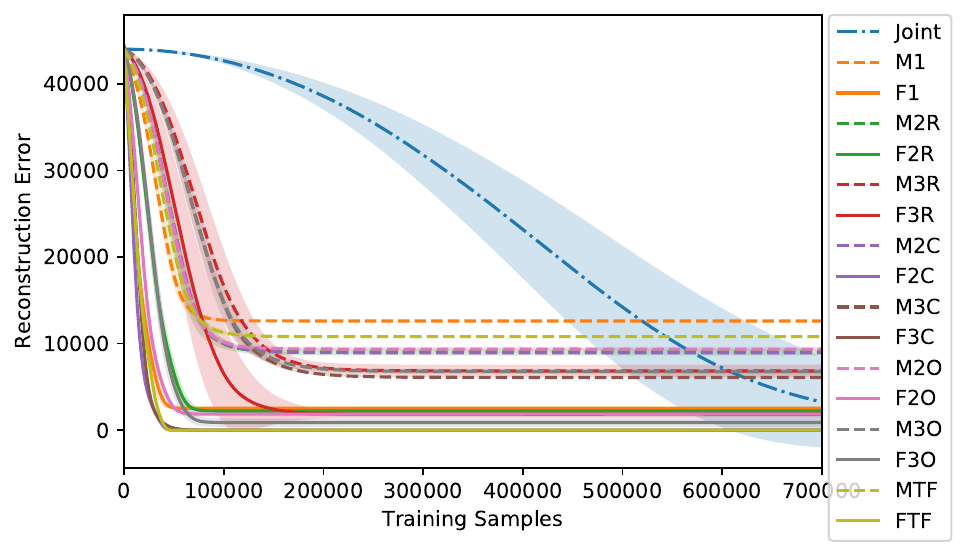}
}
\caption{Training curves for the investigated architectures on the two proposed problems.}
\label{fig:training}
\end{figure}

Another important consideration in multi-agent learning is sample complexity, as for example training data could be limited or expensive to obtain. Therefore it is a crucial aspect how efficiently we can use such data and how long does it take for a given representation to converge, especially when the system grows larger in the number of agents. We expect factored representations to improve training efficiency, reducing the number of samples required to learn a good representation, as the size of the multiple components that have to be learnt is small compared to that of the overall problem. To show the benefits of using a factored representation, here we report in Figure \ref{fig:training} the training curves for two of the proposed games, the Dispersion Game and the Generalized Firefighting, both with $n=6$ agents.

\begin{figure}[htbp]
\centering
\subfigure[$n=9$]{
\includegraphics[width=0.31\textwidth]{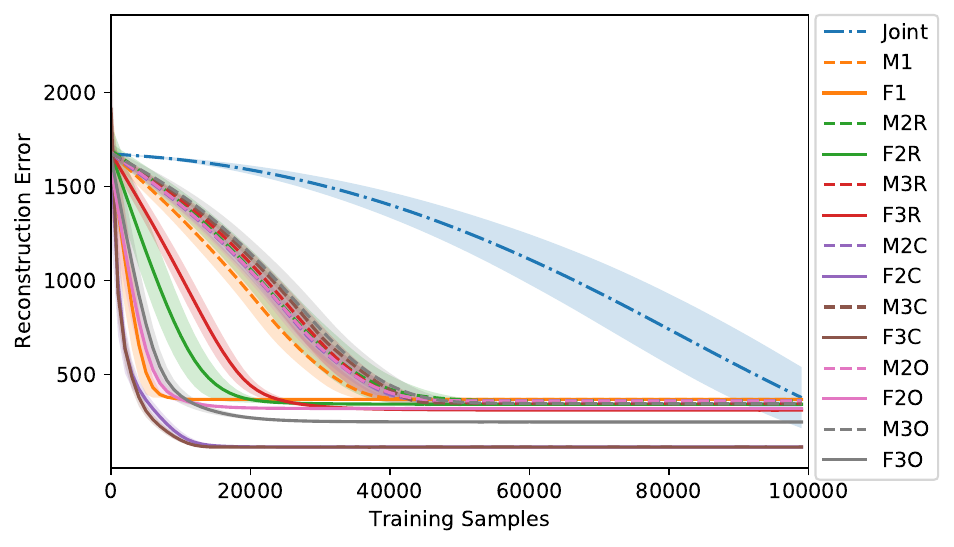}
}
\hfill
\subfigure[$n=12$]{
\includegraphics[width=0.31\textwidth]{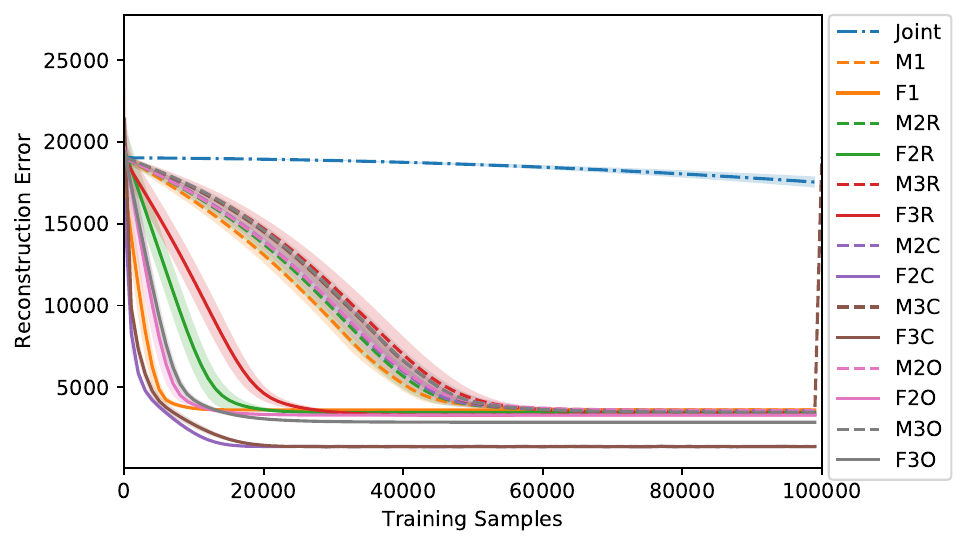}
}
\hfill
\subfigure[$n=20$]{
\includegraphics[width=0.31\textwidth]{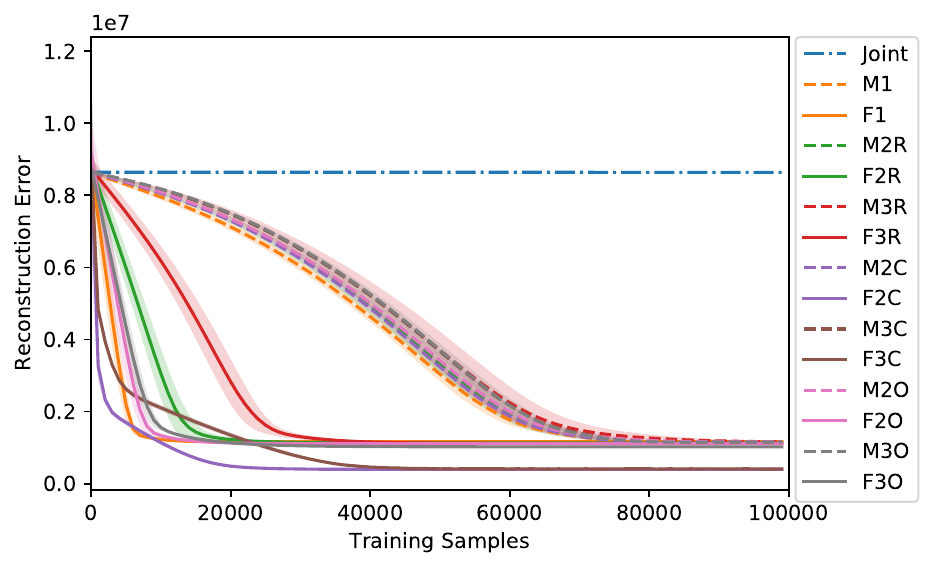}
}
\caption{Training curves for the investigated architectures on the Dispersion Game with an increasing number of agents.}
\label{fig:training_large}
\end{figure}

Even for these problems of moderate size, factored approaches achieve a stable approximation of the action-value function with just a fraction of the given training time, while a full joint learner requires many more samples to get the same results. Especially, for the Generalized Firefighting problem (that has got more than $8000$ overall joint actions), the joint learner achieves an accurate representation only after a much longer training time, while almost every factored architecture achieves a nearly perfect approximation with few samples, showing how the size of the joint action space is a critical problem that factored representations can help tackle. On one hand the mixture of experts approaches learn more slowly than the factored $Q$-function ones: each factor acts as an expert on its own, thus experiencing higher variance in the received rewards when performing a certain action. On the other hand, larger models learn more quickly, achieving the same final result as the smaller representations but with fewer samples. This could be due to the internal coordination happening inside each factor, helping the agents figure out their own contribution to the global reward, so that a stable representation is learned more easily. When the number of agents is larger, this benefit is even more apparent. Figure \ref{fig:training_large} shows the reconstruction error during the training process obtained on instances of the Dispersion Game with $n=\{9,12,20\}$ agents respectively.

We observe how the joint learner is struggling to achieve a good representation in the given training time when the size of the system increases, resulting in a higher reconstruction error. The increasingly large number of joint actions (more than $1$ million with $n=20$ agents) prevents it from converging in reasonable time, while the factored representations, although only approximating the original function, converge faster, as the size of each factor is small compared to that of the overall problem, and result in a lower reconstruction error.

\subsection{Exploratory Policy}
\label{sec:exploration}
\begin{figure}[htbp]
\centering
\includegraphics[width=0.68\textwidth]{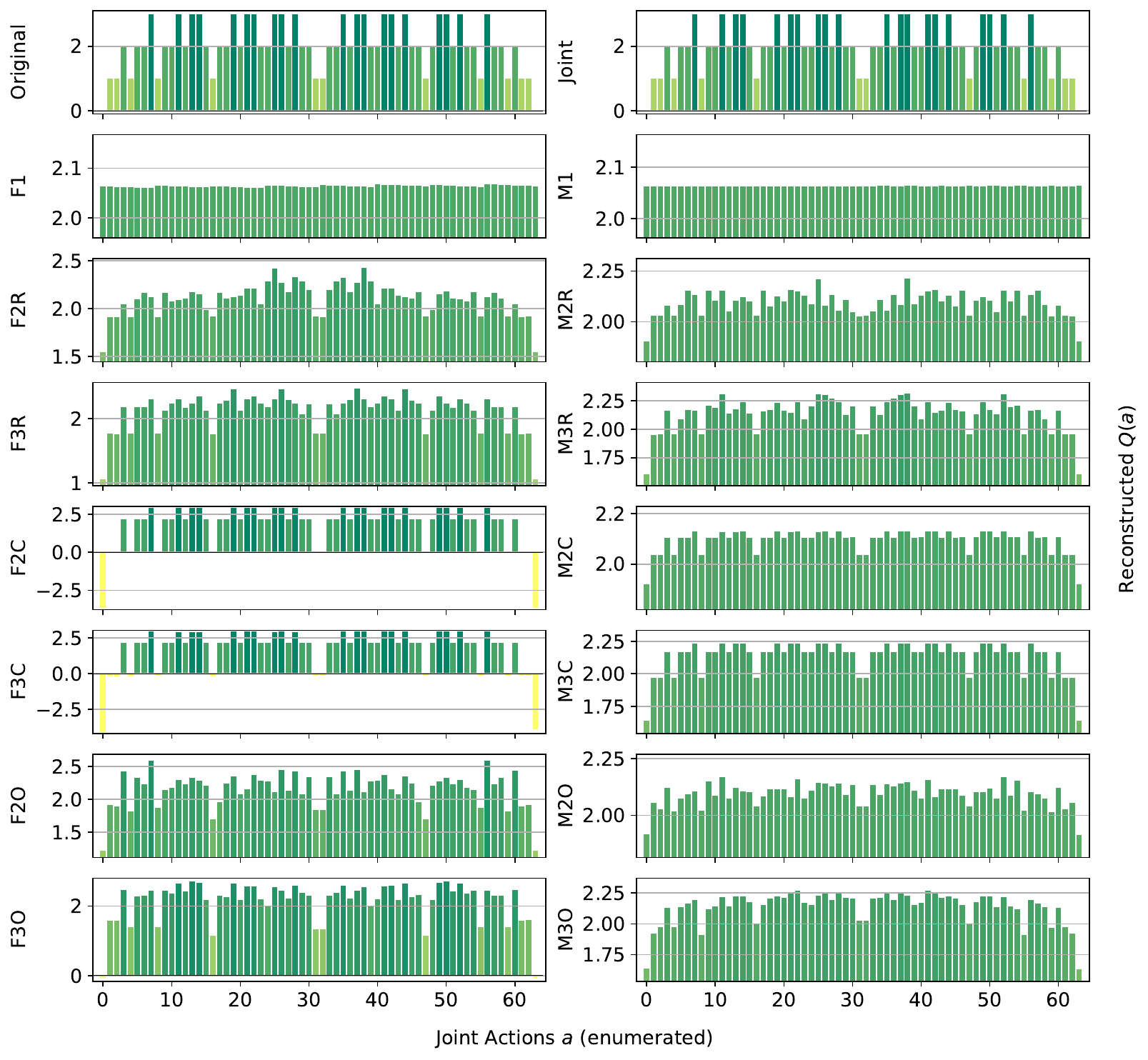}
\caption{Reconstructed $Q(a)$ for the Dispersion Game using the Boltzmann exploratory policy.}
\label{fig:boltzmann}
\end{figure}

Although we focused on a stationary uniform sampling of the actions throughout most of the paper, we also provide some preliminary results with a different, non-stationary action selection mechanism, more closely resembling those used in sequential MARL. We opted for a Boltzmann policy \cite{survey} that, given a reconstruction of the action-value function $\hat{Q}(a)$, defines the probability for each joint action $a\in A$ to be selected as:
\begin{equation}
\pi(a)=\frac{e^{\hat{Q}(a)/\tau}}{\sum_{b\in A}e^{\hat{Q}(b)/\tau}},
\end{equation}
where $\tau$ is a temperature parameter governing the exploration rate. In our experiment, we set $\tau=1$ for all methods. We choose to test these on the Dispersion Game with $n=6$ agents, as many of the methods (including the joint learner) are doing reasonably well and thus any decrease in performance would be due to the new exploratory policy. For the factored methods and the independent learners, we reconstruct $\hat{Q}(a)$ at every step and then we apply the Boltzmann policy on this reconstruction. Figure \ref{fig:boltzmann} shows the learned reconstructions on this game.

If we compare the above to Figure \ref{sub:dispersion}, we can observe how the results are very much in line with those obtained under uniform sampling. Even if the joint action space is small enough to select each action a reasonable number of time, the fact that the policy is more frequently selecting the action that looks better is not providing benefits in term of accuracy of the final representation, especially for the independent learners, that are still not able to clearly identify any of the optimal joint actions. The joint learners and the complete factorizations are still able to correctly rank all of the optimal actions, and those using the factored $Q$-function learning approach are achieving a smaller reconstruction error as under the uniform sampling. Although this is just a preliminary result, this is an important observation, as it gives more value to the previous results: if a method is not doing well under the uniform sampling of actions, it is unlikely that it can do better with a non-stationary, time-varying sampling mechanism like this.

\subsection{Summary of Results}
\label{sec:summary}
We can gain many useful insights from the analysed results: first, we observe that the factorizations using mixture of expert learning approach, although generally achieving higher reconstruction errors than the factored $Q$-function counterparts, in many cases still results in a good approximation in terms of ranking of actions, therefore being a reliable choice for decision making. For example, on the two variants of the Dispersion Game, both M2C and M3C are able to correctly rank all of the joint actions, achieving better accuracy than some smaller factorizations like F2O or F3O, even with a higher mean square error of the reconstruction. This is probably due to the higher number of factors involved in their coordination graphs, allowing for better approximation of the true action-value function and coordination amongst the agents. We can therefore deduce how the number of factors used to learn an approximation is playing a major role in achieving accurate representations in terms of coordination and actions ranking.

Also, the size of these factors is an important aspect: as expected, with more agents comprised into each factor, the resulting approximation is more reliable because the agents into each factor are able to share information and thus better coordinate. This is reflected by both learning approaches, but it is even more apparent with the mixture of experts one, with the factorizations with $3$ agents per factor usually achieving smaller reconstruction error and a better ranking of actions than their counterparts with only $2$ agents for each. However, factors that are too large (with a size very similar to that of the entire team) do not always result in a better representation, but instead can present some of the difficulties associated with joint learners. This suggests that we can find an optimal tradeoff between the totally independent learners and the full joint learner extremes that is capable of achieving a reliable representation in a reasonable training time. Of course, depending on the intended use of such a learned representation, such a tradeoff may differ: for example, if we are only interested in selecting an optimal joint action from this reconstruction after the training process (i.e. with a factored centralized joint $Q$-function agent for the entire team of agents) a smaller factorization with fewer factors and agents per factor, that is faster to train and still able to correctly rank some of the joint actions, may suffice. Conversely, if we are approximating the critic of an actor-critic method, in which the values of the selected actions are in turn influencing the policies of the agents, we may prefer a bigger factorization with a lower reconstruction error.

Generally speaking, we can observe how the value of coordination is well captured by the factored $Q$-function learning approach, that usually produce good approximations and reduces the training time with respect to a joint learner, allowing to learn even games in which there is no underlying factorization and complete team coordination is required, like the Dispersion Game. This becomes even more important with larger systems, as now more agents have to coordinate and thus more accurate representation of such coordination requirements is needed. However, when these requirements are less tight, also the mixture of experts learning approach is showing benefits, being faster to train and not requiring inter-factor communication during the training phase to optimize its objective. Overall, we showed how the choice of a suitable factorization to efficiently learn in a multi-agent system is a difficult decision that needs to be taken considering the problem structure and requirements. While many aspects can influence the learning outcome, our results have five main takeaways:

\begin{itemize}
\item There are some problematic examples, like the Platonia Dilemma, where all types of factorization with small factors result in selecting the worst possible joint action. Given that only joint learners (and certain factorizations with larger factors to some extent) seem to be able to address such problems, currently no scalable deep reinforcement learning methods for dealing with such systems seem to exist. We hypothesize that this is due to an imbalance in the frequency with which each local action for the agents leads to the optimal reward value. In the Platonia Dilemma, each agent is more frequently experiencing the positive reward if it does not send the telegram itself and leave this action to someone else. However, if all the agents do this kind of reasoning, no-one is sending the telegram, and the resulting reward is not the optimal one. Breaking this tie is possible when we behave greedily (as the first agent correctly sending the telegram will keep sending it more frequently), but learning a complete and correct representation of the entire $Q$-function remains challenging.
\item Beyond those, ``complete factorizations'' of modest factor size coupled with the factored $Q$-function learning approach yield near-perfect reconstructions and rankings of the actions, also for non-factored action-value functions. Moreover, these methods scale much better than joint learners: for a given training time, we see that these complete factorizations already outperform fully joint learners on modestly sized problems, resulting in a correct ranking of the actions and a low reconstruction error. This is a compelling property that renders these methods more suited for large multi-agent systems with a large joint action-space and justify the recent interest in factored methods from the research community.
\item For many problems with less tight coordination requirements such as Aloha and the Generalized Firefighting Problem, random overlapping factors also achieve excellent performance, comparable to those of more computationally complex methods like joint learners and complete factorizations. This suggests that such approaches are a promising direction forward for scalable deep MARL in many problem settings.
\item Factorizations with the mixture of experts approach usually perform somewhat worse than the corresponding factored $Q$-function approaches, as these are less able to capture the coordination requirements of certain problems. However, in some cases they perform better or comparably (Dispersion Game, Generalized Firefighting), in which M2R and M3R still outperform F1 (i.e., VDNs). This is promising, because the mixture of experts learning approach does not require any exchange of information between the neural networks, thus potentially facilitating learning in settings with communication constraints, and making it easier to parallelize across on multiple CPUs/GPUs.
\item Our results show that, when facing larger multi-agent systems, factored representations retain many of their benefits and can still represent the action-value function correctly in many settings (or at least identify most if not all of the optimal joint actions), while both independent learners and centralized approaches tend to quickly deteriorate, resulting in wrong or incomplete representations. The reasons however are different: while independent learners still decompose the entire function into small components that can easily be learned, these fail in representing the value of coordinated decisions when more agents are comprised into the system. On the other hand instead, a joint learner that is representing the centralized action-value function as a single component (and thus does not introduce any approximation), is now struggling to learn a correct representation because of the exponential number of joint actions. Factored representations instead learn small components easily, but take into account the value of coordination into each component, allowing for better final representations. This is an extremely desirable property that once more points out how these methods deserve attention as holding a great potential.
\end{itemize}

These observations also shed some light on the performance of independent learners in MARL, as used by many modern deep MARL algorithms \cite{vdn,qmix}: while these can outperform joint learners on large problems, the degree of independence and the final outcome is hard to predict and is affected by different factors. Designing algorithms that are able to overcome these difficulties should be a primary focus of MARL research.

\section{Discussion}
\label{sec:discussion}
The aim of our analysis was to investigate the learning capabilities of factored representations and compare them to both independent learners and joint learners, to assess eventual benefits of such representations both in terms on learning speed and final accuracy of the reconstructed representations. In order to do so, we consider different aspects of these learned representations $\hat{Q}$: we consider the optimality of the greedy joint action, which is important when using $\hat{Q}$ to select actions. We also consider the distance to the optimal value $\Delta Q=|Q-\hat{Q}|$, since verifying the optimality of the greedy action requires bounding $\Delta Q$. Although our approach focus on centralized learning, a correct representation of the joint action-value function can be used in the CTDE framework \cite{ctde} to learn improved decentralized policies. Indeed, minimising $\Delta Q$ is important for deriving good policy gradients in actor-critic architectures (for example when computing the counterfactual baseline in \cite{coma}, requiring very accurate estimates of many sub-optimal $Q$-values) and for sequential value estimation in any approach that relies on bootstrapping (such as $Q$-learning \cite{qlearning}) or message passing of local payoff values (like the max-sum algorithm \cite{maxsum}), where such values are used to update other values and thus need to be as accurate as possible.

Our analysis is focused on cooperative one-shot games. We chose these because, although they are a simpler setting than standard sequential problems, they still capture many of the aspects that can be problematic in MARL, like the exponentially large number of joint actions. Also, the shared reward observed by all the agents depends on the joint action of the whole team, and thus seems non-stationary from an agent perspective. Therefore, all of the problems that arise from these conditions are directly translated into our setting, and by removing the fostering effect of states, we are better able to analyse how the different methods can tackle these issues. Moreover, we mainly focused on using a stationary uniform sampling mechanism to select joint actions, thus not directly considering the exploration-exploitation tradeoff usually faced in MARL problems. There are multiple reasons for this: on one hand, using such a simple mechanism makes our comparison easier, as we are interested in the reconstruction of the entire action-value function (and not only the greedy action). On the other hand, if a network architecture is not suited to learn and represent accurate action-values under our stationary uniform sampling, it is extremely unlikely that the same is going to perform better under a more complex, time-changing policy as in sequential MARL. This is validated by our preliminary results with a Boltzmann policy, in which none of the method is achieving better accuracy than under the stationary uniform mechanism. However, while good performance in such one-shot settings does not necessarily imply good performance in the sequential setting, $Q$-value-based approaches aim to transform the sequential MARL problem to precisely the one-shot decision making problem that we investigate, as the sequential problem simply becomes a one-shot maximization over the action-values. This implies that any limitations we find are likely to directly transfer to such approaches.

Obviously, the opposite is not immediately true: good performance on this setting does not directly imply also good performance on general MARL problems. However, if a given representation is able to correctly capture the value of coordination and deal with the joint action set size in one-shot games, we can hypothesize that these benefits are likely to hold in sequential problems as well, where the same requirements usually brings similar challenges to our setting and thus can be tackled with similar solutions. Neural networks are known to be very good at dealing with large input spaces, but it is not so well known how these can deal with large output ones, like those resulting from exponential joint action sets. Therefore, our analysis moves an important step in this direction in the field of MARL, and has to be considered as an initial step toward a proper understanding of action-value functions in multi-agent settings. Of course, a direct investigation in full sequential settings is an interesting future direction, with key questions that are orthogonal to ours, like the presence of an input state of the exploration-exploitation tradeoff. Nonetheless, our results can still help taking informed decision in such problems as well, as they give us interesting insights and takeaways that practitioners of the field may take into account when taking informed decisions in designing multi-agent systems.

\section{Related Work}
Recently, many works have applied deep reinforcement learning techniques to multi-agent systems, achieving great performance. \citeauthor{cooperative} \cite{cooperative} compare the performance of many standard deep reinforcement learning algorithms (like DQN, DDPG and TRPO) using a variety of learning schemes (joint learners, fully independent learners, etc.) on both discrete and continuous tasks, assessing and comparing their performance. \citeauthor{madrl} \cite{madrl} present a variation of DQN capable of dealing with both competitive and cooperative settings in Pong. Applications of techniques to enhance the learning process have also been investigated: \citeauthor{lenient} \cite{lenient} apply leniency to independent double DQN learners in a coordinated gridworld problem, while \citeauthor{experience} \cite{experience} propose a novel approach to stabilize the experience replay buffer in DQNs by conditioning on when the samples were collected, thereby easing the non-stationarity of independent learners. Communication between agents has also been explored: \citeauthor{backprop} \cite{backprop} investigate the emergence of a communication mechanism that can be directly learned through backpropagation. However, none of these works compare alternative representations of $Q$-values for such networks. Another interesting line of work is that of agent definition: \citeauthor{definition} \cite{definition} analyse how different agents definitions in a multi-agent system affect how efficiently these can be learned and modelled.

Close to our idea of factorization are \cite{coordinated,maxsum,utile}. \citeauthor{coordinated} \cite{coordinated} present different reinforcement learning algorithms, based either on $Q$-learning, policy iteration or direct policy search, that exploit a coordination graph representation of the problem and the variable elimination algorithm \cite{local} to avoid enumerating all the possible joint actions. \citeauthor{maxsum} \cite{maxsum} investigate the max-sum algorithm, an approximation of variable elimination, in which agents exchange messages with an estimate of their local payoff function over a coordination graph used by other agents to update their own estimate until they all converge to a stable value. They then also introduce a modified version of $Q$-learning updates based on either an agent-based decomposition or an edge-based decomposition of the coordination graph. \citeauthor{utile} \cite{utile} instead present a method to learn a coordination graph structure for the agents to use. It works by maintaining statistics of a state future returns and the incoming transitions, and then splitting the state into a factored representation if these returns are significantly different. More recently, \citeauthor{dcg} \cite{dcg} propose Deep Coordination Graphs, published after our original work \cite{ea} and extending on it, in which different factors of a coordination graph are modelled with a single neural network through parameter sharing and uses a learned embedding of their input to differentiate. The results presented show the effectiveness of such a factored representation in cooperative sequential problems, but not a systematic analysis of the learned factored $Q$-functions.

Another line of work addressing the problem of the exponentially large joint action space exploits the paradigm of centralized learning with decentralized execution. \citeauthor{coma} \cite{coma} present COMA, an architecture based on the actor-critic framework with multiple independent actors but a single centralized critic used to efficiently estimate both $Q$-values and a counterfactual baseline to tackle the credit assignment problem and guide the agents through the learning process under partial observability. On the other hand, \citeauthor{maac} \cite{maac} propose MADDPG, expanding DDPG \cite{ddpg} to work in multi-agent systems by maintaining a different centralized critic network for every actor that estimate $Q$-values for each agents, considering the actions and observations of other agents, and apply their approach on both cooperative and competitive continuous tasks. However, these works still represents $Q(s,a)$ monolithically, and thus can experience scalability issues. \citeauthor{vdn} \cite{vdn} address the problem with value decomposition networks, training a set of independent agents collectively by using a value decomposition method that represents the original $Q$-function as the sum of local terms depending only on agent-wise information, while \citeauthor{qmix} \cite{qmix} extend this idea in QMIX by representing the $Q$-function using a monotonic non-linear combination of individual $Q$-values, weighted with an additional mixing network conditioned on the global state signal, so that the maximization step can still be performed in an efficient way. While such mixing networks may lead to more accurate $Q$-values in the sequential domain, our investigation shows that for many coordination problems, individual $Q$-components may not suffice. \citeauthor{qtran} \cite{qtran} propose QTRAN, an algorithm that transform a joint action-value function into another one with the same optimal joint action that can be easily factorized in individual components so that a monotonicity constraint is respected and thus action selection can be performed locally. \citeauthor{maven} \cite{maven} investigate the effect of poor exploration in value-based methods and propose MAVEN, an hybrid approach that improve value-based methods by conditioning them on a shared latent variable controlled by a hierarchical policy to improve over standard exploration methods, resulting in extended temporal exploration and improved performance.

\section{Conclusions}
In this work, we investigated how well neural networks can represent action-value functions arising in cooperative multi-agent systems. This is an important question since accurate representations can enable taking (near-) optimal actions in value-based approaches using algorithms like variable elimination or max-sum, thus avoiding the maximization over an exponential number of joint actions, and computing good gradient estimates in actor-critic methods. In this paper, we focused on one-shot games as the simplest setting that captures the exponentially large joint action space of multi-agent systems. We compared a number of existing and new action-value network factorizations and learning approaches.

Our results highlight the difficulty of compactly representing action values in problems that require tight coordination, but indicate that using ``higher-order'' factorizations with multiple agents in each factor can improve the accuracy of these representations substantially. We also demonstrate that there are non-trivial coordination problems, some without a factored structure, that can be tackled quite well with simpler factorizations. Intriguingly, random overlapping factors perform very well in several settings. There are also settings where the mixtures of experts approach, with its low communication requirements and amenability to parallelization, is competitive in terms of the final reconstructions.

While our results emphasize the dependence of appropriate architectural choic\-es on the problem at hand, our analysis also shows general trends that can help in the design of novel algorithms and improve general performance, highlighting how the use of factored action-value functions can be a viable way to obtain good representations without incurring in excessive costs.

As future work, we intend to extend this investigation to more complex settings and networks types. For example, investigating the learning power of factorizations using DQNs \cite{dqn}, a standard value-based choice when modelling agents with neural networks, to sequential multi-agent problems \cite{markov} seems a natural direction to try and assess what deep reinforcement learning algorithms are really learning and which problems and situations cause them to fail, while also an investigation of recurrent neural networks \cite{lstm} as a tool to deal with partial observability in the popular setting of repeated games \cite{gt} can be beneficial for the multi-agent community.

\section*{Acknowledgements}
\vspace{-3\baselineskip}
\begin{tabular}{p{.70\textwidth}p{0.3\textwidth}}
This research made use of a GPU donated by NVIDIA. F.A.O.\ is funded by EPSRC First Grant EP/R001227/1.
This project had received funding from the European Research Council (ERC) under the European Union's
Horizon 2020 research and innovation programme (grant agreement No.~758824 \textemdash INFLUENCE).
& \raisebox{-20mm}{\includegraphics[width=0.3\columnwidth]{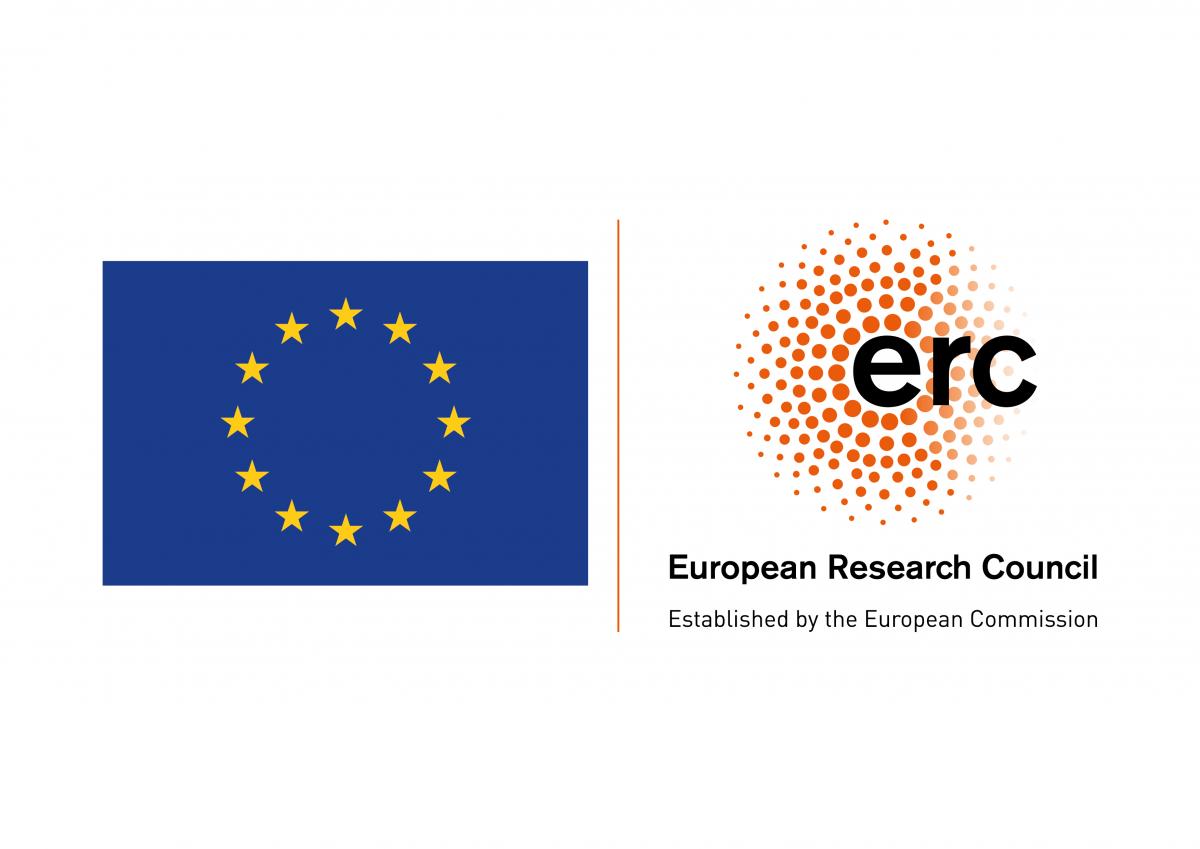}}
\end{tabular}

\bibliographystyle{plainnat}
\bibliography{Bibliography}

\clearpage
\appendix
\section{Complete Results}
Table \ref{tab:measures} presents the accuracy of the investigated representations using various measures, both in terms of action ranking and reconstruction error, as well as evaluating the action selection that these representations result in:

\begin{itemize}
\item To evaluate the reconstruction error, we compute the mean square error over all the joint actions (1st column);
\item We also compute the same measure but restricted only to those actions that are optimal in the original action-value function (2nd column);
\item We assess how many optimal actions are correctly identified by the reconstructions (3rd column);
\item The value loss (regret) obtained by following the represented value functions when doing decision making (4th column);
\item We also provide a different version of this regret, that we call \emph{Boltzmann value loss} which expresses the value loss obtained by the expected reward accrued by defining a softmax distribution over all the joint actions (5th column). This gives an indication of value loss amongst all good actions; and
\item Finally, we compute the number of correctly ranked actions (accounting for ties where needed) and the corresponding Kendall $\tau$-b coefficient \cite{rank} between the computed ranking and the original one (6th and 7th columns respectively).
\end{itemize}

A low value on the first two columns indicates that a representation is close to the original action value function: especially when the value of second column is low in combination with the value of the third one, the learned representation tends to correctly reconstruct and identify the optimal joint actions. The fourth and fifth columns tell us how reliably these representations can be used to perform decision making (i.e. when applying a greedy policy with respect to the joint action-value function during the evaluation phase) under two different kinds of policy. In fact, if the regret is low, even if the representation is not accurate, the model can still be used to pick actions for the team resulting in high rewards. Finally, the last two columns point out if the reconstructed values of the joint actions match the same hierarchy that they have in the original action-value function: even if the representation is not accurate in terms of mean squared error, a correct ranking of the actions points out that it has the same structure of the original one, and thus the two are similar except for the magnitude of the values itself. All of these measures are interlinked and express how similar a learned representation is to the original, but they all highlight different aspects that can help us analyse the benefits and drawbacks of these models.

For every method, mean values and standard errors across $10$ runs are reported. For every game, we highlight the best (green) and worst (red) performances of the investigated methods on some of the proposed measures.

\afterpage{\setlength{\tabcolsep}{0pt}
\LTcapwidth=\textwidth

\pgfplotstableset{
begin table=\begin{longtable},
alias/Model/.initial=0,
end table=\caption{Accuracy results with respect to both action ranking and reconstruction error for the different games. Best (green) and worst (red) performances for each game are highlighted.}
\label{tab:measures}
\end{longtable}
}

\rowcolors{0}{lightgray!50}{}

\begin{filecontents*}{csv/Combined.csv}
% Dispersion Game n=6 (64 joint actions, 20 optimal)
Joint,0.0,0.0,0.0,0.0,20.0,0.0,0.0,0.0,0.0,0.0,64.0,0.0,1.0,0.0
F1,0.621,0.0,0.878,0.008,4.9,2.166,1.7,1.005,0.492,0.0,21.4,2.059,-0.001,0.029
F2R,0.516,0.0,0.816,0.005,8.0,0.0,0.0,0.0,0.381,0.0,23.7,1.005,0.265,0.033
F3R,0.41,0.0,0.716,0.002,10.1,0.943,0.6,0.49,0.306,0.0,31.2,3.027,0.393,0.044
F2C,0.094,0.0,0.141,0.003,20.0,0.0,0.0,0.0,0.158,0.0,64.0,0.0,1.0,0.0
F3C,0.094,0.0,0.14,0.003,20.0,0.0,0.0,0.0,0.158,0.001,64.0,0.0,1.0,0.0
F2O,0.41,0.0,0.64,0.024,10.4,0.8,0.6,0.49,0.319,0.007,35.9,1.758,0.441,0.045
F3O,0.185,0.028,0.296,0.044,13.0,1.342,0.2,0.4,0.199,0.015,47.4,2.835,0.701,0.046
M1,0.621,0.0,0.877,0.003,6.2,1.47,1.3,0.781,0.492,0.0,24.1,2.022,0.015,0.037
M2R,0.563,0.0,0.822,0.004,8.0,0.0,0.0,0.0,0.457,0.0,23.6,1.281,0.267,0.042
M3R,0.463,0.0,0.729,0.003,10.0,0.894,0.4,0.49,0.393,0.0,31.2,2.088,0.392,0.031
M2C,0.553,0.0,0.809,0.003,20.0,0.0,0.0,0.0,0.459,0.0,64.0,0.0,1.0,0.0
M3C,0.431,0.0,0.679,0.003,20.0,0.0,0.0,0.0,0.397,0.0,64.0,0.0,1.0,0.0
M2O,0.557,0.0,0.812,0.003,10.2,1.72,0.5,0.5,0.458,0.0,35.8,3.572,0.462,0.047
M3O,0.441,0.003,0.685,0.003,11.0,1.414,0.3,0.458,0.398,0.001,40.4,3.47,0.577,0.073
% Sparse Dispersion Game n=6 (64 joint actions, 20 optimal)
Joint,0.0,0.0,0.0,0.0,20.0,0.0,0.0,0.0,0.0,0.0,64.0,0.0,1.0,0.0
F1,1.934,0.0,4.245,0.02,6.5,0.671,1.5,1.5,1.766,0.0,37.0,1.342,0.018,0.049
F2R,1.828,0.0,3.95,0.012,8.0,0.0,0.0,0.0,1.636,0.001,40.0,0.0,0.127,0.0
F3R,1.723,0.0,3.603,0.01,10.5,0.671,1.8,1.47,1.553,0.001,45.0,1.342,0.309,0.049
F2C,1.406,0.0,2.253,0.025,20.0,0.0,0.0,0.0,1.32,0.001,64.0,0.0,1.0,0.0
F3C,1.407,0.0,2.26,0.019,20.0,0.0,0.0,0.0,1.32,0.002,64.0,0.0,1.0,0.0
F2O,1.723,0.0,3.532,0.027,10.4,1.356,1.5,1.5,1.563,0.017,44.8,2.713,0.302,0.099
F3O,1.519,0.031,2.705,0.118,12.4,1.02,0.6,1.2,1.414,0.028,48.8,2.04,0.447,0.074
M1,1.934,0.0,4.271,0.011,6.4,0.917,2.1,1.375,1.766,0.0,36.8,1.833,0.011,0.067
M2R,1.875,0.0,4.128,0.008,8.0,0.0,0.0,0.0,1.729,0.0,40.0,0.0,0.127,0.0
M3R,1.773,0.0,3.878,0.009,10.0,1.342,1.5,1.5,1.657,0.0,44.0,2.683,0.273,0.098
M2C,1.865,0.0,4.112,0.008,20.0,0.0,0.0,0.0,1.731,0.0,64.0,0.0,1.0,0.0
M3C,1.741,0.0,3.83,0.009,20.0,0.0,0.0,0.0,1.663,0.0,64.0,0.0,1.0,0.0
M2O,1.869,0.0,4.125,0.008,9.8,1.166,1.2,1.47,1.731,0.0,43.6,2.332,0.258,0.085
M3O,1.751,0.002,3.836,0.007,11.5,1.025,0.6,1.2,1.665,0.001,47.0,2.049,0.382,0.075
% Platonia Dilemma n=6 (64 joint actions, 6 optimal)
Joint,0.003,0.006,0.029,0.061,6.0,0.0,0.0,0.0,0.006,0.01,64.0,0.0,1.0,0.0
F1,2.215,0.0,15.546,0.079,5.0,0.0,6.0,0.0,4.192,0.004,62.0,0.0,0.816,0.0
F2R,2.11,0.0,14.174,0.058,5.0,0.0,6.0,0.0,4.087,0.004,62.0,0.0,0.816,0.0
F3R,2.004,0.0,12.902,0.062,5.0,0.0,6.0,0.0,4.042,0.003,62.0,0.0,0.816,0.0
F2C,1.688,0.0,8.915,0.106,5.0,0.0,6.0,0.0,4.393,0.009,62.0,0.0,0.816,0.0
F3C,1.689,0.0,8.974,0.084,5.0,0.0,6.0,0.0,4.519,0.022,62.0,0.0,0.816,0.0
F2O,2.004,0.0,12.828,0.061,4.8,0.4,6.0,0.0,3.941,0.036,61.6,0.8,0.779,0.074
F3O,1.776,0.024,10.1,0.363,5.0,0.0,6.0,0.0,4.149,0.079,62.0,0.0,0.816,0.0
M1,2.8,0.001,27.009,0.035,5.0,0.0,6.0,0.0,5.145,0.001,62.0,0.0,0.816,0.0
M2R,2.533,0.001,23.931,0.037,5.0,0.0,6.0,0.0,4.928,0.001,62.0,0.0,0.816,0.0
M3R,2.276,0.004,20.514,0.075,5.0,0.0,6.0,0.0,4.64,0.005,62.0,0.0,0.816,0.0
M2C,2.522,0.001,23.903,0.028,5.0,0.0,6.0,0.0,4.922,0.001,62.0,0.0,0.816,0.0
M3C,2.245,0.002,20.536,0.032,5.0,0.0,6.0,0.0,4.623,0.002,62.0,0.0,0.816,0.0
M2O,2.543,0.008,23.933,0.058,4.3,0.458,6.0,0.0,4.921,0.001,60.6,0.917,0.687,0.084
M3O,2.276,0.009,20.604,0.068,4.4,0.49,6.0,0.0,4.61,0.009,60.8,0.98,0.706,0.09
% Climb Game n=6 (728 joint actions, 1 optimal)
Joint,0.173,0.073,18.449,4.908,0.1,0.3,2.7,0.9,1.517,0.275,727.2,0.6,0.999,0.0
F1,0.583,0.0,52.287,0.149,0.0,0.0,3.0,0.0,2.157,0.002,726.0,0.0,0.98,0.0
F2R,0.517,0.0,40.95,0.048,0.0,0.0,3.0,0.0,2.057,0.001,726.0,0.0,0.98,0.0
F3R,0.443,0.0,36.512,0.169,0.0,0.0,3.0,0.0,1.924,0.002,726.0,0.0,0.98,0.0
F2C,0.254,0.0,7.864,0.106,1.0,0.0,0.0,0.0,1.395,0.002,729.0,0.0,1.0,0.0
F3C,0.169,0.001,70.767,0.669,0.0,0.0,3.0,0.0,0.961,0.002,726.0,0.0,0.98,0.0
F2O,0.452,0.0,30.825,0.069,0.0,0.0,3.0,0.0,1.937,0.002,726.0,0.0,0.98,0.0
F3O,0.299,0.01,28.887,1.932,0.0,0.0,3.0,0.0,1.536,0.026,726.0,0.0,0.98,0.0
M1,0.709,0.0,35.914,0.012,0.0,0.0,3.0,0.0,2.363,0.0,726.0,0.0,0.98,0.0
M2R,0.629,0.0,35.768,0.042,0.0,0.0,3.0,0.0,2.304,0.0,726.0,0.0,0.98,0.0
M3R,0.529,0.001,35.336,0.118,0.0,0.0,3.0,0.0,2.2,0.001,726.0,0.0,0.98,0.0
M2C,0.624,0.0,35.765,0.027,0.0,0.0,3.0,0.0,2.303,0.0,726.0,0.0,0.98,0.0
M3C,0.511,0.001,35.298,0.071,0.0,0.0,3.0,0.0,2.197,0.001,726.0,0.0,0.98,0.0
M2O,0.63,0.002,35.741,0.038,0.0,0.0,3.0,0.0,2.304,0.0,726.0,0.0,0.98,0.0
M3O,0.521,0.003,35.313,0.069,0.0,0.0,3.0,0.0,2.199,0.001,726.0,0.0,0.98,0.0
% Penalty Game n=6 (728 joint actions, 2 optimal)
Joint,1.601,0.38,18.21,4.256,1.2,0.4,0.9,1.375,3.242,0.134,727.4,0.8,1.0,0.0
F1,2.177,0.0,63.71,0.148,0.0,0.0,3.0,0.0,3.306,0.001,722.0,0.0,0.905,0.0
F2R,1.999,0.0,65.953,0.263,0.0,0.0,3.0,0.0,3.325,0.0,722.6,0.49,0.924,0.016
F3R,1.75,0.002,66.27,0.993,0.0,0.0,3.0,0.0,3.312,0.001,723.0,0.0,0.937,0.0
F2C,1.288,0.002,79.664,0.001,0.0,0.0,6.0,0.0,3.297,0.001,722.0,0.0,0.922,0.0
F3C,0.535,0.001,82.768,0.0,0.0,0.0,3.0,0.0,2.055,0.0,724.0,0.0,0.969,0.0
F2O,1.821,0.0,68.811,0.3,0.0,0.0,6.0,0.0,3.317,0.001,723.0,0.0,0.954,0.0
F3O,1.273,0.039,73.484,1.365,0.0,0.0,6.0,0.0,3.292,0.0,723.0,0.0,0.954,0.0
M1,2.708,0.001,45.265,0.052,0.0,0.0,3.0,0.0,3.663,0.001,722.0,0.0,0.905,0.0
M2R,2.429,0.003,49.105,0.15,0.0,0.0,3.0,0.0,3.537,0.001,722.7,0.458,0.928,0.015
M3R,2.092,0.014,52.231,0.831,0.0,0.0,3.0,0.0,3.428,0.004,723.0,0.0,0.937,0.0
M2C,2.413,0.001,49.122,0.059,0.0,0.0,3.0,0.0,3.54,0.001,724.0,0.0,0.969,0.0
M3C,2.024,0.005,52.447,0.204,0.0,0.0,3.0,0.0,3.43,0.001,724.0,0.0,0.969,0.0
M2O,2.432,0.006,49.114,0.137,0.0,0.0,3.0,0.0,3.537,0.001,724.0,0.0,0.969,0.0
M3O,2.055,0.019,52.557,0.499,0.0,0.0,3.0,0.0,3.427,0.001,723.1,2.385,0.959,0.021
% Generalized Firefighting n=6 (8192 joint actions, 779 optimal)
Joint,1.293,2.452,4.958,7.21,656.2,123.081,61.6,68.194,46.416,48.364,6892.8,1475.467,0.852,0.196
F1,0.156,0.0,0.201,0.001,700.2,7.111,26.2,7.332,6.384,0.079,6235.6,37.718,0.881,0.002
F2R,0.122,0.026,0.15,0.041,722.2,18.503,16.8,8.459,4.779,1.36,6776.8,383.218,0.913,0.023
F3R,0.091,0.036,0.107,0.05,743.0,24.548,11.1,10.153,3.42,1.703,7287.6,557.77,0.944,0.034
F2C,0.0,0.0,0.0,0.0,779.0,0.0,0.0,0.0,0.0,0.0,8192.0,0.0,1.0,0.0
F3C,0.0,0.0,0.0,0.0,779.0,0.0,0.0,0.0,0.0,0.0,8192.0,0.0,1.0,0.0
F2O,0.088,0.023,0.101,0.033,746.5,17.862,8.0,7.099,3.75,1.125,7333.1,381.521,0.947,0.023
F3O,0.028,0.017,0.029,0.018,777.8,3.6,0.2,0.6,1.137,0.949,8148.6,130.2,0.997,0.008
FTF,0.0,0.0,0.0,0.0,779.0,0.0,0.0,0.0,0.0,0.0,8192.0,0.0,1.0,0.0
M1,3.548,0.004,8.347,0.027,700.4,6.344,27.8,6.4,163.836,0.073,6219.9,30.386,0.88,0.001
M2R,1.847,0.145,4.919,0.157,717.5,12.492,20.6,5.352,124.607,0.618,6602.3,300.687,0.903,0.018
M3R,1.091,0.197,2.807,0.225,738.5,32.73,14.6,13.785,88.583,2.303,7097.1,702.711,0.933,0.042
M2C,1.821,0.002,4.896,0.011,777.0,0.0,0.0,0.0,124.415,0.068,7826.0,0.0,0.974,0.0
M3C,0.848,0.001,2.577,0.008,777.5,0.5,0.0,0.0,88.094,0.104,8151.0,4.025,0.997,0.0
M2O,1.971,0.106,5.078,0.111,740.8,12.742,11.7,5.001,127.214,1.909,5627.8,249.237,0.835,0.026
M3O,1.73,1.189,4.958,3.322,737.8,54.54,8.7,14.015,97.667,10.894,5866.8,682.394,0.853,0.05
MTF,2.602,0.004,6.135,0.015,779.0,0.0,0.0,0.0,133.786,0.141,8177.2,2.227,0.999,0.0
% Aloha n=6 (64 joint actions, 2 optimal)
Joint,1.125,0.0,0.0,0.0,2.0,0.0,0.0,0.0,0.076,0.0,50.7,0.9,0.876,0.011
F1,4.782,0.0,50.927,0.107,0.0,0.0,6.0,0.0,4.036,0.001,26.8,0.748,0.673,0.006
F2R,4.054,0.439,35.002,6.975,0.0,0.0,5.0,1.342,3.69,0.375,22.2,4.445,0.696,0.025
F3R,3.164,0.531,20.644,4.614,0.0,0.0,4.2,1.4,3.234,0.856,25.7,3.874,0.741,0.04
F2C,0.914,0.0,0.138,0.007,2.0,0.0,0.0,0.0,-0.038,0.003,42.0,0.0,0.893,0.0
F3C,0.07,0.0,0.138,0.005,2.0,0.0,0.0,0.0,0.22,0.003,64.0,0.0,1.0,0.0
F2O,3.266,0.267,20.63,3.049,0.0,0.0,4.4,1.2,3.236,0.531,22.7,4.451,0.743,0.028
F3O,1.456,0.319,3.554,1.349,1.2,0.6,0.8,1.327,1.185,0.394,28.7,4.562,0.832,0.027
FTF,0.0,0.0,0.0,0.0,2.0,0.0,0.0,0.0,-0.0,0.0,64.0,0.0,1.0,0.0
M1,8.262,0.005,50.836,0.061,0.0,0.0,6.0,0.0,5.466,0.002,26.5,0.5,0.67,0.01
M2R,6.522,0.156,44.165,1.535,0.0,0.0,5.0,1.342,4.569,0.109,24.5,3.956,0.704,0.021
M3R,4.527,0.496,31.345,3.761,0.0,0.0,4.0,2.191,3.406,0.588,28.1,5.356,0.768,0.055
M2C,6.51,0.004,44.645,0.06,0.0,0.0,6.0,0.0,4.588,0.002,28.0,0.0,0.758,0.0
M3C,4.562,0.004,33.452,0.066,0.0,0.0,6.0,0.0,3.616,0.002,36.3,0.9,0.856,0.005
M2O,6.631,0.389,44.51,1.136,0.0,0.0,5.2,0.98,4.624,0.183,22.0,4.775,0.663,0.043
M3O,4.711,0.287,33.363,1.186,0.0,0.0,5.2,0.98,3.647,0.163,25.4,4.2,0.738,0.046
MTF,3.195,0.004,23.876,0.057,0.0,0.0,2.0,0.0,2.783,0.002,36.2,1.327,0.883,0.003
% Dispersion Game n=9 (512 joint actions, 252 optimal)
Joint,0.933,0.488,1.743,0.873,162.2,30.318,0.0,0.0,0.247,0.094,307.3,77.121,0.488,0.188
F1,0.736,0.0,0.528,0.006,123.9,0.7,1.1,0.943,0.471,0.0,191.3,2.648,0.017,0.032
F2R,0.661,0.0,0.53,0.004,142.8,1.6,0.0,0.0,0.395,0.001,205.9,3.048,0.193,0.01
F3R,0.568,0.0,0.512,0.003,170.6,2.154,0.4,0.49,0.318,0.0,278.5,3.384,0.366,0.01
F2C,0.063,0.0,0.032,0.002,252.0,0.0,0.0,0.0,0.089,0.0,512.0,0.0,1.0,0.0
F3C,0.063,0.0,0.033,0.002,252.0,0.0,0.0,0.0,0.088,0.001,512.0,0.0,1.0,0.0
F2O,0.568,0.0,0.492,0.01,158.5,4.031,0.2,0.4,0.324,0.004,253.6,9.468,0.349,0.018
F3O,0.355,0.022,0.332,0.028,178.1,3.506,0.1,0.3,0.208,0.011,305.0,14.276,0.537,0.027
M1,0.736,0.0,0.531,0.005,124.3,0.781,0.9,0.831,0.471,0.0,191.8,3.628,0.005,0.027
M2R,0.703,0.0,0.515,0.002,142.0,2.366,0.0,0.0,0.452,0.0,204.1,5.243,0.188,0.017
M3R,0.642,0.0,0.483,0.002,170.1,0.831,0.3,0.458,0.416,0.0,276.4,4.128,0.369,0.005
M2C,0.699,0.0,0.508,0.004,252.0,0.0,0.0,0.0,0.453,0.0,512.0,0.0,1.0,0.0
M3C,0.628,0.0,0.463,0.003,252.0,0.0,0.0,0.0,0.418,0.0,512.0,0.0,1.0,0.0
M2O,0.701,0.0,0.51,0.004,159.4,3.382,0.1,0.3,0.453,0.0,253.8,9.652,0.35,0.015
M3O,0.634,0.001,0.469,0.003,171.6,4.432,0.1,0.3,0.419,0.0,289.2,12.18,0.477,0.032
% Dispersion Game n=12 (4096 joint actions, 924 optimal)
Joint,19.477,0.688,31.715,0.897,209.6,8.476,2.3,1.005,0.796,0.001,1107.8,33.973,0.003,0.011
F1,1.168,0.0,1.818,0.016,186.0,39.426,2.4,1.281,0.797,0.0,1136.9,36.621,0.006,0.009
F2R,1.077,0.0,1.751,0.015,302.7,14.388,0.0,0.0,0.704,0.001,1635.1,16.531,0.18,0.009
F3R,0.986,0.0,1.665,0.009,350.5,6.515,0.9,0.7,0.629,0.001,1363.5,19.941,0.28,0.009
F2C,0.169,0.0,0.373,0.011,924.0,0.0,0.0,0.0,0.268,0.001,4096.0,0.0,1.0,0.0
F3C,0.198,0.044,0.433,0.096,773.5,193.822,0.0,0.0,0.267,0.003,3711.0,510.039,0.923,0.101
F2O,0.986,0.0,1.642,0.018,296.5,7.71,0.6,0.49,0.636,0.003,1403.1,55.657,0.279,0.005
F3O,0.738,0.023,1.306,0.033,344.3,12.57,0.7,0.64,0.492,0.014,1672.7,39.631,0.429,0.014
M1,1.168,0.0,1.827,0.01,186.9,30.373,2.2,1.536,0.797,0.0,1134.3,36.081,0.004,0.008
M2R,1.14,0.0,1.799,0.009,302.9,4.346,0.0,0.0,0.782,0.0,1632.6,6.375,0.179,0.005
M3R,1.088,0.0,1.746,0.006,350.3,9.306,1.0,0.632,0.752,0.0,1362.9,23.543,0.276,0.008
M2C,1.138,0.0,1.796,0.005,813.2,69.452,0.0,0.0,0.782,0.0,3831.4,245.751,0.946,0.049
M3C,1.079,0.0,1.739,0.005,920.4,5.004,0.0,0.0,0.753,0.0,4088.8,10.008,0.999,0.002
M2O,1.139,0.001,1.802,0.013,291.3,8.271,0.4,0.49,0.782,0.0,1348.2,37.344,0.271,0.012
M3O,1.083,0.001,1.737,0.011,329.2,11.574,0.6,0.49,0.753,0.0,1559.7,57.236,0.388,0.021
% Generalized Firefighting n=9 (524.288 joint actions, 17.682 optimal)
Joint,69.977,0.863,142.274,1.291,2556.3,50.257,3808.4,60.81,2339.16,1.281,94559.1,522.156,0.019,0.001
F1,0.25,0.0,0.309,0.002,14887.2,90.032,319.0,103.197,64.618,1.009,333465.5,1068.711,0.859,0.0
F2R,0.679,0.175,0.862,0.321,11985.5,545.97,621.4,64.577,325.697,50.973,257662.9,12168.296,0.782,0.017
F3R,0.172,0.038,0.199,0.052,15934.2,479.278,157.8,74.572,38.632,12.678,393625.1,30194.826,0.906,0.022
F2C,0.0,0.0,0.0,0.0,17682.0,0.0,0.0,0.0,0.0,0.0,524288.0,0.0,1.0,0.0
F3C,0.0,0.0,0.0,0.0,17682.0,0.0,0.0,0.0,0.0,0.0,524288.0,0.0,1.0,0.0
F2O,0.181,0.027,0.21,0.034,15760.9,285.83,185.7,40.485,40.583,10.054,385287.5,20998.612,0.9,0.016
F3O,0.091,0.026,0.099,0.032,16939.0,414.632,51.4,38.663,22.715,8.577,464893.1,26118.502,0.957,0.018
FTF,0.0,0.0,0.0,0.0,17682.0,0.0,0.0,0.0,0.0,0.0,524288.0,0.0,1.0,0.0
M1,6.547,0.004,17.681,0.043,14848.4,73.29,331.8,62.707,2030.738,0.866,333710.6,767.215,0.859,0.0
M2R,4.544,0.176,13.091,0.219,12443.8,530.782,596.6,121.223,1718.256,11.763,265764.8,10894.723,0.795,0.016
M3R,2.774,0.245,9.051,0.208,15596.3,428.98,209.4,59.89,1425.885,9.872,371663.9,25374.327,0.89,0.019
M2C,4.296,0.004,12.833,0.024,17680.0,0.0,0.0,0.0,1721.178,0.878,465018.6,2.375,0.954,0.0
M3C,2.706,0.003,8.99,0.014,17680.0,0.0,0.0,0.0,1422.875,0.729,465089.8,2.088,0.954,0.0
M2O,4.365,0.146,12.872,0.152,15438.6,432.853,231.4,72.577,1723.511,16.058,267534.2,15641.806,0.781,0.027
M3O,2.958,0.104,9.238,0.134,16783.2,302.897,54.0,23.921,1431.611,23.036,298182.0,26363.664,0.817,0.032
MTF,5.304,0.004,14.338,0.023,17682.0,0.0,0.0,0.0,1765.642,0.978,512022.4,810.2,0.99,0.001
% Aloha n=9 (512 joint actions, 1 optimal)
Joint,4.863,1.637,47.771,9.153,0.0,0.0,4.4,1.744,4.473,0.619,215.3,17.059,0.707,0.036
F1,6.136,0.005,130.683,0.627,0.0,0.0,10.0,0.0,6.615,0.005,127.0,1.414,0.635,0.001
F2R,6.968,0.446,123.345,26.454,0.0,0.0,9.0,1.844,6.733,0.345,100.4,11.038,0.581,0.023
F3R,5.047,0.548,87.352,15.459,0.0,0.0,8.0,2.191,5.972,0.611,132.8,7.414,0.683,0.028
F2C,2.25,0.007,9.027,0.2,1.0,0.0,0.0,0.0,2.06,0.025,186.6,1.114,0.817,0.001
F3C,1.317,0.007,18.595,0.261,0.0,0.0,2.0,0.0,1.472,0.012,364.6,1.2,0.881,0.001
F2O,4.979,0.241,82.302,7.747,0.0,0.0,6.6,1.8,5.711,0.415,135.3,9.403,0.689,0.012
F3O,3.945,0.368,49.172,9.605,0.0,0.0,7.0,2.236,5.008,0.669,138.0,7.836,0.728,0.014
FTF,0.0,0.0,0.0,0.0,1.0,0.0,0.0,0.0,0.0,0.0,512.0,0.0,1.0,0.0
M1,11.929,0.003,145.967,0.115,0.0,0.0,10.0,0.0,9.684,0.002,125.9,1.044,0.635,0.0
M2R,10.449,0.249,140.152,5.66,0.0,0.0,9.0,1.612,8.849,0.124,105.2,10.591,0.596,0.026
M3R,8.714,0.268,125.24,6.225,0.0,0.0,7.8,0.6,7.851,0.224,131.3,6.754,0.685,0.023
M2C,10.311,0.004,138.522,0.133,0.0,0.0,10.0,0.0,8.793,0.002,140.4,1.428,0.681,0.001
M3C,8.679,0.004,126.24,0.216,0.0,0.0,10.0,0.0,7.862,0.002,152.1,1.375,0.724,0.001
M2O,10.478,0.186,138.601,5.831,0.0,0.0,9.6,0.8,8.862,0.14,114.4,10.423,0.615,0.036
M3O,8.857,0.449,124.022,12.324,0.0,0.0,9.4,1.8,7.867,0.338,124.5,11.783,0.63,0.039
MTF,6.251,0.042,85.949,1.061,0.0,0.0,2.0,0.0,6.101,0.025,190.7,4.92,0.844,0.002
% Aloha n=12 (4096 joint actions, 2 optimal)
Joint,23.98,0.639,135.364,2.538,0.0,0.0,7.4,3.105,12.115,0.273,700.4,13.463,0.418,0.005
F1,7.295,0.002,231.092,0.323,0.0,0.0,12.0,0.0,7.963,0.003,861.1,3.885,0.635,0.001
F2R,6.842,0.197,199.449,11.054,0.0,0.0,10.4,1.2,7.668,0.249,851.2,33.099,0.648,0.006
F3R,6.56,0.313,182.165,15.496,0.0,0.0,9.8,2.088,7.59,0.305,860.2,47.326,0.658,0.011
F2C,2.181,0.004,10.364,0.142,2.0,0.0,0.0,0.0,2.062,0.024,1379.8,4.578,0.832,0.0
F3C,0.809,0.005,12.45,0.171,1.0,0.0,2.0,0.0,1.1,0.01,2487.5,9.521,0.914,0.0
F2O,6.322,0.267,169.464,13.094,0.0,0.0,10.2,1.661,7.286,0.329,863.6,34.021,0.665,0.01
F3O,4.707,0.539,92.222,18.581,0.0,0.0,7.6,2.332,5.941,0.727,954.4,67.353,0.725,0.021
FTF,0.0,0.0,0.0,0.0,2.0,0.0,0.0,0.0,0.0,0.0,4096.0,0.0,1.0,0.0
M1,15.374,0.029,230.648,0.367,0.0,0.0,12.0,0.0,12.449,0.015,845.0,25.838,0.632,0.01
M2R,13.846,0.087,225.179,2.281,0.0,0.0,10.8,1.327,11.605,0.051,838.0,22.204,0.647,0.01
M3R,12.378,0.177,217.039,4.976,0.0,0.0,10.0,1.789,10.762,0.122,852.8,33.811,0.658,0.014
M2C,13.826,0.012,225.327,0.238,0.0,0.0,12.0,0.0,11.596,0.006,907.6,7.283,0.671,0.001
M3C,12.273,0.009,214.924,0.177,0.0,0.0,12.0,0.0,10.69,0.004,968.6,5.851,0.705,0.001
M2O,13.938,0.182,225.436,1.462,0.0,0.0,11.6,0.8,11.614,0.105,670.6,57.535,0.555,0.031
M3O,12.562,0.308,214.012,3.247,0.0,0.0,11.4,0.917,10.846,0.193,752.3,87.455,0.609,0.04
MTF,9.637,0.048,167.677,0.614,0.0,0.0,4.0,0.0,8.824,0.017,1253.6,13.8,0.825,0.002
\end{filecontents*}
\pgfplotstabletypeset[col sep=comma,
	every head row/.style={before row=\toprule, after row=\midrule\endhead},
    every last row/.style={after row=\bottomrule},
	header=false,
	fixed,
	zerofill,
    columns/Model/.style={column type={C{0.9cm}},string type},
	columns/Mean square error/.style={string type, column type={C{1.5cm}}},
	create on use/Mean square error/.style={
		create col/assign/.code={
		\pgfmathprintpmnumber{1}{2}{2}{1}
		\pgfkeyslet{/pgfplots/table/create col/next content}\value
		}
	},
	columns/MSE on optimal actions/.style={string type, column type={C{1.7cm}}},
	create on use/MSE on optimal actions/.style={
		create col/assign/.code={
		\pgfmathprintpmnumber{3}{4}{2}{1}
		\pgfkeyslet{/pgfplots/table/create col/next content}\value
		}
	},
	columns/Optimal actions found/.style={string type, column type={C{1.4cm}}},
	create on use/Optimal actions found/.style={
		create col/assign/.code={
		\pgfmathprintpmnumber{5}{6}{0}{0}
	    \pgfkeyslet{/pgfplots/table/create col/next content}\value
		}
	},
	columns/Value loss/.style={string type, column type={C{1.7cm}}},
	create on use/Value loss/.style={
		create col/assign/.code={
		\pgfmathprintpmnumber{7}{8}{2}{1}
		\pgfkeyslet{/pgfplots/table/create col/next content}\value
		}
	},
	columns/Boltzmann value loss/.style={string type, column type={C{1.7cm}}},
	create on use/Boltzmann value loss/.style={
		create col/assign/.code={
		\pgfmathprintpmnumber{9}{10}{2}{1}
		\pgfkeyslet{/pgfplots/table/create col/next content}\value
		}
	},
	columns/Correctly ranked/.style={string type, column type={C{1.8cm}}},
	create on use/Correctly ranked/.style={
		create col/assign/.code={
		\pgfmathprintpmnumber{11}{12}{0}{0}
		\pgfkeyslet{/pgfplots/table/create col/next content}\value
		}
	},
	columns/Kendall Tau/.style={string type, column type={C{1.4cm}}, column name={Kendall $\tau$}},
	create on use/Kendall Tau/.style={
		create col/assign/.code={
		\pgfmathprintpmnumber{13}{14}{2}{1}
		\pgfkeyslet{/pgfplots/table/create col/next content}\value
		}
	},
	columns={Model,Mean square error,MSE on optimal actions,Optimal actions found,Value loss,Boltzmann value loss,Correctly ranked,Kendall Tau},
	every row no 0/.style={before row={\multicolumn{8}{c}{\textbf{Dispersion Game $n=6$ (64 joint actions, 20 optimal)}}\\\midrule}},
	every row no 15/.style={before row={\midrule\multicolumn{8}{c}{\textbf{Dispersion Game (sparse) $n=6$ (64 joint actions, 20 optimal)}}\\\midrule}},
	every row no 30/.style={before row={\midrule\multicolumn{8}{c}{\textbf{Platonia Dilemma $n=6$ (64 joint actions, 6 optimal)}}\\\midrule}},
	every row no 45/.style={before row={\midrule\multicolumn{8}{c}{\textbf{Climb Game $n=6$ (729 joint actions, 1 optimal)}}\\\midrule}},
	every row no 60/.style={before row={\midrule\multicolumn{8}{c}{\textbf{Penalty Game $n=6$ (729 joint actions, 2 optimal)}}\\\midrule}},
	every row no 75/.style={before row={\midrule\multicolumn{8}{c}{\textbf{Generalized Firefighting $n=6$ (8192 joint actions, 779 optimal)}}\\\midrule}},
	every row no 92/.style={before row={\midrule\multicolumn{8}{c}{\textbf{Aloha $n=6$ (64 joint actions, 2 optimal)}}\\\midrule}},
	every row no 109/.style={before row={\midrule\multicolumn{8}{c}{\textbf{Dispersion Game $n=9$ (512 joint actions, 252 optimal)}}\\\midrule}},
	every row no 124/.style={before row={\midrule\multicolumn{8}{c}{\textbf{Dispersion Game $n=12$ (4096 joint actions, 924 optimal)}}\\\midrule}},
	every row no 139/.style={before row={\midrule\multicolumn{8}{c}{\textbf{Generalized Firefighting $n=9$ (524.288 joint actions, 17.682 optimal)}}\\\midrule}},
	every row no 156/.style={before row={\midrule\multicolumn{8}{c}{\textbf{Aloha $n=9$ (512 joint actions, 1 optimal)}}\\\midrule}},
	every row no 173/.style={before row={\midrule\multicolumn{8}{c}{\textbf{Aloha $n=12$ (4096 joint actions, 2 optimal)}}\\\midrule}},
	every row no 0 column no 1/.style={postproc cell content/.append style={@cell content/.add={\cellcolor{green!25}}{}}},
	every row no 0 column no 3/.style={postproc cell content/.append style={@cell content/.add={\cellcolor{green!25}}{}}},
	every row no 0 column no 6/.style={postproc cell content/.append style={@cell content/.add={\cellcolor{green!25}}{}}},
	every row no 1 column no 1/.style={postproc cell content/.append style={@cell content/.add={\cellcolor{red!25}}{}}},
	every row no 1 column no 3/.style={postproc cell content/.append style={@cell content/.add={\cellcolor{red!25}}{}}},
	every row no 1 column no 6/.style={postproc cell content/.append style={@cell content/.add={\cellcolor{red!25}}{}}},
	every row no 2 column no 1/.style={postproc cell content/.append style={@cell content/.add={\cellcolor{red!25}}{}}},
	every row no 2 column no 3/.style={postproc cell content/.append style={@cell content/.add={\cellcolor{red!25}}{}}},
	every row no 2 column no 6/.style={postproc cell content/.append style={@cell content/.add={\cellcolor{red!25}}{}}},
	every row no 4 column no 1/.style={postproc cell content/.append style={@cell content/.add={\cellcolor{green!25}}{}}},
	every row no 4 column no 3/.style={postproc cell content/.append style={@cell content/.add={\cellcolor{green!25}}{}}},
	every row no 4 column no 6/.style={postproc cell content/.append style={@cell content/.add={\cellcolor{green!25}}{}}},
	every row no 5 column no 1/.style={postproc cell content/.append style={@cell content/.add={\cellcolor{green!25}}{}}},
	every row no 5 column no 3/.style={postproc cell content/.append style={@cell content/.add={\cellcolor{green!25}}{}}},
	every row no 5 column no 6/.style={postproc cell content/.append style={@cell content/.add={\cellcolor{green!25}}{}}},
	every row no 7 column no 1/.style={postproc cell content/.append style={@cell content/.add={\cellcolor{green!25}}{}}},
	every row no 8 column no 1/.style={postproc cell content/.append style={@cell content/.add={\cellcolor{red!25}}{}}},
	every row no 8 column no 3/.style={postproc cell content/.append style={@cell content/.add={\cellcolor{red!25}}{}}},
	every row no 8 column no 6/.style={postproc cell content/.append style={@cell content/.add={\cellcolor{red!25}}{}}},
	every row no 9 column no 1/.style={postproc cell content/.append style={@cell content/.add={\cellcolor{red!25}}{}}},
	every row no 9 column no 3/.style={postproc cell content/.append style={@cell content/.add={\cellcolor{red!25}}{}}},
	every row no 9 column no 6/.style={postproc cell content/.append style={@cell content/.add={\cellcolor{red!25}}{}}},
	every row no 11 column no 1/.style={postproc cell content/.append style={@cell content/.add={\cellcolor{red!25}}{}}},
	every row no 11 column no 3/.style={postproc cell content/.append style={@cell content/.add={\cellcolor{green!25}}{}}},
	every row no 11 column no 6/.style={postproc cell content/.append style={@cell content/.add={\cellcolor{green!25}}{}}},
	every row no 12 column no 3/.style={postproc cell content/.append style={@cell content/.add={\cellcolor{green!25}}{}}},
	every row no 12 column no 6/.style={postproc cell content/.append style={@cell content/.add={\cellcolor{green!25}}{}}},
	every row no 13 column no 1/.style={postproc cell content/.append style={@cell content/.add={\cellcolor{red!25}}{}}},
	every row no 15 column no 1/.style={postproc cell content/.append style={@cell content/.add={\cellcolor{green!25}}{}}},
	every row no 15 column no 3/.style={postproc cell content/.append style={@cell content/.add={\cellcolor{green!25}}{}}},
	every row no 15 column no 6/.style={postproc cell content/.append style={@cell content/.add={\cellcolor{green!25}}{}}},
	every row no 16 column no 1/.style={postproc cell content/.append style={@cell content/.add={\cellcolor{red!25}}{}}},
	every row no 16 column no 3/.style={postproc cell content/.append style={@cell content/.add={\cellcolor{red!25}}{}}},
	every row no 16 column no 6/.style={postproc cell content/.append style={@cell content/.add={\cellcolor{red!25}}{}}},
	every row no 17 column no 1/.style={postproc cell content/.append style={@cell content/.add={\cellcolor{red!25}}{}}},
	every row no 17 column no 3/.style={postproc cell content/.append style={@cell content/.add={\cellcolor{red!25}}{}}},
	every row no 19 column no 3/.style={postproc cell content/.append style={@cell content/.add={\cellcolor{green!25}}{}}},
	every row no 19 column no 6/.style={postproc cell content/.append style={@cell content/.add={\cellcolor{green!25}}{}}},
	every row no 20 column no 3/.style={postproc cell content/.append style={@cell content/.add={\cellcolor{green!25}}{}}},
	every row no 20 column no 6/.style={postproc cell content/.append style={@cell content/.add={\cellcolor{green!25}}{}}},
	every row no 23 column no 1/.style={postproc cell content/.append style={@cell content/.add={\cellcolor{red!25}}{}}},
	every row no 23 column no 3/.style={postproc cell content/.append style={@cell content/.add={\cellcolor{red!25}}{}}},
	every row no 23 column no 6/.style={postproc cell content/.append style={@cell content/.add={\cellcolor{red!25}}{}}},
	every row no 24 column no 1/.style={postproc cell content/.append style={@cell content/.add={\cellcolor{red!25}}{}}},
	every row no 24 column no 3/.style={postproc cell content/.append style={@cell content/.add={\cellcolor{red!25}}{}}},
	every row no 26 column no 1/.style={postproc cell content/.append style={@cell content/.add={\cellcolor{red!25}}{}}},
	every row no 26 column no 3/.style={postproc cell content/.append style={@cell content/.add={\cellcolor{green!25}}{}}},
	every row no 26 column no 6/.style={postproc cell content/.append style={@cell content/.add={\cellcolor{green!25}}{}}},
	every row no 27 column no 3/.style={postproc cell content/.append style={@cell content/.add={\cellcolor{green!25}}{}}},
	every row no 27 column no 6/.style={postproc cell content/.append style={@cell content/.add={\cellcolor{green!25}}{}}},
	every row no 28 column no 1/.style={postproc cell content/.append style={@cell content/.add={\cellcolor{red!25}}{}}},
	every row no 30 column no 1/.style={postproc cell content/.append style={@cell content/.add={\cellcolor{green!25}}{}}},
	every row no 30 column no 3/.style={postproc cell content/.append style={@cell content/.add={\cellcolor{green!25}}{}}},
	every row no 30 column no 6/.style={postproc cell content/.append style={@cell content/.add={\cellcolor{green!25}}{}}},
	every row no 38 column no 1/.style={postproc cell content/.append style={@cell content/.add={\cellcolor{red!25}}{}}},
	every row no 43 column no 3/.style={postproc cell content/.append style={@cell content/.add={\cellcolor{red!25}}{}}},
	every row no 43 column no 6/.style={postproc cell content/.append style={@cell content/.add={\cellcolor{red!25}}{}}},
	every row no 44 column no 3/.style={postproc cell content/.append style={@cell content/.add={\cellcolor{red!25}}{}}},
	every row no 44 column no 6/.style={postproc cell content/.append style={@cell content/.add={\cellcolor{red!25}}{}}},
	every row no 45 column no 1/.style={postproc cell content/.append style={@cell content/.add={\cellcolor{green!25}}{}}},
	every row no 49 column no 1/.style={postproc cell content/.append style={@cell content/.add={\cellcolor{green!25}}{}}},
	every row no 49 column no 3/.style={postproc cell content/.append style={@cell content/.add={\cellcolor{green!25}}{}}},
	every row no 49 column no 6/.style={postproc cell content/.append style={@cell content/.add={\cellcolor{green!25}}{}}},
	every row no 50 column no 1/.style={postproc cell content/.append style={@cell content/.add={\cellcolor{green!25}}{}}},
	every row no 53 column no 1/.style={postproc cell content/.append style={@cell content/.add={\cellcolor{red!25}}{}}},
	every row no 60 column no 3/.style={postproc cell content/.append style={@cell content/.add={\cellcolor{green!25}}{}}},
	every row no 60 column no 6/.style={postproc cell content/.append style={@cell content/.add={\cellcolor{green!25}}{}}},
	every row no 61 column no 6/.style={postproc cell content/.append style={@cell content/.add={\cellcolor{red!25}}{}}},
	every row no 64 column no 1/.style={postproc cell content/.append style={@cell content/.add={\cellcolor{green!25}}{}}},
	every row no 64 column no 6/.style={postproc cell content/.append style={@cell content/.add={\cellcolor{red!25}}{}}},
	every row no 65 column no 1/.style={postproc cell content/.append style={@cell content/.add={\cellcolor{green!25}}{}}},
	every row no 67 column no 1/.style={postproc cell content/.append style={@cell content/.add={\cellcolor{green!25}}{}}},
	every row no 68 column no 1/.style={postproc cell content/.append style={@cell content/.add={\cellcolor{red!25}}{}}},
	every row no 68 column no 6/.style={postproc cell content/.append style={@cell content/.add={\cellcolor{red!25}}{}}},
	every row no 75 column no 3/.style={postproc cell content/.append style={@cell content/.add={\cellcolor{red!25}}{}}},
	every row no 78 column no 1/.style={postproc cell content/.append style={@cell content/.add={\cellcolor{green!25}}{}}},
	every row no 79 column no 1/.style={postproc cell content/.append style={@cell content/.add={\cellcolor{green!25}}{}}},
	every row no 79 column no 3/.style={postproc cell content/.append style={@cell content/.add={\cellcolor{green!25}}{}}},
	every row no 79 column no 6/.style={postproc cell content/.append style={@cell content/.add={\cellcolor{green!25}}{}}},
	every row no 80 column no 1/.style={postproc cell content/.append style={@cell content/.add={\cellcolor{green!25}}{}}},
	every row no 80 column no 3/.style={postproc cell content/.append style={@cell content/.add={\cellcolor{green!25}}{}}},
	every row no 80 column no 6/.style={postproc cell content/.append style={@cell content/.add={\cellcolor{green!25}}{}}},
	every row no 81 column no 1/.style={postproc cell content/.append style={@cell content/.add={\cellcolor{green!25}}{}}},
	every row no 82 column no 1/.style={postproc cell content/.append style={@cell content/.add={\cellcolor{green!25}}{}}},
	every row no 82 column no 6/.style={postproc cell content/.append style={@cell content/.add={\cellcolor{green!25}}{}}},
	every row no 82 column no 3/.style={postproc cell content/.append style={@cell content/.add={\cellcolor{green!25}}{}}},
	every row no 82 column no 6/.style={postproc cell content/.append style={@cell content/.add={\cellcolor{green!25}}{}}},
	every row no 83 column no 1/.style={postproc cell content/.append style={@cell content/.add={\cellcolor{green!25}}{}}},
	every row no 83 column no 3/.style={postproc cell content/.append style={@cell content/.add={\cellcolor{green!25}}{}}},
	every row no 83 column no 6/.style={postproc cell content/.append style={@cell content/.add={\cellcolor{green!25}}{}}},
	every row no 84 column no 1/.style={postproc cell content/.append style={@cell content/.add={\cellcolor{red!25}}{}}},
	every row no 87 column no 3/.style={postproc cell content/.append style={@cell content/.add={\cellcolor{green!25}}{}}},
	every row no 87 column no 6/.style={postproc cell content/.append style={@cell content/.add={\cellcolor{green!25}}{}}},
	every row no 88 column no 3/.style={postproc cell content/.append style={@cell content/.add={\cellcolor{green!25}}{}}},
	every row no 88 column no 6/.style={postproc cell content/.append style={@cell content/.add={\cellcolor{green!25}}{}}},
	every row no 89 column no 6/.style={postproc cell content/.append style={@cell content/.add={\cellcolor{red!25}}{}}},
	every row no 90 column no 6/.style={postproc cell content/.append style={@cell content/.add={\cellcolor{red!25}}{}}},
	every row no 91 column no 1/.style={postproc cell content/.append style={@cell content/.add={\cellcolor{red!25}}{}}},
	every row no 91 column no 3/.style={postproc cell content/.append style={@cell content/.add={\cellcolor{green!25}}{}}},
	every row no 91 column no 6/.style={postproc cell content/.append style={@cell content/.add={\cellcolor{green!25}}{}}},
	every row no 92 column no 3/.style={postproc cell content/.append style={@cell content/.add={\cellcolor{green!25}}{}}},
	every row no 94 column no 6/.style={postproc cell content/.append style={@cell content/.add={\cellcolor{red!25}}{}}},
	every row no 95 column no 6/.style={postproc cell content/.append style={@cell content/.add={\cellcolor{red!25}}{}}},
	every row no 96 column no 3/.style={postproc cell content/.append style={@cell content/.add={\cellcolor{green!25}}{}}},
	every row no 97 column no 1/.style={postproc cell content/.append style={@cell content/.add={\cellcolor{green!25}}{}}},
	every row no 97 column no 3/.style={postproc cell content/.append style={@cell content/.add={\cellcolor{green!25}}{}}},
	every row no 97 column no 6/.style={postproc cell content/.append style={@cell content/.add={\cellcolor{green!25}}{}}},
	every row no 98 column no 6/.style={postproc cell content/.append style={@cell content/.add={\cellcolor{red!25}}{}}},
	every row no 99 column no 3/.style={postproc cell content/.append style={@cell content/.add={\cellcolor{green!25}}{}}},
	every row no 100 column no 1/.style={postproc cell content/.append style={@cell content/.add={\cellcolor{green!25}}{}}},
	every row no 100 column no 3/.style={postproc cell content/.append style={@cell content/.add={\cellcolor{green!25}}{}}},
	every row no 100 column no 6/.style={postproc cell content/.append style={@cell content/.add={\cellcolor{green!25}}{}}},
	every row no 101 column no 1/.style={postproc cell content/.append style={@cell content/.add={\cellcolor{red!25}}{}}},
	every row no 102 column no 6/.style={postproc cell content/.append style={@cell content/.add={\cellcolor{red!25}}{}}},
	every row no 106 column no 6/.style={postproc cell content/.append style={@cell content/.add={\cellcolor{red!25}}{}}},
	every row no 107 column no 6/.style={postproc cell content/.append style={@cell content/.add={\cellcolor{red!25}}{}}},
	every row no 109 column no 1/.style={postproc cell content/.append style={@cell content/.add={\cellcolor{red!25}}{}}},
	every row no 110 column no 3/.style={postproc cell content/.append style={@cell content/.add={\cellcolor{red!25}}{}}},
	every row no 110 column no 6/.style={postproc cell content/.append style={@cell content/.add={\cellcolor{red!25}}{}}},
	every row no 113 column no 1/.style={postproc cell content/.append style={@cell content/.add={\cellcolor{green!25}}{}}},
	every row no 113 column no 3/.style={postproc cell content/.append style={@cell content/.add={\cellcolor{green!25}}{}}},
	every row no 113 column no 6/.style={postproc cell content/.append style={@cell content/.add={\cellcolor{green!25}}{}}},
	every row no 114 column no 1/.style={postproc cell content/.append style={@cell content/.add={\cellcolor{green!25}}{}}},
	every row no 114 column no 3/.style={postproc cell content/.append style={@cell content/.add={\cellcolor{green!25}}{}}},
	every row no 114 column no 6/.style={postproc cell content/.append style={@cell content/.add={\cellcolor{green!25}}{}}},
	every row no 117 column no 3/.style={postproc cell content/.append style={@cell content/.add={\cellcolor{red!25}}{}}},
	every row no 117 column no 6/.style={postproc cell content/.append style={@cell content/.add={\cellcolor{red!25}}{}}},
	every row no 120 column no 3/.style={postproc cell content/.append style={@cell content/.add={\cellcolor{green!25}}{}}},
	every row no 120 column no 6/.style={postproc cell content/.append style={@cell content/.add={\cellcolor{green!25}}{}}},
	every row no 121 column no 3/.style={postproc cell content/.append style={@cell content/.add={\cellcolor{green!25}}{}}},
	every row no 121 column no 6/.style={postproc cell content/.append style={@cell content/.add={\cellcolor{green!25}}{}}},
	every row no 124 column no 1/.style={postproc cell content/.append style={@cell content/.add={\cellcolor{red!25}}{}}},
	every row no 124 column no 3/.style={postproc cell content/.append style={@cell content/.add={\cellcolor{red!25}}{}}},
	every row no 124 column no 6/.style={postproc cell content/.append style={@cell content/.add={\cellcolor{red!25}}{}}},
	every row no 125 column no 3/.style={postproc cell content/.append style={@cell content/.add={\cellcolor{red!25}}{}}},
	every row no 125 column no 6/.style={postproc cell content/.append style={@cell content/.add={\cellcolor{red!25}}{}}},
	every row no 127 column no 6/.style={postproc cell content/.append style={@cell content/.add={\cellcolor{red!25}}{}}},
	every row no 128 column no 1/.style={postproc cell content/.append style={@cell content/.add={\cellcolor{green!25}}{}}},
	every row no 128 column no 3/.style={postproc cell content/.append style={@cell content/.add={\cellcolor{green!25}}{}}},
	every row no 128 column no 6/.style={postproc cell content/.append style={@cell content/.add={\cellcolor{green!25}}{}}},
	every row no 129 column no 1/.style={postproc cell content/.append style={@cell content/.add={\cellcolor{green!25}}{}}},
	every row no 129 column no 3/.style={postproc cell content/.append style={@cell content/.add={\cellcolor{green!25}}{}}},
	every row no 129 column no 6/.style={postproc cell content/.append style={@cell content/.add={\cellcolor{green!25}}{}}},
	every row no 132 column no 3/.style={postproc cell content/.append style={@cell content/.add={\cellcolor{red!25}}{}}},
	every row no 132 column no 6/.style={postproc cell content/.append style={@cell content/.add={\cellcolor{red!25}}{}}},
	every row no 134 column no 6/.style={postproc cell content/.append style={@cell content/.add={\cellcolor{red!25}}{}}},
	every row no 135 column no 3/.style={postproc cell content/.append style={@cell content/.add={\cellcolor{green!25}}{}}},
	every row no 135 column no 6/.style={postproc cell content/.append style={@cell content/.add={\cellcolor{green!25}}{}}},
	every row no 136 column no 3/.style={postproc cell content/.append style={@cell content/.add={\cellcolor{green!25}}{}}},
	every row no 136 column no 6/.style={postproc cell content/.append style={@cell content/.add={\cellcolor{green!25}}{}}},
	every row no 139 column no 1/.style={postproc cell content/.append style={@cell content/.add={\cellcolor{red!25}}{}}},
	every row no 139 column no 3/.style={postproc cell content/.append style={@cell content/.add={\cellcolor{red!25}}{}}},
	every row no 139 column no 6/.style={postproc cell content/.append style={@cell content/.add={\cellcolor{red!25}}{}}},
	every row no 143 column no 1/.style={postproc cell content/.append style={@cell content/.add={\cellcolor{green!25}}{}}},
	every row no 143 column no 3/.style={postproc cell content/.append style={@cell content/.add={\cellcolor{green!25}}{}}},
	every row no 143 column no 6/.style={postproc cell content/.append style={@cell content/.add={\cellcolor{green!25}}{}}},
	every row no 144 column no 1/.style={postproc cell content/.append style={@cell content/.add={\cellcolor{green!25}}{}}},
	every row no 144 column no 3/.style={postproc cell content/.append style={@cell content/.add={\cellcolor{green!25}}{}}},
	every row no 144 column no 6/.style={postproc cell content/.append style={@cell content/.add={\cellcolor{green!25}}{}}},
	every row no 146 column no 1/.style={postproc cell content/.append style={@cell content/.add={\cellcolor{green!25}}{}}},
	every row no 146 column no 3/.style={postproc cell content/.append style={@cell content/.add={\cellcolor{green!25}}{}}},
	every row no 146 column no 6/.style={postproc cell content/.append style={@cell content/.add={\cellcolor{green!25}}{}}},
	every row no 147 column no 1/.style={postproc cell content/.append style={@cell content/.add={\cellcolor{green!25}}{}}},
	every row no 147 column no 3/.style={postproc cell content/.append style={@cell content/.add={\cellcolor{green!25}}{}}},
	every row no 147 column no 6/.style={postproc cell content/.append style={@cell content/.add={\cellcolor{green!25}}{}}},
	every row no 151 column no 3/.style={postproc cell content/.append style={@cell content/.add={\cellcolor{green!25}}{}}},
	every row no 151 column no 6/.style={postproc cell content/.append style={@cell content/.add={\cellcolor{green!25}}{}}},
	every row no 152 column no 3/.style={postproc cell content/.append style={@cell content/.add={\cellcolor{green!25}}{}}},
	every row no 152 column no 6/.style={postproc cell content/.append style={@cell content/.add={\cellcolor{green!25}}{}}},
	every row no 154 column no 3/.style={postproc cell content/.append style={@cell content/.add={\cellcolor{green!25}}{}}},
	every row no 154 column no 6/.style={postproc cell content/.append style={@cell content/.add={\cellcolor{green!25}}{}}},
	every row no 155 column no 3/.style={postproc cell content/.append style={@cell content/.add={\cellcolor{green!25}}{}}},
	every row no 155 column no 6/.style={postproc cell content/.append style={@cell content/.add={\cellcolor{green!25}}{}}},
	every row no 158 column no 6/.style={postproc cell content/.append style={@cell content/.add={\cellcolor{red!25}}{}}},
	every row no 160 column no 3/.style={postproc cell content/.append style={@cell content/.add={\cellcolor{green!25}}{}}},
	every row no 164 column no 1/.style={postproc cell content/.append style={@cell content/.add={\cellcolor{green!25}}{}}},
	every row no 164 column no 3/.style={postproc cell content/.append style={@cell content/.add={\cellcolor{green!25}}{}}},
	every row no 164 column no 6/.style={postproc cell content/.append style={@cell content/.add={\cellcolor{green!25}}{}}},
	every row no 165 column no 1/.style={postproc cell content/.append style={@cell content/.add={\cellcolor{red!25}}{}}},
	every row no 166 column no 1/.style={postproc cell content/.append style={@cell content/.add={\cellcolor{red!25}}{}}},
	every row no 166 column no 6/.style={postproc cell content/.append style={@cell content/.add={\cellcolor{red!25}}{}}},
	every row no 168 column no 1/.style={postproc cell content/.append style={@cell content/.add={\cellcolor{red!25}}{}}},
	every row no 170 column no 1/.style={postproc cell content/.append style={@cell content/.add={\cellcolor{red!25}}{}}},
	every row no 170 column no 6/.style={postproc cell content/.append style={@cell content/.add={\cellcolor{red!25}}{}}},
	every row no 173 column no 1/.style={postproc cell content/.append style={@cell content/.add={\cellcolor{red!25}}{}}},
	every row no 173 column no 6/.style={postproc cell content/.append style={@cell content/.add={\cellcolor{red!25}}{}}},
	every row no 177 column no 3/.style={postproc cell content/.append style={@cell content/.add={\cellcolor{green!25}}{}}},
	every row no 178 column no 1/.style={postproc cell content/.append style={@cell content/.add={\cellcolor{green!25}}{}}},
	every row no 178 column no 3/.style={postproc cell content/.append style={@cell content/.add={\cellcolor{green!25}}{}}},
	every row no 178 column no 6/.style={postproc cell content/.append style={@cell content/.add={\cellcolor{green!25}}{}}},
	every row no 181 column no 1/.style={postproc cell content/.append style={@cell content/.add={\cellcolor{green!25}}{}}},
	every row no 181 column no 3/.style={postproc cell content/.append style={@cell content/.add={\cellcolor{green!25}}{}}},
	every row no 181 column no 6/.style={postproc cell content/.append style={@cell content/.add={\cellcolor{green!25}}{}}},
	every row no 182 column no 1/.style={postproc cell content/.append style={@cell content/.add={\cellcolor{red!25}}{}}},
	every row no 187 column no 6/.style={postproc cell content/.append style={@cell content/.add={\cellcolor{red!25}}{}}},
	every row no 188 column no 6/.style={postproc cell content/.append style={@cell content/.add={\cellcolor{red!25}}{}}},
]{csv/Combined.csv} }
\end{document}